\documentclass{article}
\usepackage[utf8]{inputenc}
\usepackage[margin=.9in]{geometry}
\usepackage[dvipsnames]{xcolor}
\usepackage{graphicx,epstopdf}
\usepackage{amsmath,amssymb,amsthm,bm,bbm}
\usepackage{mathtools}
\usepackage{braket}
\usepackage{csquotes}
\usepackage{subcaption}
\usepackage{booktabs}
\usepackage{times}
\usepackage{multirow}
\usepackage{makecell}

\usepackage[numbers,sort&compress]{natbib}
\makeatletter \AtBeginDocument{%
\immediate\write\@auxout{\string\citation{REVTEX42Control}}%
\immediate\write\@auxout{\string\citation{apsrev42Control}}%
}
\makeatother

\usepackage[
colorlinks=true,
linkcolor=blue,
bookmarks=true,
breaklinks=true,
citecolor=blue,
urlcolor=blue
]{hyperref}

\usepackage[capitalize,nameinlink]{cleveref}
\AddToHook{cmd/appendix/before}{%
    \crefalias{section}{appendix}%
    \crefalias{subsection}{appendix}
}

\crefname{section}{Sec.}{Secs.}
\Crefname{section}{Sec.}{Secs.}

\usepackage[normalem]{ulem}
\usepackage{makecell}
\usepackage{authblk}

\theoremstyle{plain}

\newtheorem{lemma}{Lemma}
\newtheorem{corollary}{Corollary}

\theoremstyle{definition}

\newcommand{\abs}[1]{\left\lvert#1\right\rvert}
\newcommand{\norm}[1]{\left\lVert#1\right\rVert}
\newcommand{\wt}[1]{\widetilde{#1}}

\newcommand{\order}[1]{\mathcal{O}\left(#1\right)}
\DeclarePairedDelimiter{\ceil}{\lceil}{\rceil}
\DeclarePairedDelimiter{\floor}{\lfloor}{\rfloor}
\newcommand{\round}[1]{\ensuremath{\left\lfloor#1\right\rceil}}

\renewcommand{\Re}{\mathop{\mathrm{Re}}}
\renewcommand{\Im}{\mathop{\mathrm{Im}}}

\newcommand{\e}{\mathrm{e}}
\newcommand{\bme}{\bm{\mathrm{e}}}
\newcommand{\Laplacian}{\mathrm{L}}

\renewcommand{\i}{\mathrm{i}}
\renewcommand{\L}{\mathcal{L}}

\newcommand{\cost}{\mathcal{C}}
\newcommand{\gate}{\mathcal{G}}
\newcommand{\depth}{\mathcal{D}}
\newcommand{\A}{\mathcal{A}}
\newcommand{\precond}{\text{prec}}

\usepackage{dsfont}
\newcommand{\id}{\mathds{1}}

\let\originalleft\left
\let\originalright\right
\renewcommand{\left}{\mathopen{}\mathclose\bgroup\originalleft}
\renewcommand{\right}{\aftergroup\egroup\originalright}

\newcounter{todocounter}
\title{End-to-End Quantum Algorithm for Differential Equations:
Application to Simulating a Biomolecular System}

\author[1,2]{Mohsen Bagherimehrab\footnote{ Corresponding author: mohsen.bagherimehrab@gmail.com}
}
\author[2]{Daniyaal Farooqi}
\author[3,4]{Gavin K.~Brennen}
\author[5]{Barry C.~Sanders}
\author[1,2,6--9]{Al\'an Aspuru-Guzik}
\affil[1]{Department of Chemistry, University of Toronto, Toronto, Ontario, Canada}
\affil[2]{Department of Computer Science, University of Toronto, Toronto, Ontario, Canada}
\affil[3]{School of Mathematical and Physical Sciences, Macquarie University, 2109 NSW, Australia}
\affil[4]{BTQ Technologies, 16-104 555 Burrard Street, Vancouver, British Columbia, Canada V7X 1M8}
\affil[5]{Institute for Quantum Science and Technology,
University of Calgary, Calgary, Alberta, Canada}
\affil[6]{Vector Institute for Artificial Intelligence, Toronto, Ontario, Canada}
\affil[7]{Department of Chemical Engineering \& Applied Chemistry, University of Toronto, Toronto, Ontario, Canada}
\affil[8]{Department of Materials Science \& Engineering, University of Toronto, Toronto, Ontario, Canada}
\affil[9]{Canadian Institute for Advanced Research, Toronto, Ontario, Canada}

\begin{document}
\maketitle

\begin{abstract}
Quantum algorithms for differential equations have the potential for substantial algorithmic speedups over classical methods. However, they rely on certain assumptions and abstract oracle specifications that leave their practical feasibility and computational advantages for realistic applications unclear. Here we formalize the concept of an end-to-end quantum algorithm for differential equations as one that accepts classical inputs and returns classical outputs, and provide implementations for algorithmic primitives in constructing such algorithms. We present a tailored end-to-end quantum algorithm for the linearized Poisson-Boltzmann (LPB) equation, a partial differential equation (PDE) with variable coefficients widely used in simulating a biomolecular system. We explicitly implement all aspects of the algorithm, from preparing an initial state to generating a quantum solution and estimating properties of the system. In particular, we develop a new discrete representation for variable-coefficient PDEs on a grid that enables efficient implementation of the operator oracle, and build on a recent technique based on optimal wavelet preconditioning to generate a quantum solution with a cost independent of the condition number of the discretized operator. We show the algorithm can be implemented by quantum circuits of size polylogarithmic in both the number of grid points $N_g$ and the inverse of the desired accuracy~$\varepsilon$, with an overall gate cost polylogarithmic in $N_g$ and quasilinear in~$1/\varepsilon$. We also show the classical preprocessing to construct the circuits has polylogarithmic cost in~$N_g$ and~$1/\varepsilon$. Our work provides a concrete end-to-end quantum algorithm for a PDE of practical value, demonstrates an exponential improvement in scaling with~$N_g$ over classical methods, and provides a foundation for assessing its potential quantum advantage in practice.
\end{abstract}

\tableofcontents
\section{Introduction}
Partial differential equations~(PDEs) are the most universal descriptions of the natural physical laws and capture various problems in science and engineering.
This makes them an ideal case to probe the potential utility of quantum computers for various applications.
Many quantum algorithms have therefore been developed for differential equations over the years~\cite{Ber14,BCO+17,CL20,MP16,kro22,FLT23,JLY24,BNW+26,CLO21}. 
However, these algorithms mainly remain at an abstract level.
Specifically, quantum PDE solvers typically assume access to certain oracles that specify a discretized version of the PDE, such as an oracle for preparing an appropriate initial state and an oracle for encoding the discretized PDE operator. The cost of the algorithm is then assessed in terms of the number of queries to these oracles.
However, implementing these oracles from a classical specification of the PDE can introduce substantial gate costs that are not captured by the query cost.

Furthermore, unlike classical algorithms that explicitly return the solution vector of the discretized PDE, quantum PDE solvers generate a quantum state that encodes an approximation of the solution.
It is therefore assumed that physically relevant properties of the solution vector can be extracted from this state, although the cost of such post-processing of the quantum solution can be significant.
Moreover, quantum PDE solvers assume that the discretized PDE operator is well conditioned to achieve a polylogarithmic cost in the system size, whereas for typical PDEs, its condition number generally grows polynomially with the system size. 
These algorithms also commonly assume prior knowledge of additional problem-dependent parameters~\cite{JLP+25,ANB+22,CLO21,GJL22}, such as a sufficiently tight upper bound on the norm of the discretized operator, which may be computationally costly to obtain.
Considering these assumptions, the practical feasibility and the computational advantage of quantum algorithms for realistic PDEs remain unclear.

In this paper, we formalize the concept of an end-to-end quantum algorithm for differential equations, by which we mean a quantum algorithm that accepts classical inputs and returns classical outputs (\cref{fig:steps}).
To this end, we present a set of broadly applicable algorithmic primitives for realizing the main subroutines of an end-to-end quantum algorithm. In particular, we present explicit quantum circuits for loading a quantum state into a quantum register and estimating properties of a quantum solution, construct implementations for different encodings of differential operators with constant and variable coefficients, and rigorously analyze the computational costs of these primitives.
We then develop a tailored end-to-end quantum algorithm for the linearized Poisson-Boltzmann~(LPB) equation, a second-order PDE with variable coefficients that is widely used in computational biochemistry to model the electrostatic potential around a biomolecule immersed in an ionic aqueous solution. The electrostatic potential and related quantities obtained from this model are used to study how the surrounding solution affects the stability of biomolecules, such as proteins, and their binding to other molecules, such as potential drugs~\cite{AS11,FBM02}. 

A central contribution of our work is an efficient and fully explicit construction of an end-to-end quantum algorithm directly from a classical specification of the biomolecular system, without assuming access to abstract oracles or prior knowledge of additional problem-dependent parameters commonly taken as inputs to quantum algorithms. To achieve this, we exploit the mathematical and physical structure of the biomolecular system to develop suitable representations and quantum circuits tailored to the problem.
In particular, we introduce a new discrete representation of multidimensional differential operators with variable coefficients that enables us to construct the oracle for the discretized PDE operator from simple components, which can be efficiently implemented by directly exploiting the geometry and physical properties of the biomolecular system.
Similarly, we exploit the structure of the problem to develop an efficient quantum circuit for preparing an appropriate initial state from the classical specification, without explicitly loading the high-dimensional discretized data. We further identify physically relevant properties of the biomolecular
system and construct efficiently implementable observables whose expectation values with respect to a quantum encoding of the solution yield these~properties.

The differential operator of the LPB equation is a multidimensional generalization of a Sturm--Liouville operator with variable and bounded coefficients.
When discretized, operators of this form yield matrices whose condition numbers scale polynomially with the matrix size~$N_g$.
However, such operators can be optimally preconditioned using a simple diagonal wavelet preconditioner~\cite{BNW+26}, meaning that the condition number of the preconditioned matrix becomes independent of the matrix size.
We therefore build on the quantum PDE algorithm introduced in~\cite{BNW+26}, which uses wavelet preconditioning and a quantum linear-system solver to generate a distributed quantum encoding of the solution for the discretized PDE with a query cost independent of the condition number of the original discretized operator.

This favorable query cost, however, does not by itself guarantee an efficient quantum implementation, as the overall cost also depends on the normalization and gate cost of implementing a block encoding of the preconditioned matrix.
The algorithm of~\cite{BNW+26} assumes access to a suitably normalized block encoding of the discretized PDE operator, without accounting for the normalization and gate cost associated with constructing such an encoding from the classical specification of a particular PDE.
A key technical contribution of our work is that we construct an explicit and efficient block encoding of the wavelet-preconditioned matrix needed by this algorithm.
To this end, we exploit the multiscale structure of the wavelet-transformed matrices to achieve a block encoding with a logarithmic normalization and polylogarithmic gate cost in the matrix size.
Beyond constructing the required oracles, we make the quantum algorithm for generating the distributed quantum solution state explicit by implementing fixed-point amplitude amplification using a known lower bound on the success probability, and provide a detailed analysis of the computational cost and error of this procedure.
These constructions enable an efficient gate-level realization of the algorithm and allow the favorable scaling of its query cost to be translated into the overall gate-cost scaling.

In constructing our algorithm for the biomolecular system, we treat the number of grid points~$N_g$ and the desired algorithmic error~$\varepsilon$ as independent parameters.
Here,~$\varepsilon$ denotes the error tolerance in estimating a solution property relative to its ideal value for the discretized problem. It accounts for errors in generating a quantum solution state and estimating the expectation value of the corresponding observable, but not the error in spatial discretization of the continuous PDE.
This treatment is motivated by the physical structure of the biomolecular system. Specifically, in spatial discretization of such systems, the grid resolution is constrained by the molecular geometry, particularly by the positions and van~der~Waals radii of the atoms. Consequently, the grid resolution, characterized by the number of grid points~$N$ per spatial dimension, is not determined as a function of the desired accuracy~$\varepsilon$, as is done in~\cite{CLO21} to control the discretization error for second-order PDEs with constant coefficients.
In our application, $N$ can thus be much larger than~$1/\varepsilon$. We therefore treat~$N_g$ and~$\varepsilon$ as separate parameters and analyze the computational cost of our algorithm with respect to both parameters.

We rigorously analyze the gate cost and error of the complete end-to-end algorithm using several methods for property estimation. 
Our analysis reveals different tradeoffs between circuit size, ancilla count, and the number of circuit runs.
In particular, depending on the method used for property estimation, the end-to-end algorithm can be implemented using quantum circuits of size polylogarithmic in both~$N$ and~$1/\varepsilon$, but estimating a desired property to accuracy~$\varepsilon$ needs a number of circuit runs that scales quadratically with~$1/\varepsilon$.
Alternatively, methods using larger circuits can reduce the number of runs and achieve an overall gate cost quasilinear in~$1/\varepsilon$, with the circuit size remaining polylogarithmic in~$N$.
We also account for the classical preprocessing needed to construct the quantum circuits for the end-to-end algorithm directly from the classical specification of the biomolecular system. We show that, for a fixed number of atoms, the computational cost of this preprocessing is at most polylogarithmic in both~$N$ and~$1/\varepsilon$, and therefore does not eliminate the favorable scaling of the gate cost of the quantum algorithm.

Our results thus demonstrate an exponential improvement in scaling with the number of grid points compared with known classical algorithms that solve the same discretized problem. By providing explicit circuit constructions and accounting for their costs, our work establishes a concrete realization of an end-to-end quantum algorithm for a realistic PDE of practical value and provides a foundation for assessing its potential quantum advantage in practice.

\paragraph{Organization.}
The remainder of the paper organized as follows.
We begin~\cref{sec:algorithm} by presenting a general framework for an end-to-end quantum algorithm for differential equations. We first discuss the classical specification of a general linear PDE and introduce a new discrete representation for variable-coefficient PDEs that is suitable for efficient implementation of various oracles of the discretized operator. 
We then present several algorithmic primitives, which have applications beyond the scope of this work, and provide a detailed analysis of their computational costs.
In particular, we present an explicit approach for loading a quantum state onto a quantum register, construct various implementations for different oracular encodings of matrices obtained from discretizing a differential equation, and discuss several methods for estimating poperies of a quantum state.

In~\cref{sec:tailored}, we present our tailored end-to-end quantum algorithm for the LPB equation used in simulating a biomolecular system. We describe this system and develop tailored implementations for preparing the initial state, generating a quantum solution, and estimating properties of interest.
We analyze the computational cost of these these implementations and provide a detailed cost analysis of the complete end-to-end algorithm.
We conclude with a discussion of our results in~\cref{sec:dicuttion}.
Technical proofs and detailed analyses are
provided in the appendices.

\section{End-to-end quantum algorithm for differential equations}
\label{sec:algorithm}
\begin{figure}[ht]
    \centering
    \includegraphics[width=.6\linewidth]{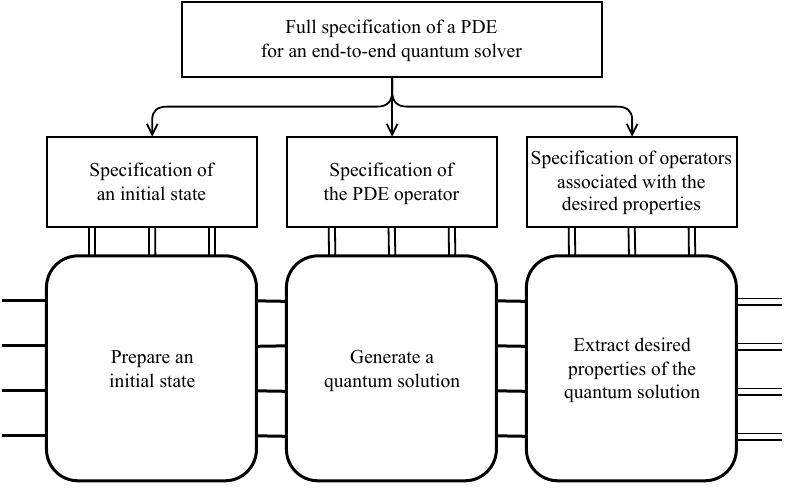}
    \caption{
    Common steps of an end-to-end quantum algorithm.
    The algorithm accepts classical inputs and returns classical outputs, denoted by double lines.
    In the first step, an initial state is prepared. The initial state is then evolved to a final state.
    The final step is to perform a measurement to obtain an expectation value of an observable with respect to the final state. These steps may need to be repeated to obtain multiple samples for more accurate computation of the desired expectation value.}
    \label{fig:steps}
\end{figure}

An end-to-end quantum algorithm for a PDE accepts classical inputs and returns classical outputs.
The classical inputs fully specify the PDE together with the desired properties of its solution, and the classical outputs are estimates of those properties (see \cref{fig:steps}).
In \cref{subsec:specification}, we elaborate the full specification of a general linear PDE and the solution properties that serve as classical inputs for an end-to-end quantum PDE solver.
We present a new representation for PDEs with variable coefficients in \cref{subsec:representation} and discuss how this representation provides a simple and efficient implementation for the oracles of the PDE operator.
We then use this representation to implement various oracle encoding of operators in~\cref{subsec:oracle_implementation} and analyze the cost of these implementations.
We conclude in~\cref{subsec:property_estimation} by presenting several approaches for extracting properties of the quantum solution state with a rigorous cost analysis.

\subsection{Specification of the discretized PDE}
\label{subsec:specification}
When discretized, a differential equation together with its boundary conditions, i.e., a boundary-value problem, becomes a finite-dimensional algebraic problem.
For a linear PDE $\A u(x)=b(x)$, the discretization leads to a system of linear equations $A\bm{u}=\bm{b}$.
The vector $\bm{u}$ contains the discrete degrees of freedom used to represent the approximate solution. These degrees of freedom could be, for example, the values of $u(x)$ at grid points in the finite-difference method, or the unknown expansion coefficients in the spectral or wavelet method, where the solution is expressed as a linear combination of basis functions of a finite-dimensional space.
The matrix $A$ represents the action of the differential operator $\A$ on this space for the chosen discretization, and the vector $\bm{b}$ represents the corresponding discretization of $b(x)$.
The boundary conditions are encoded in the linear system through appropriate modifications to~$A$ and~$\bm{b}$, for example by adding boundary rows in $A$ with the corresponding boundary value in $\bm{b}$, or by wrap-around entries in~$A$ for periodic boundary conditions.

The discretized PDE problem is therefore fully specified by the matrix $A$ and the vector $\bm{b}$.
Quantum algorithms for PDEs often assume quantum encodings of $A$ and $\bm{b}$ are given as input.
For the vector $\bm{b}$, the quantum encoding is often a procedure that prepares a quantum state that encodes elements of $\bm{b}$ on its amplitudes.
More specifically, it is assumed that the quantum algorithm takes as an input a procedure $\mathcal{P}_b$ that prepares the initial state $\ket{\bm{b}}:=\sum_j \bar{b}_j\ket{j}$ with the normalized coefficients $\bar{b}_j:=b_j/\norm{\bm{b}}$. That is to say,~$\mathcal{P}_b$ maps the all-zero state to
the state $\ket{\bm{b}}$.
The underlying assumption for the procedure~$\mathcal{P}_b$ is that it can be constructed by access to a quantum oracle $O_b\ket{j}\ket{0}\mapsto\ket{j}\ket{\bar{b}_j}$ that returns coefficients~$\bar{b}_j$ of the initial state $\ket{\bm{b}}$. 
Having an implementation for this oracle, $\mathcal{P}_b$ can be constructed, for example, using the Grover method~\cite{GR02} for quantum-state preparation or by the inequality-test method~\cite{SLSB19}, which significantly reduces the quantum arithmetic cost of the Grover method.

The oracle $O_b$ or the procedure $\mathcal{P}_b$ can be regarded as a quantum specification of $\bm{b}$. For an end-to-end algorithm, a classical specification of $\bm{b}$ is needed as input for implementing $O_b$ and $\mathcal{P}_b$.
In general, $\bm{b}$ needs to be sparse or structured to efficiently prepare the state $\ket{\bm{b}}$, as otherwise the quantum speedup of a full quantum PDE solver may be eliminated by the cost of the initial-state preparation.
The classical specification could be the nonzero entries of~$\bm{b}$ if it is sparse or an efficient description of the structure of~$\bm{b}$. For example, if elements of $\bm{b}$ follow a Gaussian distribution, then $\bm{b}$ is fully specified by the mean and standard deviation of the Gaussian distribution, and these parameters will be the classical specification of~$\bm{b}$. 
In \cref{sec:tailored}, we discuss the classical inputs that specify the initial state of the LPB equation and provide a tailored implementation for preparing the initial state.

The common quantum oracle encodings for the matrix $A$ are the sparse encoding and the block encoding.
The sparse encoding is specified by two quantum oracles that yield the location and the value of nonzero elements in $A$. Specifically, let $r\left(\ell,j\right)$ be a function that returns the row index of the $\ell$th nonzero element in the $j$th column of~$A$.
The sparse encoding of $A$ is then described by two oracles
\begin{equation}
\label{eq:sparse_oracles}
    \texttt{LOC} \ket{\ell} \ket{j}
    =\ket{\ell} \ket{r(\ell,j)},\enspace
    \texttt{VAL} \ket{0} \ket{\ell}
    \ket{j}=
    R(\theta(\ell,j))\ket{0}
    \ket{\ell} \ket{j}
\end{equation}
where $R(\theta)$ is the single-qubit rotation gate around the $y$-axis with angle~$\theta$,
$R(\theta)\ket{0}=\cos(\theta/2)\ket{0}+\sin(\theta/2)\ket{1}$, and where $\cos(\theta(\ell,j)/2)=A_{r(\ell,j),j}/\norm{A}_\text{max}$ with $\norm{\cdot}_\text{max}$ the max norm of a matrix. The location oracle $\texttt{LOC}$ yields the location (i.e., the row index) of the $\ell$th nonzero element in the $j$th column of~$A$ and the value oracle $\texttt{VAL}$ encodes the value of the nonzero element in the rotation angle of the single-qubit~register.
The sparse encoding strongly depends on the function $r\left(\ell,j\right)$, which itself depends on the matrix structure.

In the block encoding, the nonunitary matrix $A$ is embedded as a block of a larger unitary matrix as 
\begin{equation}
    U_A=\left[\begin{matrix}A/\alpha&\ast\\ \ast&\ast\\ \end{matrix}\right],
\end{equation}
where $\alpha\geq \norm{A}$, and the blocks denoted by $\ast$ could be any matrix as long as $U_A$ is a unitary matrix. 
The block encoding does not need to be exact and can be accurate up to an error $\varepsilon>0$. More specifically, an $(\alpha, a, \varepsilon)$-block-encoding of an $n$-qubit matrix $A$ is an $(n+a)$-qubit unitary $U_A$ such that 
$\norm{A-\alpha (\bra{0^a}\otimes \id_n)U_A(\ket{0^a}\otimes\id_n)}\leq \varepsilon$.

The sparse and block encodings described above provide a quantum specification of the discretized PDE matrix~$A$.
In practice, however, implementing these encodings requires a suitable classical specification of~$A$.
The classical specification of $A$ is the classical data needed to implement these quantum encodings.
The form of this classical specification depends on the structure of~$A$. 
For a banded or sparse matrix with a fixed stencil pattern, it can be the list of distinct nonzero entries and their locations, or an efficiently computable function for generating them. If~$A$ is represented as a linear combination of Pauli operators, then the relevant classical data is a description of the Pauli strings and their coefficients. More generally, the classical specification should describe~$A$ in a form that allows efficient implementation of a quantum encoding of~$A$. Otherwise, the overhead of constructing the quantum encoding could dominate the overall cost of a quantum PDE solver.

The last step of an end-to-end quantum PDE solver is to extract physically or computationally relevant properties of the solution vector~$\bm{u}$.
In general, a solution property can be expressed as the expectation value $\braket{\bm{u}|M|\bm{u}}$ of a Hermitian operator $M$ with respect to the quantum state $\ket{\bm{u}}$ that encodes~$\bm{u}$. 
The operator $M$ encodes the desired property and it is often a physical observable of the system under study. More generally, the desired property can be a function of the solution vector, for example a weighted average of the solution vector. In such cases, the measurement operator~$M$ is constructed so that its expectation value with respect to~$\ket{\bm{u}}$ yields the desired property.
Quantum algorithms for estimating expectation values are often based on a quantum encoding of the measurement operator~$M$. Therefore, similar to the matrix~$A$, a suitable classical specification of~$M$ is needed to construct this quantum encoding efficiently.

\subsection{Loading an initial state onto a quantum register: the binary-tree approach}
\label{subsec:loading}
\begin{figure}[ht]
    \centering
    \includegraphics[width=.86\linewidth]{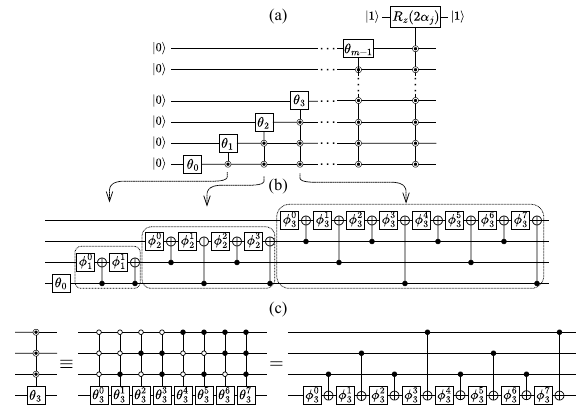} 
    \caption{
    Quantum circuit for loading a general $m$-qubit initial state.
    (a)~The circuit first loads the state with magnitude of the amplitudes using uniformly controlled rotations. The phases of the amplitudes are then added using a uniformly controlled $z$-rotation with angle $2\alpha_j$ on an ancilla qubit initiated at~$\ket{1}$.
    (b)~The circuit for loading the four-qubit magnitude state using rotation and \texttt{CNOT} gates; all rotation angles are classically precomputable.
    (c)~An example of a uniformly controlled rotation implemented using an alternating sequence of a single-qubit rotation and \texttt{CNOT} gates.}
    \label{fig:state_loading}
\end{figure}

In this section, we present a procedure for loading a general $m$-qubit initial state onto an $m$-qubit quantum register using only a single ancilla qubit.
The approach we present here is suitable when the amplitudes of the quantum state are given explicitly as classical data, as in our application later in~\cref{subsec:initial_state}, not implicitly by a function or an oracle. 
Specifically, we consider a general $m$-qubit~state
\begin{equation}
    \ket{\psi}:=\sum_{j=0}^{2^m-1} a_j \ket{j},
    \enspace
    \sum_j \abs{a_j}^2=1,
\end{equation}
whose amplitudes~$\{a_j\}$ are given explicitly as an array of numbers.
The binary-tree approach we present here provides a quantum circuit for loading this state using only CNOT and single-qubit rotation gates with classically precomputed rotation angles~(\cref{fig:state_loading}). 
First, we construct the circuit and then show the number of gates in the circuit scales as $\Theta(2^m)$. We also present a procedure for classical pre-computation of the rotation angles used in the circuit with a cost of~$\Theta(2^m)$.
This gate cost is expected for loading an arbitrary quantum state; note that just reading the input amplitudes already costs~$\Omega(2^m)$. However, the construction here gives an explicit quantum circuit and a recipe for computing the angles, and is advantageous for small register size~$m$, where $m\in \order{1}$ or scales logarithmically with some input parameter.

Writing each amplitude as $a_j=\abs{a_j}\exp{(\i\alpha_j)}$, we load the state in two steps. In the first step, we load the magnitude state whose amplitudes are $\{\abs{a_j}\}$. In the second step, we incorporate the phase factors by implementing an operation that maps each basis state as~$\ket{j}\mapsto\exp{(\i\alpha_j)}\ket{j}$, thereby loading the phases of the original amplitudes.

The magnitude-loading step is based on a binary tree of probability weights.
Let us define the probabilities $p_j:=\abs{a_j}^2$ and write the computational basis as $\ket{j}=\ket{j_{m-1}\ldots j_1j_0}$, where qubits are ordered from right to left, $j_0$ is the least significant bit in binary representation of~$j$, and~$\ket{j_\ell}$ is the state of the $\ell$th qubit $q_\ell$.
Consider a binary tree with $m+1$ levels $\ell=0,\dots,m$, where the root node of the tree is at the top level~$(\ell=0)$ and the leaf nodes of the tree are at the bottom level~$(\ell=m)$.
At level~$\ell<m$, we set the state of qubit~$q_{\ell}$ by applying a rotation gate.
Let $x=x_{\ell-1}\ldots x_0$ be the bitstring of the rightmost $\ell$ qubits $q_{\ell-1}\ldots q_0$ already prepared before level~$\ell$; note that no qubit is prepared at~$\ell=0$.
We use $x=0,\ldots,2^{\ell}-1$ as the index of the tree nodes at level~$\ell$ and assign the probability weight
\begin{equation}
    w(\ell, x) = \sum_{j:\,\text{right $\ell$ bits of $j=x$}} p_j
    = \sum_{j:j\bmod 2^\ell=x} p_j
\end{equation}
for the node~$(\ell,x)$. Here $w(\ell,x)$ is the total probability weight of all basis states~$\ket{j}$ whose lower~$\ell$ bits~$j_{\ell-1}\cdots j_0$ are equal to~$x$.
For instance, we have nodes $x=x_0\in\{0,1\}$ at level $\ell=1$ and
\begin{equation}
    w(1,0)=\sum_{j:j_0=0} p_j = \sum_{j\, \text{even}} p_j, \enspace
    w(1,1)=\sum_{j:j_0=1} p_j = \sum_{j\, \text{odd}} p_j,
\end{equation}
where $w(1,x_0)$ is the total probability weight of all basis states~$\ket{j}$ with $j_0=x_0$.
Nodes at level~$\ell=2$ are $x=x_1x_0\in\{00,01,10,11\}$, and the probability weights are
\begin{equation}
    w(2,0)=\sum_{j:j_1j_0=00} p_j, \enspace w(2,1)=\sum_{j:j_1j_0=01} p_j,
\end{equation}
and so on. Note that the root node has weight~$w(0,0)=\sum_jp_j=1$, and the leaf nodes have weights equal to the target probabilities, i.e., we have~$w(m,j)=p_j$.

For each node~$x$ at level~$\ell\geq1$, we use the bitstring~$x=x_{\ell-1}\cdots x_0$ of the node to define its two children nodes located at level~$\ell+1$ as~$0x:= \text{$x$ preceded by $0$}$ and~$1x:= \text{$x$ preceded by $1$}$, where~$0x$ and $1x$ are $(\ell+1)$-bit numbers. The indices of these nodes in decimal are~$x$ and~$x+2^\ell$, and we have the recursive formula
\begin{equation}
\label{eq:recursive_formula}
    w(\ell,x) = w(\ell+1,x) + w(\ell+1,x+2^\ell)
\end{equation}
for the probability weight of the nodes, as the children nodes~$(\ell+1,x)$ and~$(\ell+1,x+2^\ell)$ partition the probability weight of the parent node~$(\ell,x)$ based on the next bit~$x_{\ell}$ (one bit to the left).

Using the probability weights of the nodes, let us now define the rotation angle
\begin{equation}
\label{eq:rotation_angle}
    \theta(\ell,x) = 2 \arccos\sqrt{\frac{w(\ell+1,x)}{w(\ell,x)}}
\end{equation}
for all $\ell<m$ and $x=0,\ldots,2^{\ell}-1$.
We set the qubit~$q_\ell$ using a rotation gate that is coherently controlled by previous qubits.
Specifically, we apply the $y$-rotation gate $R(\theta(\ell,x))$ on~$q_\ell$ controlled on the state~$\ket{x}$ of the rightmost~$\ell$~qubits (note that the gate on~$q_0$ is uncontrolled).
Let
\begin{equation}
    \ket{\psi(\ell)}:= \sum_{x\in\{0,1\}^\ell} \sqrt{w(\ell,x)} \ket{0\cdots0}_{m-\ell} \ket{x}_{\ell}
\end{equation}
be the state of the $m$-qubit register at level~$\ell$ before applying the rotation gate.
Here~$\sqrt{w(\ell,x)}$ is the amplitude of the basis state~$\ket{0\cdots0\,x}$ whose rightmost~$\ell$ bits are~$x$ and whose leftmost $m-\ell$ bits are all zero.
After the rotation gate, the state of qubit~$q_\ell$ is transformed as
\begin{equation}
    \ket{0}\ket{x}_\ell
    \mapsto \left(\cos(\theta(\ell,x)/2)\ket{0}+ \sin(\theta(\ell,x)/2)\ket{1}\right) \ket{x}_\ell
\end{equation}
and the amplitude of the basis state is split as
\begin{equation}
    \sqrt{w(\ell,x)} \ket{0\cdots0}_{m-\ell}\ket{x}_\ell \mapsto
    \sum_{x_\ell\in\{0,1\}}
    \sqrt{w(\ell+1,x+x_\ell2^\ell)}\ket{0\cdots0}_{m-\ell-1}\ket{x_\ell x}_{\ell+1}
\end{equation}
by the definition of $\theta(\ell,x)$.
Therefore, the state~$\ket{\psi(\ell+1)}$ is obtained after the rotation gate at level~$\ell$.
By induction, we have the state~$\ket{\psi(m)}$ after the rotation at level~$m-1$, where each basis state $\ket{j}$ with the full bitstring $j=j_{m-1}\ldots j_0$ will have the amplitude $\sqrt{w(m,j)}=\sqrt{p_j}=\abs{a_j}$.

Loading the magnitude state is therefore achieved by applying a sequence of uniformly controlled rotations as shown in~\cref{fig:state_loading}. Specifically, we apply the $y$-rotation gate $R(\theta(\ell,x))$ at level $\ell=0,\ldots,m-1$ controlled on the state~$\ket{x}$ of previously prepared~$\ell$ qubits.
Each uniformly controlled rotation gate with~$\ell$ control qubits can be implemented by~$\Theta(2^\ell)$ single-qubit rotations and \texttt{CNOT} gates~\cite{MVB+04}, so the total gate cost of loading the magnitude state is~$\sum_\ell\Theta(2^\ell)=\Theta(2^m)$.
This described approach is closely related to the probability-splitting idea used in the Grover-Rudolph method for state preparation~\cite{GR02}. However, the approach presented here gives an explicit and implementable quantum circuit using only \texttt{CNOT}s and single-qubit rotation gates.

\paragraph{Classical precomputation of the angles.}
We use the binary tree of probability weights to precompute the rotation angles classically before running the quantum circuit.
To this end, we start from the leaves of the tree with weights~$w(m,j)=p_j$ and recursively compute the weights of the internal nodes from leaves upward to the tree root.
Specifically, we use~\cref{eq:recursive_formula} to compute~$w(\ell,x)$ for $\ell=m-1,\ldots,0$ and each~$x=0,\ldots,2^\ell-1$.
As the tree has~$2^m-1$ internal nodes, computing all weights costs~$\Theta(2^m)$.
Using the weights, we then compute the rotation angle for each node by~\cref{eq:rotation_angle}.
Therefore, the cost of computing the angles of the uniformly controlled rotation gates is~$\Theta(2^m)$.

To decompose each uniformly controlled rotation gate into single-qubit rotations and \texttt{CNOT} gates, an additional classical computation is needed.
A uniformly controlled rotation with~$\ell$ control qubits has~$2^\ell$ conditional rotation angles.
Following the construction in Ref.~\cite{MVB+04}, the angles of the single-qubit rotations in the decomposition are obtained by applying a Walsh--Hadamard transform to the vector of the conditional rotation angles.
Using the fast Walsh--Hadamard transform, this operation costs~$\Theta(\ell2^\ell)$ for a uniformly controlled rotation with~$\ell$ control qubits. Summing over all levels $(\ell=0,\ldots,m-1)$ we obtain $\sum_\ell \Theta(\ell2^\ell)=\Theta(m2^m)$ for the cost of computing the angles.
The overall cost of the classical precomputation is therefore~$\Theta(m2^m)$.

\paragraph{Phase factors.}
We have so far described loading the magnitude state. To add the phase factors, we use an ancilla qubit initialized in~$\ket{1}$ state and apply the map $\ket{1}\ket{j}\mapsto \exp(\i\alpha_j)\ket{1}\ket{j}$.
This map can be achieved by applying the $z$-rotation gate $R_z(2\alpha_j):=\operatorname{diag}(\e^{-\i\alpha_j},\e^{\i\alpha_j})$ on the ancilla qubit, uniformly controlled by the~$m$-qubit register being in the state~$\ket{j}$.
As before, this uniformly controlled rotation can be decomposed into~$\Theta(2^m)$ single-qubit rotations and \texttt{CNOT} gates~\cite{MVB+04}.
Therefore, the gate cost for adding the phase factors to the magnitude state scales as~$\Theta(2^m)$.
The additional classical precomputation of the rotation angles used in implementing the phase factors has the same scaling~$\Theta(m2^m)$ as in the classical precomputation of the magnitude-loading step.

Finally, we note that if the amplitudes~$a_j$ are signed real numbers, the phase factors are~$\pm1$. In this case, we can apply the~$Z$ gate on the ancilla qubit, controlled on basis state~$\ket{j}$ for which $a_j<0$. If only a small fraction of the amplitudes are negative, direct application of the multi-controlled $Z$ gates may be simpler than implementing the uniformly controlled rotation.

\subsection{Discrete representation for PDEs with variable coefficients}
\label{subsec:representation}
We now provide a new discrete representation for variable-coefficient PDEs.
The discrete representation we discuss here is suitable for implementing the oracles that specify the discretized differential operator of the PDE.
We consider PDEs of the form
\begin{equation}
\label{eq:variable_coeff_PDE}
    \L u(x) = -\nabla\cdot\left(\varepsilon(x)\nabla u(x)\right) + k(x) u(x) = b(x) \quad \forall x\in \Omega=(0,L)^d
\end{equation}
where $\varepsilon(x)$ and $k(x)$ are variable coefficients of the differential operator $\L$.
For convenience, we take the domain~$\Omega$ as a hypercube with side length~$L$ and consider periodic boundary conditions. However, the analysis below can be generalized to a hyperrectangle domain and to Dirichlet or Neumann boundary conditions.

We discretize the domain on a staggered uniform grid with periodic boundaries.
This grid has two types of uniformly spaced points: node points and face points (see \cref{fig:lattice}).
We use $N$ node points with spacing $h=L/N$ for each dimension.
The node points are used to store the value of the solution~$u(x)$ and any variable coefficients that are multiplied to~$u(x)$.
The face points are located at the midpoint between two node points and are used to store the variable coefficients that are multiplied to~$\nabla u(x)$.

We begin with the 1D case and then extend the analysis to higher dimensions.
In 1D, the node points are $x_j=jh$ and the face points are $x_{j+1/2}$ $\forall j\in\{0,\ldots,N-1\}$.
The solution~$u(x)$ and the coefficient~$k(x)$ on node points are $u_j=u(x_j)$ and $k_j=k(x_j)$, respectively, and the coefficient $\varepsilon(x)$ on face points are $\varepsilon_{j+1/2}:=\varepsilon(x_{j+1/2})$.
The discrete derivative at the face point $j+1/2$ is
\begin{equation}
    (\nabla u)_{j+1/2} \approx (D u)_{j+1/2} = \frac1h \left(u_{j+1}-u_j\right).
\end{equation}
From this expression, the discrete derivative matrix on the grid whit periodic boundaries is
\begin{equation}
\label{eq:derivative_matrix}
    D=\frac1h(S^\dagger-\id),
    \enspace
    S\ket{j}:=\ket{j+1 \bmod N},
\end{equation}
where~$S$ is the cyclic shift matrix that shifts the basis vector $\ket{j}$ as above.
Note that the discrete derivative is second-order accurate at face points.
To see this, we use the Taylor expansion of $u(x)$ around the face point $x_m$ with $m=j+1/2$ as
\begin{align}
    u_{j+1} &= u(x_m+h/2) = u(x_m) + (h/2)u'(x_m) + (h^2/8)u''(x_m) + \order{h^3}\\
    u_j &= u(x_m-h/2) = u(x_m) - (h/2)u'(x_m) + (h^2/8)u''(x_m) + \order{h^3},
\end{align}
so the discrete derivative is second-order accurate at face points; that is, $(Du)_{j+1/2}=u'(x_{j+1/2})+\order{h^2}$.

As~$D$ is the discrete version of $\nabla$, the discrete version of $-\nabla$ is simply $D^T$. The discrete version of the differential operator~$\L$ in 1D is then the matrix
\begin{equation}
\label{eq:A1D}
    A := D^TED + K,
\end{equation}
with the diagonal matrices
\begin{equation}
    E=\text{diag}(\varepsilon_{1/2}, \ldots, \varepsilon_{N-1/2}),
    \enspace
    K=\text{diag}(k_0,\ldots,k_{N-1}),
\end{equation}
where~$E$ encodes the value of $\varepsilon(x)$ on the face points and~$K$ encodes the value of $k(x)$ on the node points.
Specifically, for each node point $j$ we have
\begin{equation}
    \left(\L u(x)\right)_j \approx \left(A u\right)_j = \frac1{h^2}
    \left(
    -\varepsilon_{j-1/2} u_{j-1}
    +\left(\varepsilon_{j-1/2}+\varepsilon_{j+1/2}\right)u_j
    -\varepsilon_{j+1/2} u_{j+1}
    \right) + k_j u_j.
\end{equation}
Observe that when $\varepsilon(x) = c$ is constant, we have $A=-(c/h^2)\textsc{tridiag}(1,-2,1) + K$, where $\textsc{tridiag}(1,-2,1)$ is the familiar tridiagonal matrix form of the 1D Laplacian in the central finite-difference method.

For the higher-dimensional problems, we use row-major ordering to label the Cartesian coordinates of the grid points by a 1D index.
In 2D, this means we label the node $(i,j)$ by the 1D index $\ell=i+Nj$, which corresponds to ordering grid points from left to right within each row, and rows from bottom to top.
With this ordering, the discrete 2D Laplacian has the known form $L_\text{2D} = \id_N \otimes L_\text{1D} + L_\text{1D}\otimes \id_N$.
For our variable-coefficient operator~$\L$ in 2D, the discretized operator is
\begin{equation}
\label{eq:A2D}
    A:=D^T_x E_x D_x + D^T_yE_yD_y + K
\end{equation}
with
\begin{equation}
\label{eq:Dxy}
    D_x = \id_N \otimes D,
    \enspace
    D_y = D\otimes \id_N.
\end{equation}
Here $E_x$ and $E_y$ are diagonal matrices that encode the value of $\varepsilon(x,y)$ on face points and $K$ is the diagonal matrix that encodes the value of $k(x,y)$ on node points; i.e.,
\begin{equation}
    (E_x)_{\ell\ell} = \varepsilon_{i+1/2,j},
    \enspace
    (E_y)_{\ell\ell} = \varepsilon_{i,j+1/2},
    \enspace
    K_{\ell\ell} = k_{i,j}
    \enspace\text{for}\enspace
    \ell = i+ Nj.
\end{equation}
For the 3D problem, we label the node $(i,j,k)$ by the 1D index $\ell=i+N(j+Nk)$.
The discretized $\L$ in 3D becomes
\begin{equation}
\label{eq:A3D}
    A :=
    D^{T}_x E_x D_x +
    D^{T}_y E_y D_y +
    D^{T}_z E_z D_z + K,
\end{equation}
where
\begin{equation}
\label{eq:Dxyz}
    D_x = \id_N \otimes \id_N \otimes D,
    \enspace
    D_y = \id_N \otimes D \otimes \id_N,
    \enspace
    D_z = D\otimes \id_N \otimes \id_N,
\end{equation}
and the diagonal matrices are 
\begin{equation}
\label{eq:diagonal_encodings}
    (E_x)_{\ell\ell} = \varepsilon_{i+1/2,j,k},
    \enspace
    (E_y)_{\ell\ell} = \varepsilon_{i,j+1/2,k},
    \enspace
    (E_z)_{\ell\ell} = \varepsilon_{i,j,k+1/2},
    \enspace
    K_{\ell\ell} = k_{i,j,k}
    \enspace\text{for}\enspace
    \ell=i+N(j+Nk).
\end{equation}
The discrete representation of the $d$-dimensional problem with $d>3$ follows similarly.

\subsection{Oracle implementations for differential operators}
\label{subsec:oracle_implementation}

Here we present several approaches for implementing sparse and block encodings of matrices obtained from discretizing differential operators and analyze their gate costs and ancilla qubit requirements.
We begin with implementing sparse encoding of Toeplitz matrices, which frequently arise in the discretization of differential operators with constant coefficients.
We then present several approaches for constructing a block encoding of these matrices.
Finally, we utilize our discrete representation of variable-coefficient differentiation operators and construct block encodings for this class of operators.

\subsubsection{Sparse encoding for Toeplitz matrices}
\label{subsec:sparse}
A Toeplitz matrix is a matrix whose entries are constant along each diagonal.
Formally, an $N\times N$ matrix $A=[A_{ij}]$ is Toeplitz if $A_{ij}=A_{i-1,j-1}$, or equivalently if $A_{ij}=a_{i-j}$ for some sequence $\{a_m\}$ of constant values.
We consider Toeplitz matrices of the form

\begin{equation}
A =
\begin{pmatrix}
a_0      & a_{-1}      & \cdots & a_{-w}      &        &        &        & a_{w-N}   & \cdots & a_{1-N} \\
a_1   & a_0      & a_{-1} &          & \ddots &        &        &           & \ddots & \vdots  \\
\vdots   & \ddots   & \ddots & \ddots   &        & \ddots &        &           &        & a_{w-N} \\
a_w   &          & \ddots & a_0      & \ddots &        & a_{-w}    &           &        &  \\
         & \ddots   &        & \ddots   & \ddots & \ddots &        & \ddots    &        &          \\
         &          & \ddots &          & \ddots & \ddots & \ddots &        & \ddots &        \\
         &          &        & a_w   &        & \ddots & a_0    & \ddots    &        & a_{-w}      \\
a_{N-w} &          &        &          & \ddots &        & \ddots & \ddots    & \ddots & \vdots   \\
\vdots   & \ddots   &        &          &        & \ddots &        & a_1    & a_0    & a_{-1}      \\
a_{N-1} & \cdots   & a_{N-w} &        &        &        & a_w & \cdots    & a_1 & a_0
\end{pmatrix},
\end{equation}
where $2w+1 \leq N$. For cases where~$w$ is a constant, we have a banded Toeplitz matrix, 
which arises naturally in the discretization of differential operators.
The sparse encoding we construct here applies to any valid~$w$ and is not limited to banded matrices. We refer to the bands on the top-right and bottom-left corners of $A$ as the boundary bands, and the bands in the middle as the interior bands.

\begin{figure}[ht]
    \centering
    \includegraphics[width=.86\linewidth]{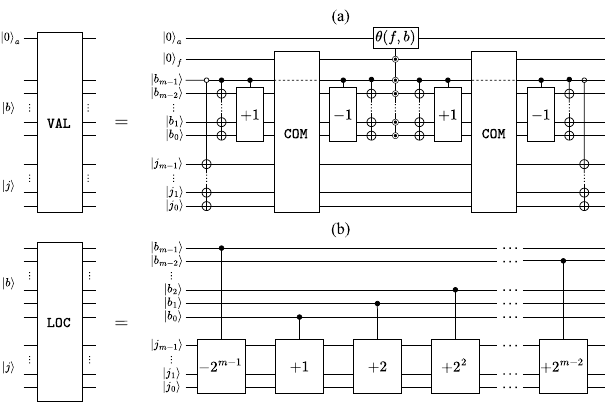}
    \caption{Quantum circuits for implementing the value and location oracles in the sparse encoding of Toeplitz matrices.
    }
    \label{fig:sparse_encoding}
\end{figure}

We exploit the structure of this matrix to construct the location and value oracles of the sparse encoding in \cref{eq:sparse_oracles}. Note that the location of a nonzero entry is independent of its value.
We utilize this feature in constructing the location oracle. Specifically, to construct this oracle, we treat the matrix as if it is a periodic matrix (note that we are not assuming the matrix is periodic) and use the signed integer $b\in\{-w,\ldots,0,\ldots,w\}$ as the index for bands of the matrix with cyclic location.
We use $b=0$ for the main diagonal with the constant $a_0$; $b>0$ for the $b$th band below the main diagonal with the constant $a_b$ and the band in the top-right corner with the constant $a_{b-N}$; and $b<0$ for the $\abs{b}$th band above the main diagonal with the constant $a_{-\abs{b}}$ and the band in the bottom-left corner with the constant $a_{N-\abs{b}}$.
With this cyclic indexing for bands, the row index $r(b,j)$ (i.e., the location) of the nonzero value in a given column index $j$ and band index $b$ is
\begin{equation}
r(b,j)=
    \begin{cases}
        j+b & 0 \leq j+b <N\qquad\;\,
        \text{(interior bands)}\\
        j+b-N & b>0\,\text{and}\ j+b\geq N\,
        \text{(top-right bands)}\\
        j-\abs{b}+N  & b<0\,\text{and}\ j+b< 0\;\;
        \text{(bottom-left bands)},
    \end{cases}
\end{equation}
which simplifies to $r(b,j)=(j+b) \bmod N$.
The location oracle is then
\begin{equation}
\label{eq:O_loc}
    \texttt{LOC} \ket{b}\ket{j} = \ket{b} \ket{(j+b)\bmod N},
\end{equation}
where the band register with the state $\ket{b}$ has $m=\ceil{\log_2(2w+1)}$ qubits ($\forall w\geq 1$) and encodes the signed integer~$b$, and the data register with the state $\ket{j}$ has $n$ qubits ($N=2^n$).
This oracle then can be implemented by applying modular addition of two registers.
However, we use a bit-weighted method to implement this addition for banded matrices (where~$w$ is a constant number and so $m\ll n$) that typically arise in the discretization. 
In this method, we use the two's complement binary representation\footnote{We use two's complement instead of sign-magnitude representation to reduce the number of control bits in implementing the value oracle.}
\begin{equation}
    b=\sum_{k=0}^{m-2} 2^k b_k - 2^{m-1}b_{m-1}
\end{equation}
of the $m$-bit number~$b$ to implement the location oracle as a product of bit-controlled constant additions.
Specifically, let
\begin{equation}
\label{eq:cadd}
    \texttt{CADD}(a) \ket{b_k} \ket{j}:= \ket{b_k} \ket{j+b_k a \bmod N}
\end{equation}
be the controlled-addition operation that adds the constant $a$ to $j$ controlled on the bit value $b_k$, then
\begin{equation}
    \texttt{LOC} =
    \left(\prod_{k=0}^{m-2}
    \texttt{CADD}\left(+2^k\right)
    \right) \texttt{CADD}\left(-2^{m-1}\right),
\end{equation}
from which we obtain the quantum circuit in~\cref{fig:sparse_encoding}(b) for implementing the location oracle.

To implement the value oracle, note that the nonzero values of~$A$ depend only on the band index~$b$. However, as we used a cyclic indexing for bands, care must be taken for the boundary bands. To this end, we use the function
\begin{equation}
    f(b,j) =
    \begin{cases}
        0, & 0\leq j+b<N\\
        1, & b>0\ \text{and}\ \bar{j}<b\\
        1, & b<0\ \text{and}\ j<\abs{b}
    \end{cases}
\end{equation}
as a flag to indicate if the band~$b$ in column $j$ is an interior or boundary band.
Here $\bar{j}=N-1-j$ is the bitwise complement of~$j$, which is achieved by applying bitwise NOT on bits of $j$ (NOT gate on each qubit of $\ket{j}$).
The value oracle is then implemented by applying a $y$-rotation gate on the ancilla qubit with an appropriate rotation angle based on values of~$f$ and~$b$.
Specifically, we perform the multi-controlled rotation
\begin{equation}
    \texttt{CROT}: \ket{0}_a \ket{f} \ket{b}
    \mapsto R_y(\theta(f,b))\ket{0}_a \ket{f} \ket{b}
\end{equation}
controlled on the values of $f$ and $b$, with the rotation angle
\begin{equation}
    \theta(f,b) =
    \begin{cases}
        2\arccos(a_b/\beta) & f=0\\
        2\arccos(a_{b-N}/\beta) & f=1\ \text{and}\ b>0\\
        2\arccos(a_{b+N}/\beta) &f=1\ \text{and}\ b<0
    \end{cases}
\end{equation}
where $\beta=\norm{A}_\text{max}=\max_d \abs{a_d}$ is the maximum magnitude of diagonals in $A$.
Let $\ket{f,b}:=\ket{f}\ket{b}$, then notice that
\begin{equation}
    \texttt{CROT} = \sum_{b,f}
    R_y(\theta(f,b))
    \otimes \ket{f,b}\!\!\bra{f,b}
\end{equation}
is simply a uniformly controlled rotation over the joint control register comprised of the flag and band registers.
That is, we treat these registers as an address register for selecting one angle.
As shown in~\cref{fig:state_loading}(c), this uniformly controlled rotation can be decomposed into an alternating sequence of \texttt{CNOT} and single-controlled rotation gates.

The value oracle thus can be implemented by computing the flag bit~$f$, performing \texttt{CROT} and uncomputing~$f$.
We use a comparator operation to compute the flag.
Let us write $\ket{b}=\ket{s}\ket{x}$, where $s$ is the sign bit of $b$ in two's complement and the lower $r=m-1$ bits of $b$ contain the integer $x=2^r-\abs{b}$.
Then we compute the flag by the following steps. 
\begin{enumerate}
    \item Temporarily turn $j$ into its bitwise complement $\bar{j}$ if $b>0$. This is achieved by applying qubitwise NOT on qubits of~$\ket{j}$ controlled on $s=0$. We then have $\ket{j}\mapsto\ket{u_s(j)}$, where $u_s(j)=\bar{j}$ if $s=0$ and $u_s(j)=j$ if $s=1$.
    \item Temporarily turn the lower bits of $b$ into $\abs{b}$ if $b<0$. To this end, controlled on $s=1$, we apply qubitwise \texttt{NOT} on qubits of $\ket{x}$ to obtain bitwise complement $\bar{x}=2^r-1-x=\abs{b}-1$ and then perform $\ket{\bar{x}}\mapsto\ket{\bar{x}+1}$ by the increment operator.
    \item Apply the quantum-quantum comparator
    \begin{equation}
    \label{eq:quantum_quantum_comparator}
        \texttt{COM}
        \ket{0}_f\ket{\abs{b}}\ket{u_s(j)} =
        \begin{cases}
            \ket{0}_f\ket{\abs{b}}\ket{u_s(j)} &\text{if}\ \abs{b}\leq u_s(j)\\
            \ket{1}_f\ket{\abs{b}}\ket{u_s(j)} &\text{if}\ \abs{b}>u_s(j).
        \end{cases}
    \end{equation}
    \item Apply the inverse of the operations in Step~2 to restore $b$ in two's complement.
\end{enumerate}
The next step to implement the value oracle is to apply \texttt{CROT}. As this operation does not depend on $j$, we do not restore~$j$ before \texttt{CROT}.
To uncompute the flag, we then apply the operations in Step~2 above to have $\abs{b}$ in the lower~$r$ bits of the band register, followed by \texttt{COM} in Step~3. Finally, we restore $b$ to two's complement, and restore $j$ by applying the inverse of the operation in Step~1.
Altogether, we have the quantum circuit in~\cref{fig:sparse_encoding}(a) for implementing the value oracle.

We note that for the particular class of Toeplitz matrices where~$A$ is periodic with elements $A_{ij}=A_{(i-j)\bmod N}$, we have $a_{b\pm N}=a_{b}$ for any $b\neq 0$, and $A$ is a circulant matrix.
The flag qubit and comparison operation are not needed in this case, as the nonzero values depend only on the band index $b$ modulo $N$. The value oracle is simply a uniformly controlled rotation with the band register as the only control register.

\textbf{Cost analysis.}
Let $\cost(\text{flag})$ be the cost of computing the flag and let $\cost(\texttt{CROT})$ be the cost of implementing \texttt{CROT}, where the cost is either the gate cost or the circuit depth. The cost of implementing the value oracle is $\cost(\texttt{VAL})=\cost(\texttt{CROT})+2\cost(\text{flag})$.
\texttt{CROT} is a single-qubit rotation uniformly controlled on the flag qubit and the $m$-qubit band register.
As in~\cref{fig:state_loading}(c), this uniformly controlled rotation can be implemented by a circuit with~$\Theta(2^m)$ gates and~$\Theta(2^m)$ depth.

To analyze the cost of computing the flag by the above steps, we use the following lemmas for the cost of the controlled increment operator in Step~2 and the quantum-quantum comparator in~\cref{eq:quantum_quantum_comparator} in Step~3.

\begin{lemma}[{\cite[Cor.~7]{vandaele26}}]
\label{lem:multi_controlled_increment}
    The $k$-controlled increment operator on~$n$ qubits can be implemented using one dirty ancilla qubit and a circuit with $\Theta(k+n)$ gates from $\{X,\mathrm{CNOT}, \mathrm{Toffoli}\}$ and $\Theta(\log (kn))$ circuit depth.
\end{lemma}
\begin{lemma}[{\cite[Thm.~2]{vandaele26}}]
\label{lem:quantum_quantum_comparator}
    The $n$-qubit quantum-quantum comparator can be implemented without any ancilla qubits by a circuit with~$\Theta(n)$ gates from $\{X,\mathrm{CNOT}, \mathrm{Toffoli}\}$ and $\Theta(\log n)$ depth.
\end{lemma}
\begin{itemize}
    \item Step~1: The qubitwise NOT on $n$ qubits in this step is just the fan-out operator that can be implemented using a circuit with $\Theta(n)$ gates and $\Theta(\log n)$ depth~\cite[Lem.~2]{vandaele26}.
    \item Step~2: The qubitwise NOT on $m-1$ qubits can be implemented by a circuit with $\Theta(m)$ gates and $\Theta(\log m)$ depth.
    By~\cref{lem:multi_controlled_increment}, the 1-controlled increment acting on $r=m-1$ qubits can be implemented using a dirty ancilla and a circuit with $\Theta(m)$ gates and $\Theta(\log m)$ depth.
    \item Step~3:
    By~\cref{lem:quantum_quantum_comparator}, this step can be implemented by a circuit with~$\Theta(n)$ gates and $\Theta(\log n)$ depth.
\end{itemize}
Step~4 is the inverse of Step~2 with the same cost. Altogether, the \texttt{VAL} oracle can be implemented by a circuit with~$\Theta(n)+\Theta(2^m)$ gates and $\Theta(\log n)+\Theta(w)$ depth.
As shown in \cref{fig:sparse_encoding}(b), the \texttt{LOC} oracle is simply $m$ single-controlled additions of a known classical value to an $n$-qubit register.
By the following lemma, this oracle can be implemented using one dirty ancilla and a circuit with $\Theta(mn\log n)$ gates and $\Theta(m\log^2n)$ depth.
Together with the cost of \texttt{VAL}, we have the total gate cost~$\Theta(mn\log n)+\Theta(2^m)$ and the total circuit depth~$\Theta(m\log^2 n)+\Theta(2^m)$ for the sparse encoding. 

\begin{lemma}[{\cite[Cor.~8]{vandaele26}}]
\label{lem:classical_quantum_adder}
    The $k$-controlled classical-quantum adder on~$n$ qubits can be implemented using one dirty ancilla qubit and a circuit with~$\Theta(k+n\log n)$ gates from $\{X,\mathrm{CNOT}, \mathrm{Toffoli}\}$ and $\Theta(\log k +\log^2n)$ depth.
\end{lemma}
The total number of clean ancillas used is~$m+2$; $m$ ancillas used for the band register, one ancilla as the flag qubit, and one ancilla for value encoding. Only one dirty ancilla is used for the implementation.
With~$m=\ceil{\log_2(2w+1)}$, we obtain the resource cost reported in~\cref{tab:Toeplitz_encoding_cost} for the sparse encoding of a Toeplitz matrix.

\begin{table}[htb]
\centering
\setlength{\tabcolsep}{4.8pt}
  \begin{tabular}{cccccc}
    \toprule
    &&&& \\
    Matrix type & Encoding approach & Gate cost & Circuit depth & \multicolumn{2}{c}{Ancilla qubits}\\
    \cmidrule{5-6}
    &&&&Clean & \makecell{Dirty}\\
    \midrule
    Generic Toeplitz
    &\makecell{Sparse encoding\\
    or block encoding\\ by sparse encoding}
    &$\Theta(n\log n\log w)$
    &\makecell{$\Theta(w)+$\\
    $\Theta(\log^2 n\log w)$}
    &$\ceil{\log_2(2w+1)}+2$ &1\\
    \midrule
    \multirow{4}{*}{
    \makecell{Periodic Toeplitz\\(circulant matrix)}
    }
    &\makecell{Sparse encoding\\
    or block encoding\\ by sparse encoding}
    &$\Theta(n\log n\log w)$
    &\makecell{$\Theta(w)+$\\
    $\Theta(\log^2 n\log w)$}
    &$\ceil{\log_2(2w+1)}+1$
    &1\\
    &\makecell{Block encoding by LCU}
    &$\Theta(n\log n\log w)$
    &\makecell{$\Theta(w)+$\\$\Theta(\log^2 n\log w)$}
    &$\ceil{\log_2(2w+1)}$
    &1\\
    &\makecell{Block encoding\\
    by diagonalization}
    &$\Theta(2^n)$&$\Theta(2^n)$&1&0\\
    \bottomrule
  \end{tabular}
  \caption{\label{tab:Toeplitz_encoding_cost}
  Resource scaling of various approaches for encoding a Toeplitz matrix. 
  }
\end{table}
\begin{figure}[ht]
    \centering
    \includegraphics[width=.86\linewidth]{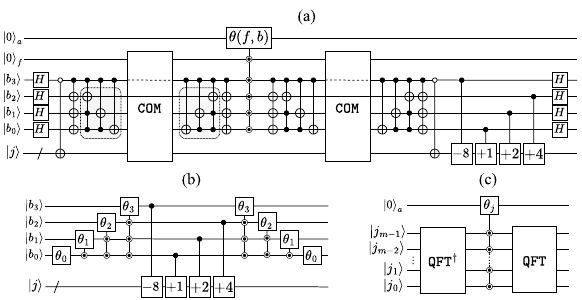}
    \caption{Quantum circuits for various block encodings of Toeplitz matrices.}
    \label{fig:blockencoding_constant_PDEs}
\end{figure}

\subsubsection{Block encodings for Toeplitz matrices and differential operators with constant coefficients}
\label{subsec:BE}
We now present several approaches to construct a block encoding for Toeplitz matrices. These matrices arise in discretizing differential operators with constant coefficients, so the block encodings we construct here apply to these operators.

\textbf{Block encoding by sparse encoding.}
A generic approach to block-encode a matrix is via its sparse encoding~\cite{GSL+19,CLV+24}.
Our first approach is to utilize the oracles we constructed for the sparse encoding to build a block encoding for a Toeplitz matrix.
Specifically, we block-encode $A$ as
\begin{equation}
\label{eq:BE_sparse}
    U_A=\left(\id_1\otimes H^{\otimes m} \otimes\id_n\right)
    \left(\id_1\otimes \texttt{LOC}\right)\ \texttt{VAL}\ \left(\id_1\otimes H^{\otimes m} \otimes\id_n\right),
\end{equation}
where $H$ is the Hadamard gate.
This block encoding can be verified by its action on basis states
\begin{equation}
    \bra{0}\bra{0^m} \bra{i} U_A \ket{0}\ket{0^m}\ket{j} = \frac1\alpha A_{ij},
\end{equation}
where $\alpha=2^m \beta = (2w+1)\norm{A}_\text{max}$ is the normalization factor. Observe that
\begin{align}
    \bra{0}\bra{0^m} \bra{i} U_A \ket{0}\ket{0^m}\ket{j}
    &= \frac1{2^m}\sum_{b'b}
    \bra{0}\bra{b'}\bra{i}
    \left(\id_1\otimes \texttt{LOC}\right)\ \texttt{VAL}
    \ket{0}\ket{b}\ket{j}\\
    &= \frac1{2^m}\sum_{b'b} \bra{0} R(\theta(b,j))\ket{0}
    \bra{b'}\bra{i} \id_1\otimes \texttt{LOC}\ket{b}\ket{j}\\
    &= \frac1{2^m\beta}\sum_{b'b}
    A_{r(b,j),j}\, \langle b'|b \rangle 
    \langle i|r(b,j) \rangle\\
    &= \frac1{2^m\beta}\sum_{b}
    A_{r(b,j),j}\, \delta_{i, r(b,j)}
    = \frac{A_{ij}}{2^m\beta},
\end{align}
where the last equality follows because $i=r(b,j)$ only for one band index~$b$. 
Note that this block encoding applies to a generic Toeplitz matrix, as the sparse encoding oracles apply to a generic Toeplitz matrix.
We also note that we have omitted the flag qubit used in the value oracle here as it starts and ends in $\ket{0}$ state.
The explicit quantum circuit for block encoding based on this approach, including the flag qubit, is given in \cref{fig:blockencoding_constant_PDEs}(a).
Note that the resource cost of this approach is similar to that of sparse encoding reported in~\cref{tab:Toeplitz_encoding_cost}.

\textbf{Block encoding by LCU.}
We now provide another approach for block encoding of a periodic Toeplitz matrix (i.e., a circulant matrix) using the Linear Combination of Unitaries~(LCU) technique~\cite{CW12}.
In this approach, we use the structural properties of the matrix and decompose it as a linear combination of unitary matrices as $A=\sum_{b=-w}^w{a_b S_{b}}$, where $S_b$ is the cyclic shift matrix with the action $S_b:\ket{j}\mapsto\ket{(j+b)\bmod N}$ on basis state $\ket{j}$, and construct the block encoding as
\begin{equation}
    U_A={(\texttt{PREP}}^\dag\otimes \id_n)\, \texttt{SELECT}\, (\texttt{PREP}\otimes \id_n),
\end{equation}
where
\begin{equation}
    \texttt{PREP} \ket{0}^{\otimes m} := \frac{1}{\sqrt{a}}\sum_{b=-w}^w \sqrt{\abs{a_b}} \ket{b}
\end{equation}
with $a:=\sum_b\abs{a_b}$ and
\begin{equation}
    \texttt{SELECT} \ket{b} \ket{j}:= \ket{b} S_b\ket{j} = \ket{b}\ket{(j+b)\bmod N}.
\end{equation}
Note that \texttt{SELECT} here is identical to \texttt{LOC} in~\cref{eq:O_loc}, and so we use the bits of $b$ to implement it as in \cref{fig:blockencoding_constant_PDEs}(b).
The cost of \texttt{SELECT} is also identical to the cost of \texttt{LOC} discussed before.
\texttt{PREP} can be implemented by a circuit with $\Theta(2^m)$ gates and $\Theta(2^m)$ depth by the loading procedure discussed in~\cref{subsec:loading}.
Altogether, we obtain the resource cost reported in~\cref{tab:Toeplitz_encoding_cost} for this encoding.

\textbf{Block encoding by diagonalization.}
Yet another approach to block encoding of a periodic Toeplitz matrix is using the fact that it can be diagonalized by a unitary transformation.
This approach does not require ancilla qubits as the control register in the previous approaches and could be appealing when the number of qubits is limited.
Specifically, we diagonalize a circulant matrix~$A$ by the Fourier transform as $A=F\Lambda F^\dag$, where $\Lambda$ is a diagonal matrix with diagonals~$\{\lambda_j\}$.
This decomposition yields the block encoding $U_A = (\id_1 \otimes \texttt{QFT}) O_\Lambda (\id_1 \otimes \texttt{QFT}^\dagger)$, where \texttt{QFT} is the quantum Fourier transform and $O_\Lambda:= \sum_{j} R(\theta_j) \otimes \ket{j}\!\bra{j}$ is a block encoding for the diagonal matrix~$\Lambda$, which is simply a uniformly controlled rotation with the angle $\theta_j$ where $\cos(\theta_j/2)= \lambda_j/\lambda_\text{max}$ and $\lambda_\text{max}=\max_j\abs{\lambda_j}$. 
This block encoding is shown in \cref{fig:blockencoding_constant_PDEs}(c).

The standard implementation for $n$-qubit QFT uses~$\Theta(n^2)$ elementary gates.
Note that~$O_\Lambda$ is simply a single-qubit rotation uniformly controlled on~$n$ qubits.
As in~\cref{fig:state_loading}(c), this uniformly controlled rotation can be implemented by a circuit with $\Theta(2^n)$ gates and $\Theta(2^n)$ depth.
Altogether, we obtain the resource cost reported in~\cref{tab:Toeplitz_encoding_cost} for this encoding.

\subsubsection{Block encodings for differential operators with variable coefficients}
\label{subsec:block_encoding_variable_operators}
We now utilize our representation for the variable-coefficient differentiation operators in~\cref{subsec:representation} and construct a block encoding for such operators. We first consider the 1D case and then extend the construction to higher dimensions.

Our discrete representation for these operators takes the form of a sum and a product of matrices. For instance, the discretized operator has the form $A=D^TED+K$ in \cref{eq:A1D} for the 1D case.
A straightforward approach to block encode $D^TED$ is to construct a block encoding for each factor, $D^T$, $E$ and $D$, and then combine them using the standard product rule for block encodings. However, this approach does not exploit the structure of the matrices involved in the product.
Instead of block encoding each factor in the product separately, we only block-encode the diagonal factor~$E$ and use it to build a block encoding of the matrix product by simple controlled-shift operations. This approach leads to a simple and more efficient construction than the direct product of individual block encodings.

To elaborate our approach, we use the decomposition $D=(S-\id)/h$ in terms of the shift operator~$S$ in~\cref{eq:derivative_matrix} and express the matrix product~$\wt{A}:=h^2 D^TED$ as
\begin{equation}
    \wt{A}= S^\dag ES -S^\dag E - ES + E.
\end{equation}
Now suppose $U_E$ is an $\alpha_E$ block encoding of $E$. We prepare two ancilla qubits in the state~$\ket{\xi}=\ket{-}_L\ket{-}_R$ so that the left/right qubit is in the state $\ket{-}:=(\ket{0}-\ket{1})/\sqrt{2}$.
The right qubit~$R$ controls whether to apply~$S$ before $U_E$, and the left qubit~$L$ controls whether to apply $S^\dag$ after $U_E$. Note that~$S$ and~$S^\dag$ act only on data qubits (not on the ancilla) of $U_E$.
Specifically, the right- and left-controlled shifts are
\begin{align}
    C_R &:= \ket{0}\!\!\bra{0}_R \otimes\id_{n+1}
    + \ket{1}\!\!\bra{1}_R\otimes\id_1\otimes S,\\
    C_L &:= \ket{0}\!\!\bra{0}_L \otimes\id_{n+1}
    + \ket{1}\!\!\bra{1}_L\otimes\id_1\otimes S^\dag,
\end{align}
where $\id_1$ is the identity on the ancilla of~$U_E$.
Let $\Xi:=ZH\otimes ZH$ be the operator that prepares~$\ket{\xi}$.
Then the unitary
\begin{equation}
    W := \Xi^\dag C_L (\id_2\otimes U_E)C_R \Xi
\end{equation}
block encodes $\wt{A}$ with three ancilla qubits, i.e.,
\begin{equation}
    (\bra{0^3}\otimes\id_n)W(\ket{0^3}\otimes\id_n) = \frac{\wt{A}}{4\alpha_E},
\end{equation}
and with the block-encoding normalization factor of $4\alpha_E$.
The key point here is that~$U_E$ is applied only once and is not controlled by qubits~$L$ and~$R$.
Note that~$W$ block encodes the scaled matrix $\wt{A}/h^2$ as well, but with the normalization factor $4\alpha_E/h^2$.

\begin{figure}[ht]
    \centering
    \includegraphics[width=.86\linewidth]{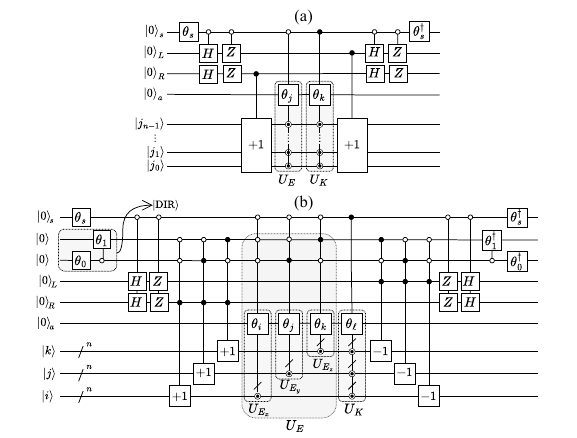}
    \caption{Quantum circuits for block encoding of differential operators with variable coefficients.}
    \label{fig:blockencoding_variable_PDEs}
\end{figure}

Now let $U_K$ be an $\alpha_K$ block encoding for the diagonal matrix~$K$.
Then a block encoding for $A=\wt{A}/h^2+K$ can be constructed using the block encodings $W$ and $U_K$ following the standard LCU approach.
However, we instead use the following approach to reduce the number of control qubits for the shift operations in the standard approach.
Let us use a new ancilla qubit labeled as the `summation' qubit~$s$ and let
\begin{equation}
\label{eq:1D_ancilla_state}
    \Omega \ket{0^3} = \cos(\theta/2)\ket{0}_s\ket{\xi} +
    \sin(\theta/2)\ket{1}_s\ket{00},
\end{equation}
where the rotation angle satisfies
\begin{equation}
\label{eq:1D_ancilla_angle}
    \cos(\theta/2) = \frac2h \sqrt{\frac{\alpha_E}{\alpha_A}},
    \enspace
    \alpha_A:=\frac{4\alpha_E}{h^2}
    +\alpha_K,
\end{equation}
then
\begin{equation}
\label{eq:1D_UA}
    U_A = \Omega^\dag C_L (\ket{0}\!\!\bra{0}_s\otimes U_E + \ket{1}\!\!\bra{1}_s\otimes U_K) C_R \Omega
\end{equation}
is a block encoding for~$A$ with the normalization factor~$\alpha_A$.
Note that here~$U_E$ and~$U_K$ share a single ancilla to block encode the diagonals~$E$ and~$K$, so four ancilla qubits are used in total for~$U_A$.
Also, note that $\Omega$ is simply a rotation $R(\theta)$ applied on qubit $s$ followed by a $\ket{0}_s$-controlled preparation of $\ket{\xi}$ on qubits~$L$ and~$R$.
When qubit~$s$ is in the~$\ket{1}_s$ state, the left~$L$ and right~$R$ qubits are in the zero state, and so~$C_R$ and~$C_L$ act as identity on system qubits.
Altogether, our approach yields the quantum circuit in~\cref{fig:blockencoding_variable_PDEs}(a) for block encoding of~$A$.

\textbf{Two dimensions.}
To extend our construction to two dimensions, we decompose the derivative matrix in~\cref{eq:Dxy} in the $\nu\in\{x,y\}$ direction as $D_\nu = (S_\nu-\id_{2n})/h$, where $S_x:=\id_n\otimes S$ and $S_y:=S\otimes\id_n$ with $S$ the cyclic shift matrix as before, and express the discretized 2D operator in \cref{eq:A2D} as $A=\wt{A}/h^2+K$, where
\begin{align}
    \wt{A}
    &= D^T_x E_x D_x + D^T_y E_y D_y\\
    &= (S^\dag_x-\id_{2n}) E_x (S_x-\id_{2n})
     + (S^\dag_y-\id_{2n}) E_y (S_y-\id_{2n}).
\end{align}
First we construct a block encoding for~$\wt{A}$, from which we build a block encoding for~$A$.
Suppose $U_{E_\nu}$ is an $\alpha_{E_\nu}$ block encoding of $E_\nu$.
We use a fresh ancilla labeled as the `direction' qubit~$\nu$ and prepare it in the direction state~$\ket{\textsc{dir}}:=R(\theta)\ket{0}= \cos(\theta/2)\ket{0} + \sin(\theta/2)\ket{1}$, where
\begin{equation}
\label{eq:2D_ancilla_angle}
    \cos(\theta/2)=\sqrt{\frac{\alpha_{E_x}}{\alpha_E}}
    \enspace \text{with}\enspace \alpha_E:=\alpha_{E_x}+\alpha_{E_y},
\end{equation}
to select the $x\leftrightarrow\ket{0}$ or $y\leftrightarrow\ket{1}$ direction based on the value of the direction qubit.
Then for each direction~$\nu$, the right qubit~$R$ (left qubit~$L$) selects whether to apply $S_\nu$~($S_\nu^\dag$) before~(after)
\begin{equation}
\label{eq:2D_UE}
    U_E:= \ket{0}\!\!\bra{0}_\nu\otimes U_{E_x} +
    \ket{1}\!\!\bra{1}_\nu\otimes U_{E_y},
\end{equation}
the controlled application of $U_{E_\nu}$ based on the direction qubit~$\nu$.
The described procedure block encodes~$\wt{A}$ with the normalization factor $4\alpha_E$.
The block encoding for~$A$ follows similarly as in the 1D case.
Using the summation qubit~$s$, we prepare the state in~\cref{eq:1D_ancilla_state}, where the angle is obtained from~\cref{eq:1D_ancilla_angle} but with $\alpha_E$ given in~\cref{eq:2D_ancilla_angle}.
The block encoding for~$A$ in 2D is analogous to the 1D case in~\cref{eq:1D_UA} but with~$U_E$ in~\cref{eq:2D_UE} and with~$\Omega$ that also prepares the direction state~$\ket{\textsc{dir}}$. Additionally, $C_{L}$ and $C_R$ are controlled by the direction qubit~$\nu$. Overall, five ancillas are used to construct $U_A$ in 2D.

\textbf{Higher dimensions.}
The same construction extends to three and higher dimensions.
In general, for $d$-dimensional discretized operator with~$d>1$ we write $A=\wt{A}/h^2+K$, where
\begin{equation}
    \wt{A}=\sum_{\nu=0}^{d-1}D_\nu^T E_\nu D_\nu
    = \sum_{\nu=0}^{d-1} (S^\dag_\nu-\id_{dn}) E_\nu (S_\nu-\id_{dn})
\end{equation}
and $S_\nu$ is the shift operator on direction~$\nu$.
In 3D, for instance, we have $S_0 = S_x = \id_n\otimes\id_n\otimes S$, $S_1=S_y=\id_n\otimes S\otimes\id_n$ and $S_2=S_z=S\otimes \id_n\otimes\id_n$.
We use $\ceil{\log_2d}$ direction qubits prepared on the state
\begin{equation}
    \ket{\textsc{dir}}:=\sum_\nu \sqrt{\frac{\alpha_{E_\nu}}{\alpha_E}}\ket{\nu}
    \enspace
    \text{with}
    \enspace
    \alpha_E:=\sum_\nu \alpha_{E_\nu}
\end{equation}
to select the direction. For example, in 3D, $x\leftrightarrow\ket{00}$, $y\leftrightarrow\ket{01}$, $z\leftrightarrow\ket{10}$ and $\ket{11}$ is unused.
Then for each direction~$\nu$, the right qubit~$R$ (left qubit~$L$) selects whether to apply $S_\nu$~($S_\nu^\dag$) before~(after)
\begin{equation}
    U_E = \sum_{\nu} \ket{\nu}\!\!\bra{\nu}\otimes U_{E_\nu}.
\end{equation}
The block encoding for~$A$ follows similarly as in the 2D case. As all $U_{E_v}$ share a single block-encoding ancilla, the total ancilla qubits used to block encode the $d>1$ dimensional operator is $4+\ceil{\log_2d}$.
\cref{fig:blockencoding_variable_PDEs}(b) shows the quantum circuit for block encoding of~$A$ in $d=3$ dimensions.

\subsection{Computing properties of a quantum solution}
\label{subsec:property_estimation}
In this section, we present several approaches for extracting desired properties of the quantum solution state~$\ket{\psi}$, which encodes an approximate solution of the discretized PDE. By a property of the quantum solution, we mean the expectation value of a given observable~$M$ with respect to this state, namely $\mu:=\bra{\psi}M\ket{\psi}$.
Depending on~$M$, this value can represent a physically or mathematically relevant feature of the solution.
The purpose of this section is to make explicit how such information can be extracted from the quantum solution using a unitary encoding of observable~\cite{KOS07,ANB+22} and, more importantly, to analyze and compare the end-to-end complexities when the costs of preparing the solution state and estimating a desired property are considered.
We will later use these results in~\cref{sec:tailored} to obtain the full cost of the end-to-end algorithm tailored to the Poisson-Boltzmann equation.

\subsubsection{Unitary embedding and reduction to nonnegative overlap}
\label{subsec:reduction}
We begin by embedding the expectation value of the generally nonunitary observable~$M$ into an expectation value of a unitary operator, and then show how it can be reduced to a nonnegative overlap.
Without loss of generality, we assume that $\norm{M}\leq1$. Because~$M$ is Hermitian,~$\mu$ is real and satisfies $\mu\in[-1,1]$.
We assume~$M$ is given as an exact block encoding~$U_M$; we will later construct such a block encoding for our application.
That is to say, $U_M$ acts on an $a$-qubit ancilla register and the $n$-qubit solution register and satisfies 
\begin{equation}
    (\bra{0^a}\otimes\id_n)U_M (\ket{0^a}\otimes\id_n)=M.
\end{equation}
Defining
\begin{equation}
    \ket{\phi} := \ket{0^a}\otimes\ket{\psi},
\end{equation}
we obtain
\begin{equation}
    \bra{\phi}U_M\ket{\phi} = \bra{\psi} M \ket{\psi} = \mu.
\end{equation}
Thus, the expectation value of the unitary~$U_M$ on the extended space embeds the expectation value of~$M$.
We note that an alternative unitary embedding follows from Hamiltonian simulation~\cite{KOS07}. In particular, we have
\begin{equation}
    \frac1t \Im \left(\bra{\psi}\e^{\i Mt}\ket{\psi}\right) =
    \frac1t \left(\bra{\psi}\sin(Mt)\ket{\psi}\right) = \mu + \order{t^2},
\end{equation}
or more precisely
\begin{equation}
    \abs{\mu - \frac1t\Im \left(\bra{\psi}\e^{\i Mt}\ket{\psi}\right)} \leq \frac{t^2}6.
\end{equation}
One can therefore estimate~$\mu$ by simulating~$\exp(\i Mt)$ for a sufficiently small~$t$. However, this approach has a small-$t$ approximation error. The block-encoding approach avoids this approximation, so we use it in the remainder of the~section.

To reduce the expectation value~$\mu$ to a nonnegative overlap, let us define the controlled block encoding
\begin{equation}
    U^c_M :=\ket{0}\!\!\bra{0}\otimes \id + \ket{1}\!\!\bra{1}\otimes U_M,
\end{equation}
and let
\begin{equation}
\label{eq:chi_state}
    \ket{\chi} := \ket{+}\otimes\ket{\phi} = \ket{+}\otimes\ket{0^a}\otimes\ket{\psi},
\end{equation}
where $\ket{+}:=(\ket{0}+\ket{1})/\sqrt{2}$.
We then obtain
\begin{equation}
\label{eq:nonnegative_overlap}
    u:= \bra{\chi}U^c_M\ket{\chi} = \frac{1+\bra{\phi}U_M\ket{\phi}}{2} =\frac{1+\mu}{2}.
\end{equation}
Note that $u\in[0,1]$ as $\mu\in[-1,1]$.
Thus, estimating the observable expectation value is reduced to estimating the nonnegative overlap~$\bra{\chi}U^c_M\ket{\chi}=u$, from which~$\mu$ is obtained as $\mu=2u-1$.

\begin{table}[htb]
\centering
\renewcommand{\arraystretch}{1.26}
\setlength{\tabcolsep}{4.8pt}
  \begin{tabular}{cccccc}
    \toprule
    && \multicolumn{3}{c}{Cost per run}\\
    \cmidrule{3-5}
    Approach & Ancilla qubits &Uses of&Uses of& Additional& Number of runs\\
    &&$U_\psi$ or~$U^\dagger_\psi$ &$U_M$ or $U^\dagger_M$& gates&\\
    \midrule
    Hadamard test
    &1
    &1
    &1
    &2
    &$\ceil*{\frac2{\varepsilon^2}\ln\left(\frac2\delta\right)}$\\
    Phase estimation
    &$t_\textsc{pe}=\ceil*{\log_2\left(\frac4\varepsilon\right)}$
    &1
    &$2^{t_\textsc{pe}}-1$
    &$\Theta\left(t^2_\textsc{pe}\right)$
    &$\ceil*{\frac8{\varepsilon^2}\ln\left(\frac2\delta\right)}$\\
    \multirow{2}{9em}{Amplitude estimation}
    &$t_\textsc{ae}=\ceil*{\log_2\left(\frac{2\pi}\varepsilon\right)}$
    &$2^{t_\textsc{ae}}-3$
    &$2(2^{t_\textsc{ae}}-1)$
    &$\Theta\left(t^2_\textsc{ae}\right) +\Theta(a+n)$
    &$\ceil*{\frac{\ln(1/\delta)}{2\left(\frac8{\pi^2}-\frac12\right)^2}}$\\
    & $t_\textsc{ae}=\ceil*{\log_2
    \left[
    \frac{2\pi}{\varepsilon}\left(2+\frac1{2\delta}\right)
    \right]}$
    &$2^{t_\textsc{ae}}-3$
    &$2(2^{t_\textsc{ae}}-1)$
    &$\Theta\left(t^2_\textsc{ae}\right) +\Theta(a+n)$
    &1\\
    \bottomrule
  \end{tabular}
  \caption{\label{tab:cost}
  Cost of an end-to-end algorithm based on three different approaches for estimating a property of the quantum solution with accuracy~$\varepsilon$ and success probability at least~$1-\delta$.
  Here $U_\psi$ is the unitary that prepares the $n$-qubit quantum solution state~$\ket{\psi}$ and $U_M$ is a block encoding of an observable~$M$ using~$a$ qubits.
  Ancilla qubits here are only those used for property estimation.
  The overall cost is obtained by the cost of a single run times the number of runs.
  }
\end{table}

\subsubsection{The Hadamard-test approach}
In this approach, we directly estimate the desired expectation value~$\mu=\bra{\psi}M\ket{\psi}$ using the expectation value~$\bra{\phi}U_M\ket{\phi}$ of the block encoding~$U_M$ of~$M$.
To this end, a test qubit is initialized in the state~$\ket{0}$ and transformed to~$\ket{+}$ by a Hadamard gate.
This qubit is then used as the control qubit to apply~$U_M$ on the target register containing the block-encoding ancillas and the quantum solution register. Another Hadamard is then applied to the test qubit, after which the qubit is measured in the computational basis~[\cref{fig:offloading}~(b)].
The measurement outcome is either~$0$ or~$1$, and the probability of obtaining~$0$ is
\begin{equation}
    p_0 = \frac{1+\Re(\bra{\phi}U_M\ket{\phi})}2 = \frac{1+\mu}{2}.
\end{equation}
Now define a random variable~$X$ that takes the value~$+1$ when the outcome is~$0$ and the value~$-1$ when the outcome is~$1$. The average of~$X$ based on the possible measurement outcomes is then $\mathbb{E}(X)=p_0-(1-p_0)=\mu$.
Therefore, the sample average of independent Hadamard-test outcomes provides an unbiased estimator of~$\mu$.
Specifically, let $X_1,\ldots,X_R$ be the outcomes of~$R$ independent Hadamard tests. Then the sample average
$\tilde{\mu}:=\sum_r X_r/R$ is an estimate of~$\mu$.

For a finite number of measurements, the error bound $\abs{\tilde{\mu}-\mu}\leq\varepsilon$ does not always hold true.
Because the measurement outcomes are random, the estimate~$\tilde{\mu}$ is also random. That is to say, different sets of~$R$ measurements generally give slightly different estimates.
The probability that the statistical error from finitely many measurements exceeds the given error tolerance~$\varepsilon$ can be bounded using Hoeffding’s inequality as
\begin{equation}
\label{eq:Hoeffding}
    \Pr\left(\abs{\tilde{\mu}-\mathbb{E}(X)}\geq\varepsilon\right) \leq 2\exp(-R\,\varepsilon^2/2).
\end{equation}
Therefore, $R\geq \ceil{2\varepsilon^{-2}\ln(2/\delta)}$ independent repetitions are sufficient to estimate~$\mu=\mathbb{E}(X)$ to additive accuracy~$\varepsilon$, with failure probability at most~$\delta$.

\textbf{Complexity.}
Each repetition needs a new preparation of the solution state~$\ket{\psi}$ and one controlled application of the block encoding~$U_M$. Let $C_\psi(\varepsilon)$ denote the cost of preparing the quantum solution state to accuracy~$\varepsilon$. Also, let $C_M$ denote the cost of implementing the block encoding $U_M$, or its controlled version; the additional control only changes the implementation cost by at most a constant factor.
Then we have
\begin{equation}
    C_{\textsc{hta}} = \left(C_\psi(\varepsilon) + C_M\right)\times \ceil*{\frac2{\varepsilon^2}\ln\left(\frac2\delta\right)}
\end{equation}
for the end-to-end complexity of the Hadamard-test approach for estimating solution properties.

\subsubsection{The phase-estimation approach}
Another approach for estimating the desired expectation value is to apply the quantum phase estimation~(QPE) to the block encoding~$U_M$ and use samples from eigenphases of~$U_M$ to obtain an estimate of~$\bra{\phi}U_M\ket{\phi}$.
This approach has two types of errors, one systematic error due to an approximation of the eigenphases and the other is statistical error due to a finite number of samples.
For an accurate estimation, we need to bound both of these errors.
To this end, let
\begin{equation}
    U_M = \sum_j \e^{2\pi\i\theta_j} \ket{v_j}\!\!\bra{v_j},
    \enspace \theta_j \in [0,1),
\end{equation}
be the spectral decomposition of~$U_M$ and expand~$\ket{\phi}$ in the eigenbasis of~$U_M$ as~$\ket{\phi}=\sum_j\alpha_j\ket{v_j}$.
Then we have
\begin{equation}
    \mu= \bra{\phi} U_M\ket{\phi}
    = \sum_j \abs{\alpha_j}^2\e^{2\pi\i\theta_j}
    = \sum_j \abs{\alpha_j}^2 \cos(2\pi\theta_j)
\end{equation}
for the real number~$\mu$.
The phase-estimation algorithm~[\cref{fig:offloading}~(a)] acts on $t$-qubit phase register initially prepared in the all-zero state, and a target register that encodes~$\ket{\phi}$. Each qubit of the phase register is first transformed into the~$\ket{+}$ state. Then~$U^{2^k}_M$ is applied to the target register controlled on the $k$th qubit of the phase register.
The inverse quantum Fourier transform~(QFT) is then applied on the phase qubits, after which they are measured in the computational basis.

The result of measurement is a random $t$-bit number~$B\in\{0,\ldots,2^t-1\}$. Each measurement outcome $B=b$ provides an approximation~$\tilde{\theta}_j=b/2^t$ to one of the phases~$\theta_j$, sampled with probability$\abs{\alpha_j}^2$.
When $\theta_j$ is not exactly representable by~$t$ bits, the QPE algorithm returns a $t$-bit estimate of an individual phase~$\theta_j$ only with a constant success probability.
This success probability for each individual phase can be increased to $1-\delta$ by adding~$\order{\log(1/\delta)}$ qubits to the phase register. However, for estimating an expectation value, it is not necessary to obtain individually accurate phases.
That is to say, even though the $t$-bit phase is obtained only with constant probability, the expectation value of outputs can be accurate with high probability without requiring individually accurate phases.
Moreover, increasing the number of phase qubits~$t$ only reduces the systematic error from the finite-bit approximation of the eigenphases. The scaling of the number of runs remains unchanged due to the statistical error associated with the eigenphase sampling.

More specifically, let us define the random variable $X:=\cos(2\pi B/2^t)$ using the random integer~$B$ obtained from measuring the phase register and let $\mathbb{E}(X)$ be the expectation value of all measurement outcomes irrespective of which phase component~$j$ is selected. Then the systematic error due to the $t$-bit approximation of the phases is bounded as
\begin{equation}
\label{eq:systematic_error}
    \abs{\mathbb{E}(X)-\mu} \leq 2^{1-t}.
\end{equation}
See~\cref{subsec:systematic_error} for a proof.
Thus, choosing $t\geq\ceil{\log_2(2/\varepsilon_\textsc{qpe})}$ suffices to make the systematic error at most~$\varepsilon_\textsc{qpe}$.
Now suppose $X_1,\ldots,X_R$ are outcomes of~$R$ independent runs of QPE and let $\tilde{\mu}:=\sum_r X_r/R$ be their average. 
We can use Hoeffding’s inequality to bound the probability that the statistical error $\abs{\tilde{\mu}-\mathbb{E}(X)}\geq \varepsilon_\textsc{stat}$, for some error~$\varepsilon_\textsc{stat}$, from which we obtain a sufficient value for~$R$.
For simplicity, we assume $\varepsilon_\textsc{qpe}=\varepsilon_\textsc{stat}=\varepsilon/2$, where~$\varepsilon$ is the given error tolerance.
Then by Hoeffding’s inequality, \cref{eq:Hoeffding} with~$\varepsilon\to\varepsilon/2$, choosing $R\geq \ceil{8\varepsilon^{-2}\ln(2/\delta)}$ is sufficient to obtain $\abs{\tilde{\mu}-\mathbb{E}(X)}\leq\varepsilon/2$ with success probability at least~$1-\delta$. 
When $\abs{\tilde{\mu}-\mathbb{E}(X)}\leq\varepsilon/2$, the triangle inequality gives the bound
\begin{equation}
    \abs{\tilde{\mu}-\mu} \leq  \abs{\tilde{\mu}-\mathbb{E}(X)} + \abs{\mathbb{E}(X)-\mu} \leq \frac\varepsilon2+\frac\varepsilon2=\varepsilon
\end{equation}
for the total error.
Therefore, the chosen values for~$t$ and~$R$ ensure that we have an $\varepsilon$-accurate estimation of~$\mu$ with high probability, i.e., $\Pr(\abs{\tilde{\mu}-\mu}\leq \varepsilon)\geq 1-\delta$.

\textbf{Complexity.}
Each run of QPE needs a new preparation of the solution state~$\ket{\psi}$, one application of the inverse QFT on~$t$ qubits, and $2^t-1$ controlled applications of the block encoding~$U_M$.
Let $C_\textsc{qft}(t)$ denote the cost of $t$-qubit QFT. Then we have
\begin{equation}
    C_{\textsc{pea}} = \left(C_\psi(\varepsilon) + \left(2^t-1\right)C_M + C_\textsc{qft}(t)\right)\times R
\end{equation}
with
\begin{equation}
    t=\ceil*{\log_2\left(\frac4{\varepsilon}\right)},
    \enspace
    R = \ceil*{\frac8{\varepsilon^2}\ln\left(\frac2\delta\right)}
\end{equation}
for the end-to-end complexity of estimating solution properties by the phase-estimation approach.
As~$t$ is logarithmic in~$1/\varepsilon$, we can use the standard implementation for QFT that uses~$\Theta(t^2)$ elementary gates.

\subsubsection{The amplitude-estimation approach}
\label{subsec:AEA}
We now build on the reduction discussed in~\cref{subsec:reduction} to compute the expectation value~$\mu$ by estimating an amplitude.
Notice that the nonnegative overlap $u=(1+\mu)/2$ in \cref{eq:nonnegative_overlap} is equal to the amplitude
$\abs{\bra{\chi}U^c_M\ket{\chi}}$.
To estimate this amplitude, we construct a rotation operator~$S$ so that the state $\ket{\chi}$ is an equal superposition of two eigenstates of~$S$ with eigenvalues $\exp{(\pm2\pi\i\theta)}$, where~$\cos(\pi\theta)=u$.
A single run of QPE applied to~$S$ with target state~$\ket{\chi}$ then yields an accurate estimate of~$\theta$, and hence of~$u$ and~$\mu$, with constant success probability.
A single run of QPE applied to~$S$ requires~$\order{1/\varepsilon}$ controlled applications of~$S$ to obtain~$\theta$ with accuracy~$\varepsilon$, so this approach provides a quadratic improvement over the~$\order{1/\varepsilon^2}$ cost of previous approaches.

To elaborate this approach, let us define two $r$-qubit state vectors $\ket{\psi_0}:=\ket{\chi}$ and $\ket{\psi_1}:=U^c_M\ket{\chi}$ where~$r=1+a+n$.
Then we can express the overlap $u=\braket{\psi_0|\psi_1}$ in terms of the angle~$\alpha$ between these vectors as $\braket{\psi_0|\psi_1}=\cos\alpha$, where $\alpha\in[0,\pi/2]$ by~$u\in[0,1]$.
Now let~$V$ be the $r$-qubit unitary such that $V\ket{0^r}=\ket{\psi_0}$, i.e., $V$ is the unitary that prepares $\ket{\psi_0}$ from the all-zero state. Also let $R_\psi:=\id-2\ket{\psi}\!\!\bra{\psi}$ be the reflection operator about the state~$\ket{\psi}$.
The product of two reflections $S:=R_{\psi_1}R_{\psi_0}$ is a rotation operator in the 2D space spanned by~$\ket{\psi_0}$ and~$\ket{\psi_1}$. This is because a reflection about one vector followed by a reflection about another vector is a rotation by twice the angle between the vectors.
The eigenvalues of~$S$ are therefore~$\lambda_\pm=\exp(\pm2\i\alpha)$.
It also follows that the target state~$\ket{\psi_0}$ is an equal superposition of two eigenstates of~$S$, i.e., $\ket{\psi_0}=(\ket{v_+}+\ket{v_-})/\sqrt{2}$.

In the QPE convention, these eigenvalues are $\lambda_+=\exp(2\pi\i\theta)$ and $ \lambda_-=\exp(2\pi\i(1-\theta))$, where $\theta:=\alpha/\pi\in[0,1/2]$.
Thus, QPE applied to~$S$ with target state~$\ket{\psi_0}$ yields an estimate close to either~$\theta$ or~$1-\theta$, from which we obtain an estimate of~$\theta$.
Specifically, let $b\in\{0,\ldots,2^t-1\}$ be the $t$-bit integer obtained from measuring $t$-qubit phase register, then $\tilde{\theta}=\min(b/2^t, 1-b/2^t)$ is an estimate of~$\theta$ obtained from the single run of QPE.
From this estimate we have~$\tilde{u}=\cos(\pi\tilde{\theta})$ and $\tilde{\mu}=2\tilde{u}-1$.

As conventional in QPE, if we consider the nearest integer~$b=\lfloor2^t\phi\rceil$ to~$2^t\phi$ as the successful measurement outcome, where $\phi$ is either of the eigenphases~$\theta$ or~$1-\theta$, then we have~$|\tilde{\theta}-\theta|\leq1/2^{t+1}$ with a success probability that is at least~$4/\pi^2$.
We therefore have
\begin{equation}
    \abs{\tilde{\mu}-\mu}=2|\cos(\pi\tilde{\theta})-\cos(\pi\theta)|\leq 2\pi|\tilde{\theta}-\theta|\leq \frac\pi{2^t},
\end{equation}
with probability $\geq 4/\pi^2\approx0.4$.
As this probability is smaller than~$1/2$, it cannot be amplified to an arbitrarily high value by repetition.
As shown in~\cref{subsec:combined_success}, if we allow the phase error to be at most~$1/2^t$ instead of~$1/2^{t+1}$, then the success probability, i.e., the probability that~$|\tilde{\theta}-\theta|\leq2^{-t}$, becomes at least~$8/\pi^2$.
In this case, $\abs{\tilde{\mu}-\mu}\leq 2\pi/2^t$ and thus choosing $t\geq\ceil{\log_2(2\pi/\varepsilon)}$ is sufficient to obtain an $\varepsilon$-accurate $\mu$ with probability at least~$8/\pi^2$.

We can now amplify the success probability to a high value of~$1-\delta$ by repetition. To this end, let~$\tilde{\mu}_1,\ldots,\tilde{\mu}_R$ be estimates obtained from~$R$ independent runs and let $\tilde{\mu}$ be the median of these estimates.
Each run is successful if $\abs{\tilde{\mu}_r-\mu}\leq\varepsilon$, so the median can fail only if at most half of the runs are successful.
To bound the failure probability, let us use~$f_r\in\{0,1\}$ as a flag to indicates whether run~$r$ is successful $f_r=1$ or not $f_r=0$.
The median~$\tilde{\mu}$ then fails if the average of successful runs $\bar{f}:=\sum_r f_r/R$ satisfies $\bar{f}\leq 1/2$.
Note that we have the bound $\mathbb{E}(\bar{f})\geq 8/\pi^2$ on the expected value of the average for different sets of~$R$ estimates because the expected value of each flag is $\mathbb{E}(f_r) = \Pr(\abs{\tilde{\mu}_r-\mu}\leq\varepsilon)\geq8/\pi^2$.
The failure probability of the median can therefore be bounded as
\begin{align}
    \Pr(\abs{\tilde{\mu}-\mu}\geq \varepsilon)
    &\leq \Pr(\bar{f}\leq1/2)\\
    &\leq \Pr(\bar{f}-\mathbb{E}(\bar{f})\leq-c)\\
    &\leq\e^{-2Rc^2},
\end{align}
where $c:=8/\pi^2-1/2$, the second line follows from $\mathbb{E}(\bar{f})\geq 8/\pi^2$, and the third line is obtained from Hoeffding’s inequality. Therefore, choosing $R\geq\ceil{\ln(1/\delta)/(2c^2)}$ would be sufficient to obtain an~$\varepsilon$-accurate~$\mu$ with probability at least~$1-\delta$.
For a unique median estimate, we choose the smallest odd integer for~$R$ here.

\begin{figure}[ht]
    \centering
    \includegraphics[width=.94\linewidth]{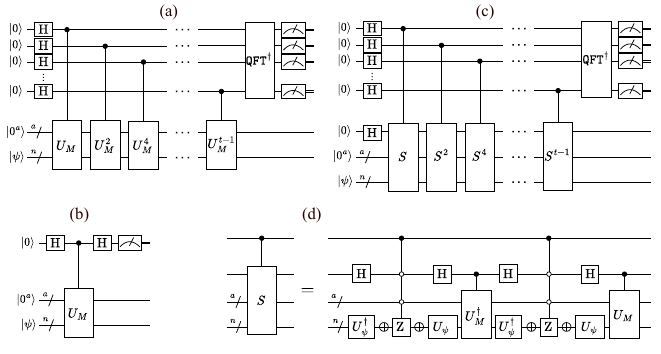}
    \caption{Quantum circuits for estimating a property of the quantum solution state~$\ket{\psi}$ by three approaches: (a)~phase estimation, (b)~Hadamard test and (c)~amplitude estimation.
    (d)~Implementation of the controlled-$S$ , with~$S$ give in~\cref{eq:S}, in the amplitude-estimation approach}
    \label{fig:offloading}
\end{figure}

\textbf{Complexity.}
Each run of QPE here needs a new preparation of the $r$-qubit state~$\ket{\psi_0}=\ket{\chi}$ with~$r=1+a+n$, one application of the inverse QFT on~$t$ qubits, and $2^t-1$ controlled applications of
\begin{equation}
\label{eq:S}
    S=U^c_MR_{\psi_0}U^{c\dagger}_MR_{\psi_0}
    = U^c_MVR_{0^r}V^\dagger U^{c\dagger}_MV R_{0^r}V^\dagger
\end{equation}
on $r$ qubits~[\cref{fig:offloading}~(c)].
Here  $R_{0^r}$ is the reflection about~$\ket{0^r}$ and~$V$ is the $r$-qubit unitary that prepares $\ket{\chi}$ from the all-zero state, i.e., $V\ket{0^r}=\ket{\chi}=\ket{+}\ket{0^a}\ket{\psi}$. The cost of~$V$ is then simply the cost~$C_\psi(\varepsilon)$ of unitary~$U_\psi$ for preparing the $n$-qubit solution state~$\ket{\psi}$.
Note that for the controlled-$S$ operation, we only need to replace~$R_{0^r}$ with its controlled version~$R^c_{0^r}$. This is because if the control qubit is at~$\ket{0}$, then~$R^c_{0^r}$ becomes identity and all unitaries in~$S$ cancel out and if the control qubit is at~$\ket{1}$, $S$ is applied~[\cref{fig:offloading}~(d)].
The cost of controlled $S$ is therefore $C_S= 2C_M+4C_\psi(\varepsilon) + 2C_R(r)$, where $C_R(r)$ denotes the cost of controlled reflection about the $r$-qubit zero state.
We therefore have
\begin{equation}
    C_{\textsc{aea}} = \left(C_\psi(\varepsilon) + \left(2^t-1\right)C_S + C_\textsc{qft}(t)\right)\times R
\end{equation}
with
\begin{equation}
    t=\ceil*{\log_2\left(\frac{2\pi}{\varepsilon}\right)},
    \enspace
    R= \ceil*{\frac{\ln(1/\delta)}{2\left(\frac8{\pi^2}-\frac12\right)^2}}
\end{equation}
for the end-to-end complexity of estimating solution properties by the amplitude-estimation approach.
The controlled reflection about~$\ket{0^r}$ can be implemented using~$\Theta(r)$ elementary gates (\cref{subsec:controlled_reflection}).

Finally, we note that unlike the phase-estimation approach, the number of runs~$R$ in the amplitude-estimation approach can be reduced to $R=1$ by increasing the size of the phase register.
This is possible because the input state here is an equal superposition of two eigenstates with phases~$\theta$ and~$1-\theta$, and so a single run of QPE provides an $\varepsilon$-accurate estimate of~$\theta$.
By the standard analysis, if we increase the number of phase qubits to
\begin{equation}
    t=\ceil*{\log_2
    \left[
    \frac{2\pi}{\varepsilon}\left(2+\frac1{2\delta}\right)
    \right]},
\end{equation}
the success probability becomes at least~$1-\delta$.
Although this approach has a single run, it needs more ancillas and running a deeper circuit as~$2^t\in\order{1/\delta\varepsilon}$.
The approach based on repetition is therefore preferable, as its circuit depth is independent of~$\delta$ and the number of runs scales only logarithmically with~$1/\delta$. 

\section{Tailored end-to-end quantum algorithm for the Poisson-Boltzmann equation}
\label{sec:tailored}
In this section, we construct a tailored quantum algorithm for the Poisson-Boltzmann equation in simulating a biomolecular system.
We begin by describing the PDE in~\cref{subsec:LPBE}.
In~\cref{subsec:initial_state}, we discuss the appropriate initial state of the system and present a tailored implementation for its oracle with a detailed cost analysis.
We then present a quantum algorithm for generating a quantum solution of the PDE and provide an implementation for the operator oracle.
Finally, in~\cref{subsec:obs}, we identify several properties of interest of the biomolecular system, construct a suitable observable for them and provide a tailored implementation for the observable oracle.

\begin{figure}[ht]
    \centering
    \includegraphics[width=\linewidth]{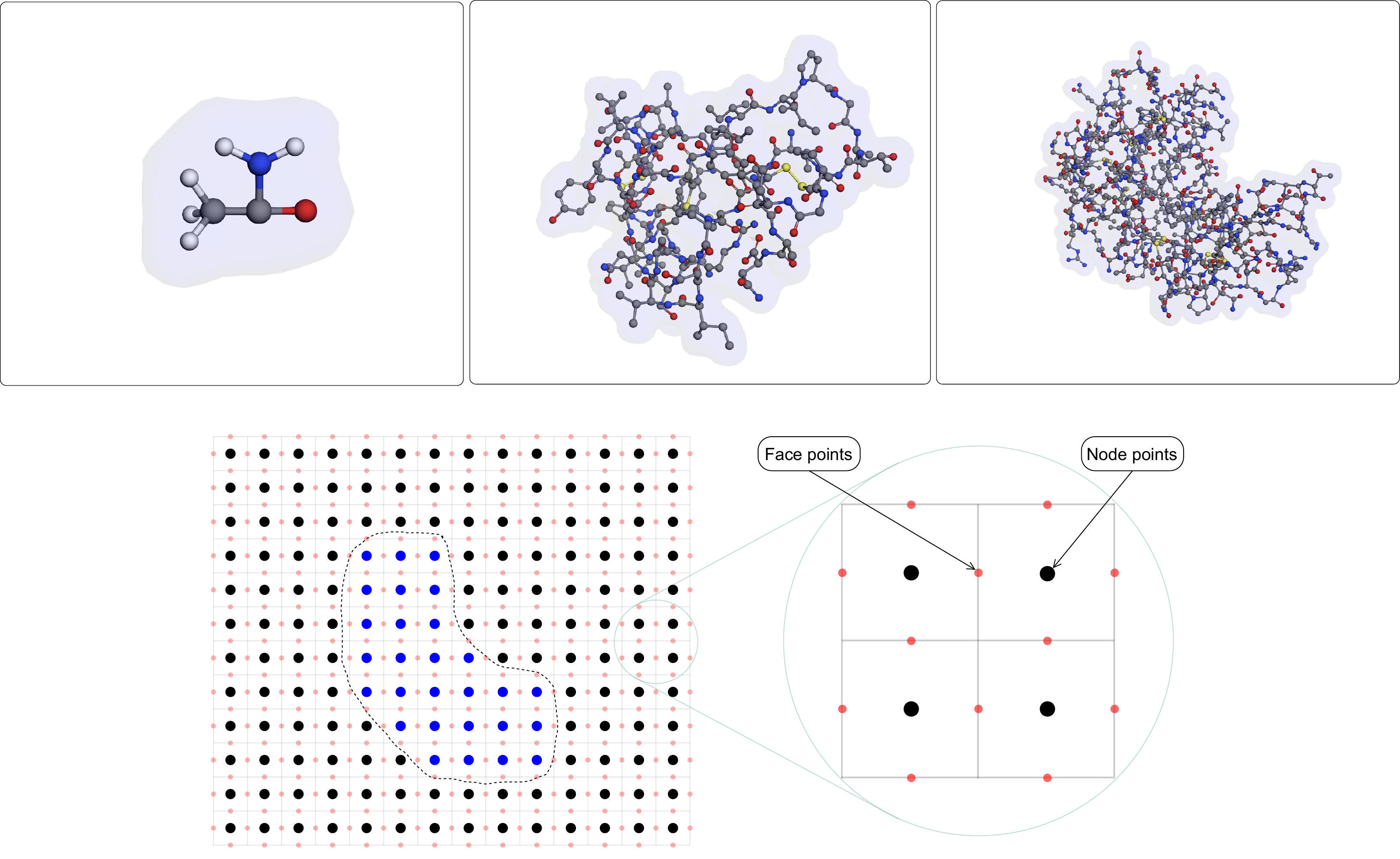}
    \caption{Examples of small (acetamide), medium (Crambin), and large (lysozyme) biomolecule typically used in PBE solvers. The bottom panel shows the staggered grid used for discretization. The biomolecule is inside the cavity shown by dotted loop. }
    \label{fig:lattice}
\end{figure}

\subsection{The linearized Poisson-Boltzmann equation}
\label{subsec:LPBE}
The Poisson-Boltzmann equation~(PBE) is a nonlinear elliptic PDE that models the electrostatic potential in a system comprising a biomolecule, such as a protein, immersed in an ionic solution consisting of a solvent, such as water, and dissolved ions~(see~\cref{fig:lattice}). The biomolecule is represented by a collection of fixed atomic charges, whereas the solvent is treated as a dielectric continuum specified by its permittivity rather than explicitly representing its individual molecules. The dissolved ions are also described by continuous concentration distributions rather than by explicitly representing individual ions, and they are assumed to follow the Boltzmann distribution at equilibrium.
This Poisson-Boltzmann equation has the form~\cite{QSM19,BSH+01,KFS+20, BHW00, HBW00}
\begin{equation}
\label{eq:PBE}
    -\nabla\cdot\left(
    \varepsilon(\bm{x})\nabla u(\bm{x})\right)
    +\bar{\kappa}^2(\bm{x})\sinh \left(u(\bm{x})\right)=
    \frac{4\pi \e^2}{K_BT}
    \sum_{a=0}^{N_\text{atm}-1}
    z_a\delta\left(\bm{x}-\bm{x}_a\right)
    \quad
    \bm{x} \in \Omega= (0,L)^3 \subset \mathbb{R}^3,
\end{equation}
where~$N_\text{atm}$ is the number of atomic point charges of the biomolecule, $z_a$ is the partial charge of atom~$a$ at position~$\bm{x}_a$ as a fraction of the elementary charge~$\e$, and~$u(\bm{x})=\e\phi(\bm{x})/K_BT$ is the electrostatic potential~$\phi(\bm{x})$ in the dimensionless form, with~$K_BT$ the thermal energy.
The position-dependent function~$\varepsilon(\bm{x})$ is the dielectric coefficient, and~$\bar{\kappa}^2(\bm{x})$ is the Debye–H\"uckel screening parameter, which depends on the concentration of the dissolved ions and determines their effect on the electrostatic potential.

The domain~$\Omega$ here is a cubic computational domain with side length~$L$, chosen sufficiently large to contain the biomolecule and the surrounding solvent. The physical problem is naturally on the unbounded domain~$\mathbb{R}^3$, with the potential vanishing far from the molecule, so~$\Omega$ represents a finite truncation of this domain.
Biomolecules such as proteins and enzymes typically have sizes ranging from about~10 to over~100 angstroms, and a value of~$L$ several times larger than the molecular size is commonly used~\cite{Holst94}.

A conventional approach~\cite{QSM19,JNM+23} to model this system is to divide the domain~$\Omega$ into two regions~$\Omega= \Omega_\text{cav} \cup \Omega_\text{out}$, where~$\Omega_\text{cav}$ is the region occupied by the biomolecule and~$\Omega_\text{out}=\Omega\setminus\Omega_\text{cav}$ is the region outside the molecular region occupied by the ionic solution.
The molecular region is modeled by the van der Waals cavity as the union of overlapping atom-centered balls
\begin{equation}
\label{eq:cav}
    \Omega_\text{cav} = \bigcup_{a=0}^{N_\text{atm}-1} B(\bm{x}_a, r_a),
\end{equation}
where
\begin{equation}
    B(\bm{x}_a, r_a):=\{\norm{\bm{x}-\bm{x}_a}\leq r_a\}
\end{equation}
is the ball centered at position~$\bm{x}_a$ of atom~$a$ with the van der Waals radius of~$r_a$. Hereafter, we refer to~$\Omega_\text{cav}$ as the molecular cavity.
In this model, the dielectric coefficient~$\varepsilon(\bm{x})$ and the scaled screening parameter~$k(\bm{x}):=\bar{\kappa}^2(\bm{x})$ are piecewise constant functions of the form
\begin{equation}
\label{eq:epsKappa}
    \varepsilon(\bm{x}) = 
    \begin{cases}
        \varepsilon_\text{cav} &\bm{x} \in \Omega_\text{cav}\\
        \varepsilon_\text{out} &\bm{x} \in \Omega_\text{out}
    \end{cases}
    \quad
    k(\bm{x}) = 
    \begin{cases}
        0 & \bm{x} \in\Omega_\text{cav} \\
        \varepsilon_\text{out}\,\kappa^2_\text{out} & \bm{x} \in \Omega_\text{out}, 
    \end{cases}
\end{equation}
where~$\varepsilon_\text{cav}$ and~$\varepsilon_\text{out}$ are the dielectric  permittivity of the biomolecule and solvent, respectively, and~$\kappa_\text{out}$ is the Debye–H\"uckel screening constant of the ionic solution.
Numerical values of these parameters depend on the solvent and temperature, but~$\varepsilon_\text{out}$ is two orders of magnitude greater than~$\varepsilon_\text{cav}$. For reference, at room temperature, representative values are $\varepsilon_\text{cav}=1$ or~$2$, $\varepsilon_\text{out}=78.55$ for water as the solvent, and~$\kappa_\text{out}=0.104\text{\AA}^{-1}$~\cite{QSM19}; note that $k(\bm{x})$ is dimensionless.

The nonlinearity of the PBE is due to $\sinh{(u)}$ term, which describes the response of the dissolved ions to the electrostatic potential.
In the regime where $\abs{u}\ll 1$, the approximation $\sinh{(u)} = u + \order{|u|^3}$ is used to replace the nonlinear ionic response by a linear term, and we obtain the linearized PBE.
For clarity and convenience, we write the linearized PBE compactly as $\L u(\bm{x})=b(\bm{x})$, where
\begin{equation}
\label{eq:differential_operator}
    \L:= -\nabla\cdot\left(\varepsilon(\bm{x})\nabla\right)+k(\bm{x})
\end{equation}
and the inhomogeneity is
\begin{equation}
\label{eq:b(x)}
    b(\bm{x}) :=c_b
    \sum_{a=0}^{N_\text{atm}-1}
    z_a\delta\left(\bm{x}-\bm{x}_a\right),
    \enspace
    c_b:= \frac{4\pi \e^2}{K_BT},
\end{equation}
where $c_b$ is a constant.
The set of atom positions $\{\bm{x}_a\}$, partial charges~$\{z_a\}$, and the parameters in~$c_b$ specify~$b(\bm{x})$. The cavity region~$\Omega_\text{cav}$ is determined from the set of atom positions~$\{\bm{x}_a\}$ and the van der Waals radii~$\{r_a\}$.
These sets along with constants~$\varepsilon_\text{cav}$, ~$\varepsilon_\text{out}$, and~$\kappa_\text{out}$ in \cref{eq:epsKappa} specify the functions~$\varepsilon(\bm{x})$ and~$k(\bm{x})$, and hence the differential operator~$\L$.

\subsection{The initial state and tailored implementation for its oracle}
\label{subsec:initial_state}

The initial state to prepare is a quantum state $\ket{\bm{b}}$ whose amplitudes encode a discretized version of the function $b(\bm{x})$ in~\cref{eq:b(x)}.
A standard discretization approach used in classical Poisson-Boltzmann solvers is to map the Dirac delta associated with each point charge centered at position~$\bm{x}_a$ onto nearest-neighbor grid points with weights that depend on the atom position.
We adopt the same approach, but first consider the simpler case in which each atomic position is mapped onto its nearest-neighbor grid points with equal weights. We present our state-preparation algorithm for this case and then extend it to the case with position-dependent weights.
We also extend our state-preparation approach to discretizations where each Dirac delta is mapped onto a larger set of neighboring grid points.

\subsubsection{The initial state with uniform charge distribution}
For clarity, we first describe this state in 1D and then extend it to higher dimensions. For the 1D problem, we map the Dirac delta associated with atom~$a$ centered at position~$x_a$ equally onto two nearest-neighbor grid points $\mathrm{x}_a=\floor{x_a/h}$ and $\mathrm{x}_a+1$, where $h$ is the grid spacing. Specifically, we map the Dirac delta as
\begin{equation}
\label{eq:1D_delta_equal_weights}
    \delta(x-x_a)\mapsto \frac1{2h}
    \sum_{s\in\{0,1\}}\delta_{j,\mathrm{x}_a+s},
\end{equation}
so that the same weight~$1/2$ is assigned to each neighboring point. 
The factor $1/h$ here is included so that the discretized Delta preserves the continuum normalization $\int_\Omega\mathrm{d}x\, \delta(x-x_a)=1$.
Similarly, for the $d$-dimensional problem, we map the Dirac delta as
\begin{equation}
    \delta(\bm{x}-\bm{x}_a)\mapsto \frac1{2^dh^d}
    \sum_{\bm{s}\in\{0,1\}^d}\delta_{\bm{q},\bm{\mathrm{x}}_a+\bm{s}},
\end{equation}
with equal weight of~$1/2^d$ for each neighboring point. Here $\bm{q}$ is the index of grid points and we refer to $\bm{\mathrm{x}}_a$ as the position of atom~$a$ in the grid coordinates. For the 3D problem, we have $\bm{q}=(i,j,k)$ and $\bm{\mathrm{x}}_a=(\mathrm{x}_a,\mathrm{y}_a, \mathrm{z}_a)$, where $\mathrm{x}_a=\floor{x_a/h}$, $\mathrm{y}_a=\floor{y_a/h}$ and $\mathrm{z}_a=\floor{z_a/h}$.

The discretized version of $b(\bm{x})$ in~\cref{eq:b(x)} is then
\begin{equation}
    b(\bm{x})\mapsto
    b(\bm{q}) :=
    \frac{c_b}{2^dh^d}
    \sum_{a,\bm{s}}z_a\delta_{\bm{q},\bm{\mathrm{x}}_a+\bm{s}}.
\end{equation}
Let us define $\bm{b}:=b(\bm{q})/\norm{b(\bm{q})}$, then the normalized quantum state to prepare is
\begin{equation}
\label{eq:initial_state_uniform}
    \ket{\bm{b}} =
    \frac1{\norm{\bm{z}}} \sum_{a=0}^{N_\text{atm}-1} z_a \ket{\psi_a},
    \enspace
    \ket{\psi_a}:= \frac1{\sqrt{2^d}}
    \sum_{\bm{s}\in\{0,1\}^d}
    \ket{\bm{\mathrm{x}}_a+\bm{s}},
\end{equation}
where the overall prefactor in $b(\bm{q})$ is removed by normalization.
We refer to the state $\ket{\psi_a}$ as the weighted atom-position state.
In~$d$ dimensions, this state is simply a tensor product of 1D states as
\begin{equation}
    \ket{\psi_a}= \ket{\psi_{a,0}}\otimes\cdots\otimes
    \ket{\psi_{a,d-1}},
    \enspace
    \ket{\psi_{a,\nu}} = \frac1{\sqrt{2}}\sum_{s_\nu\in\{0,1\}}\ket{\mathrm{x}_{a,\nu}+s_\nu},
\end{equation}
where $\ket{\psi_{a,\nu}}$ is the atom-position state in direction~$\nu$.
The explicit state for the 3D problem is
\begin{equation}
    \ket{\psi_a} = 
    \frac1{\sqrt{2^3}}
    \sum_{s_x,s_y,s_z\in\{0,1\}}
    \ket{\mathrm{x}_a+s_x}
    \ket{\mathrm{y}_a+s_y}
    \ket{\mathrm{z}_a+s_z}.
\end{equation}

An important assumption we use here for the initial state~$\ket{\bm{b}}$ to be a normalized state is that we assume the neighboring points associated with different atoms do not overlap.
Specifically, let 
\begin{equation}
    \mathrm{supp}(a):= \{\bm{\mathrm{x}}_a+\bm{s}:
\bm{s}\in\{0,1\}^d\}
\end{equation}
be the support of atom~$a$, defined as the set of atom~$a$'s neighboring grid points. We assume atoms have disjoint supports, i.e.,
\begin{equation}
    \mathrm{supp}(a)\cap \mathrm{supp}(a') = \varnothing
\end{equation}
for each pair~$(a,a')$ of atoms. In this case, the atom-position states $\ket{\psi_a}$ are mutually orthogonal and $\ket{\bm{b}}$ is a normalized state.
The support of each atom depends on the grid spacing~$h$; this is also explicitly noted in the documentation of the Adaptive Poisson–Boltzmann Solver~(APBS)~\cite{APBS}, which is a widely used software package for solving the PB equation.

As the grid spacing is a discretization parameter, the assumption of disjoint supports can always be satisfied by sufficiently refining the grid. That is, by choosing a sufficiently small grid spacing so that the support of different atoms do not overlap. 
A sufficient condition for disjoint supports is separation of atoms by at least~$2h$ in at least one spatial direction. A simple and sufficient condition in~$d$ dimensions is then choosing~$h$ so that $\norm{\bm{x}_a-\bm{x}_{a'}}\geq 2h\sqrt{d}$. Typical bond lengths in biomolecules are in the range of~$1$--$1.5$\AA, and the grid spacing of $h=0.5$\AA\ is commonly used in classical PBE solvers~\cite{HBF13}. Therefore, reducing the commonly used grid spacing to~$h=0.25$\AA\ gives a separation of $2\sqrt{3}h\approx 0.9$\AA\ in 3D, which is below the typical bond lengths, and so it is sufficient to have disjoint supports.

\subsubsection{Loading the initial sate with uniform charge distribution}
\label{subsubsec:loading_uniform}
We now present our approach for loading the initial state~$\ket{\bm{b}}$ in~\cref{eq:initial_state_uniform}. To this end, first we prepare the state
\begin{equation}
\label{eq:ketz}
    \ket{\bm{z}} :=
    \sum_{a=0}^{N_\text{atm}-1} \bar{z}_a \ket{a},
    \enspace \bar{z}_a:= z_a/\norm{z},
\end{equation}
where $\ket{a}$ is the basis state encoding the index~$a$ of atoms.
This state can be simply prepared using the binary-tree approach discussed in~\cref{subsec:loading}.
Using the state~$\ket{\bm{z}}$ and the classically known atomic positions~$\{\bm{\mathrm{x}}_a\}$ in the grid coordinates, we then prepare
\begin{equation}
\label{eq:ket_pos}
    \ket{\text{pos}} := \sum_a \bar{z}_a\ket{\bm{\mathrm{x}}_a},
\end{equation}
which we call the atom-position state.
To prepare this state, first we perform the map
\begin{equation}
\label{eq:pos}
    \texttt{POS}:\ket{a}_\texttt{idx}
    \ket{0^n}^{\otimes d}_\texttt{sys} \mapsto \ket{a}_\texttt{idx}\ket{\bm{\mathrm{x}}_a}_\texttt{sys}
\end{equation}
by adding the atom position~$\bm{\mathrm{x}}_a$ to the $dn$-qubit system register \texttt{sys} controlled on the value of the $m$-qubit atom-index register \texttt{idx} with $m=\ceil{\log_2(N_\text{atm})}$. Then we erase \texttt{idx} using values of the atom positions~$\bm{\mathrm{x}}_a$. To erase~\texttt{idx}, for each atom~$a$, we test if the value encoded in \texttt{sys} is equal to~$\bm{\mathrm{x}}_a$ and write the result into a flag qubit~$f$. That is, we perform the equality test
\begin{equation}
U_\text{test}: \ket{0}_f \ket{\bm{q}}_\texttt{sys} \mapsto
    \begin{cases}
        \ket{1}_f\ket{\bm{q}}_\texttt{sys}
        &\text{if } \bm{q}= \bm{\mathrm{x}}_a\\
        \ket{0}_f\ket{\bm{q}}_\texttt{sys}
        &\text{else }
    \end{cases}
\end{equation}
for atom~$a$, where $\bm{q}$ is the value of register \texttt{sys}.
Then controlled on~$\ket{1}_f$, we add $-a$ into $m$-qubit register \texttt{idx}, where~$a$ is the classically known $m$-bit atom index; this is simply controlled-addition operation $\texttt{CADD}(-a)$ in~\cref{eq:cadd} on $m$ qubits. We then erase the flag by reapplying the equality test. Let
\begin{equation}
\label{eq:U_erase}
    U_\text{erase}:= U_\text{test}\,\texttt{CADD}\, U_\text{test}
\end{equation}
be the sequence of operations that erase the index register and let $U_{\bm{z}}$ be the unitary that prepares the state~$\ket{\bm{z}}$ in~\cref{eq:ketz}.
Also, let us define
\begin{equation}
\label{eq:U_pos}
    U_\text{pos}:= U_\text{erase}\,\texttt{POS}\, U_{\bm{z}},
\end{equation}
then by the described steps
\begin{equation}
\label{eq:U_pos_action}
    U_\text{pos}\ket{0}_f\ket{0^m}_\texttt{idx}\ket{0^n}^{\otimes d}_\texttt{sys} =
    \ket{0}_f\ket{0^m}_\texttt{idx}
    \ket{\text{pos}}_\texttt{sys},
\end{equation}
and so the atom-position state $\ket{\text{pos}}$ is prepared on the system register.

We now perform the map
$\ket{\bm{\mathrm{x}}_a}\mapsto\sum_{\bm{s}}\ket{\bm{\mathrm{x}}_a+\bm{s}}$.
To this end, we prepare a $d$-qubit ancilla register, which we call the selector register and denote it by \texttt{sel}, in the uniform state
\begin{equation}
    \ket{+}_\texttt{sel}^d
    =\frac1{\sqrt{2^d}} \sum_{\bm{s}\in\{0,1\}^d} \ket{\bm{s}}_\texttt{sel},
\end{equation}
by applying~$d$ Hadamard gates. Then we apply the shift operation
\begin{equation}
\label{eq:Ushift}
    U_{\text{shift}}: \ket{\bm{s}}_\texttt{sel}
    \ket{\bm{\mathrm{x}}_a}_\texttt{sys}
    \mapsto 
    \ket{\bm{s}}_\texttt{sel}
    \ket{\bm{\mathrm{x}}_a+\bm{s}}_\texttt{sys}
\end{equation}
followed by $d$ Hadamard gates on the selector register.
Let us denote these operations by
\begin{equation}
\label{eq:W}
    W:=(H^{\otimes d}\otimes\id)\,
    U_\text{shift}\,
    (H^{\otimes d}\otimes\id),
\end{equation}
then we have
\begin{equation}
\label{eq:W_action}
    W \ket{0^d}_\texttt{sel}\ket{\text{pos}}_\texttt{sys}
    = \frac1{\sqrt{2^d}}
    \ket{0^d}_\texttt{sel}\ket{\bm{b}}_\texttt{sys} + \ket{\perp}_{\texttt{sel}\otimes\texttt{sys}}
\end{equation}
with $(\bra{0^d}\otimes\id)\ket{\perp}=0$.
This transformation can be verified as
\begin{align}
    (\bra{0^d}\otimes\id) W\ket{0^d}\ket{\text{pos}}
    &= (\bra{0^d}\otimes\id) (H^{\otimes d}\otimes\id) \frac1{\sqrt{2^d}}
    \sum_{\bm{s}}\sum_a \bar{z}_a \ket{\bm{s}}\ket{\bm{\mathrm{x}}_a+\bm{s}}
    \nonumber\\
    &=\frac{1}{2^d}\sum_{\bm{s},\bm{s}'}\sum_a \bar{z}_a
    \langle \bm{s}'|\bm{s}\rangle
    \ket{\bm{\mathrm{x}}_a+\bm{s}}
    \nonumber\\
    &=\frac1{\sqrt{2^d}}\ket{\bm{b}}.
\end{align}
Observe that, condition on success, i.e., the $d$-qubit selector register being in the zero state, the generated state on the second register is the desired initial state~$\ket{\bm{b}}$. The success probability is also known and is exactly~$1/2^d$.
This probabilistic state preparation can be made deterministic using the amplitude-reduction technique~\cite[Sec.~2.3]{BA24} followed by amplitude amplification. 

If the initial success probability is $p=\sin^2\theta$, then the success probability becomes $\sin^2[(2r+1)\theta]$ after~$r$ rounds of amplitude amplification. Perfect success is therefore obtained when initial probability is $p^\star_r=\sin^2[(\pi/2)/(2r+1)]$.
If the initial probability is larger than~$p^\star_r$, we deliberately reduce it to~$p^\star_r$ with smallest~$r$ by appending one ancilla qubit prepared in the state
$R(\alpha)\ket{0}= \cos(\alpha/2)\ket{0}+\sin(\alpha/2)\ket{1}$
and defining success to additionally require this ancilla to be in~$\ket{0}$. The new probability then becomes $p\cos^2(\alpha/2)$, so choosing $\cos^2(\alpha/2)=p^\star_r/p$, makes the initial success probability~$p^\star_r$, and it then can be amplified to exactly~1 by~$r$ rounds of amplitude amplification.

For our $3$D problem~($d=3$), the initial success probability is~$p=1/8>p^\star_2=\sin^2(\pi/10)=(3-\sqrt{5})/8$, so we choose the rotation angle
\begin{equation}
\label{eq:alpha}
    \alpha =2\arccos\left(\sqrt{3-\sqrt{5}}\right)
\end{equation}
for the ancilla qubit to reduce the initial success probability to~$\sin^2(\pi/10)$.
As the flag qubit in \cref{eq:U_pos_action} is returned to~$\ket{0}$, we use it as the ancilla qubit for amplitude reduction.
Let us define
\begin{equation}
\label{eq:U_prob}
    U_\text{prob} := R(\alpha) W U_\text{pos},
\end{equation}
then the probabilistic state preparation for the 3D problem can be written as
\begin{equation}
\label{eq:U_prob_action}
    U_\text{prob}\ket{0}_f\ket{0^3}_\texttt{sel}
    \ket{0^m}_\texttt{idx}\ket{0^n}^3_\texttt{sys} = \sqrt{p_\text{succ}}\ket{\text{G}} + \sqrt{1-p_\text{succ}}\ket{\text{B}},
\end{equation}
where the success probability is $p_\text{succ}=\sin^2(\pi/10)$, the good state is
\begin{equation}
    \ket{\text{G}} = \ket{0}_f\ket{0^3}_\texttt{sel}\ket{0^m}_\texttt{idx}
    \ket{\bm{b}}_\texttt{sys},
\end{equation}
and the bad state that satisfies $\langle\text{B}|\text{G}\rangle=0$.
After just two rounds of amplitude amplification, the good state, and hence the initial state, is generated with unit success probability. That is,
\begin{equation}
    \mathcal{A}^2U_\text{prob} \ket{0}_f\ket{0^3}_\texttt{sel}
    \ket{0^m}_\texttt{idx}\ket{0^n}^3_\texttt{sys} =
    \ket{0}_f\ket{0^3}_\texttt{sel}\ket{0^m}_\texttt{idx}
    \ket{\bm{b}}_\texttt{sys}
\end{equation}
where $\mathcal{A}$ denotes the amplitude-amplification operator.

\begin{figure}[ht]
    \centering
    \includegraphics[width=.94\linewidth]{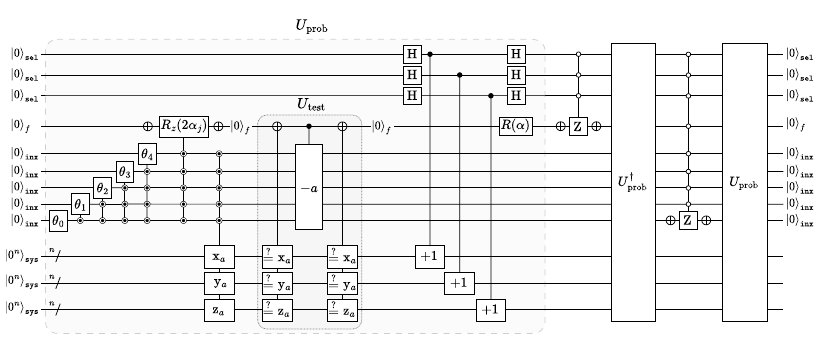}
    \caption{Quantum circuit for loading the initial state with uniform charge distribution. The index register with $m=5$ qubits is shown for illustration. The boxes with $\stackrel{?}{=}$ represent equality testing.}
    \label{fig:loading}
\end{figure}

As the input state is the fixed all-zero state, we use ordinary amplitude amplification with the amplification operator
\begin{equation}
    \mathcal{A} = U_\text{prob}\, R_t\, U^\dagger_\text{prob}\,
    R_\text{G},
\end{equation}
where $R_t:= 2\ket{0^t}\!\!\bra{0^t}-\id_t$
is the reflection about the input state $\ket{0^t}$ with $t=3n+m+4$ the total number of qubits of the input state,
and $R_\text{G}=R_4\otimes\id_{t-4}$ is the reflection about the good state. Notice that $R_\text{G}$ is simply a reflection about the flag and three ancilla qubits, because the index register \texttt{idx} is zero in both good and bad states.
The quantum circuit based on the described approach for loading the initial state with uniform charge distribution is given in~\cref{fig:loading}.

\paragraph{Cost analysis.}
By the described approach, the total cost~$\cost_\text{load}$ of loading the initial state is
\begin{equation}
    \cost_\text{load}=
    5\,\cost(U_\text{prob})+
    2\,\cost(R_\text{G})+
    2\,\cost(R_t),
\end{equation}
where $\cost(U)$ denotes the cost of implementing the unitary~$U$, with cost being either the gate cost~$\gate(U)$ or the circuit depth~$\depth(U)$. 
By~\cref{eq:U_prob} and~\cref{eq:W}
\begin{equation}
    \cost(U_\text{prob})
    = 2d+1+ \cost(U_\text{shift})+\cost(U_\text{pos})
\end{equation}
and by~\cref{eq:U_pos}
\begin{equation}
    \cost(U_\text{pos})
    = \cost(U_{\bm{z}})+ N_\text{atm}\left[
    \cost(\texttt{POS})+\cost(U_\text{erase})
    \right],
\end{equation}
where we used that \texttt{POS} and $U_\text{erase}$ are applied for each atom.
Also, by~\cref{eq:U_erase}, we have
\begin{equation}
    \cost(U_\text{erase})= 2\,\cost(U_\text{test})
    +\cost(\texttt{CADD}).
\end{equation}
We now analyze the cost of the primitive operations:~$U_{\bm{z}}$, \texttt{POS},  $U_\text{test}$, \texttt{CADD}, $U_\text{shift}$, and the reflections~$R_\text{G}$ and $R_t$.
We consider $d=3$ in our cost analysis as our application is naturally a 3D problem, but the generalization to higher~$d$ is straightforward.
We use the following lemma and its corollary in our cost analysis. Proof of~\cref{cor:multi_controlled_reflection} is provided in~\cref{subsec:controlled_reflection}.

\begin{lemma}[{\cite{NZS24} and \cite[Lem.~1]{vandaele26}}]
\label{lem:multi_controlled_not}
    The $k$-controlled $\mathrm{NOT}$ gate, denoted by $C^k(X)$, can be implemented using one dirty ancilla qubit and a circuit with $\Theta(k)$ gates from $\{X,\mathrm{CNOT}, \mathrm{Toffoli}\}$ and $\Theta(\log k)$ depth.
\end{lemma}
\begin{corollary}
\label{cor:multi_controlled_reflection}
    The $k$-controlled reflection operator about the $n$-qubit zero state~$\ket{0^n}$ can be implemented using one dirty ancilla qubit and a circuit with $\Theta(k+n)$ gates from
    $\{X,\mathrm{CNOT}, \mathrm{Toffoli}\}$ and $\Theta(\log (k+n))$ depth.
\end{corollary}

\begin{itemize}
    \item Using the binary-tree approach in~\cref{subsec:loading} for preparing~$\ket{\bm{z}}$, $U_{\bm{z}}$ can be implemented using one clean ancilla and a circuit with~$\Theta(N_\text{atm})$ gates and $\Theta(N_\text{atm})$ depth.
    The cost of classical precomputation is $\Theta(N_\text{atm}\log(N_\text{atm}))$.
    \item \texttt{POS} writes the classically known atom position $\bm{\mathrm{x}}_a$ in grid coordinates into the $3n$-qubit register \texttt{sys} controlled on the $m$-qubit register \texttt{idx}.
    This is a $m$-controlled classical-quantum addition that, by~\cref{lem:classical_quantum_adder}, can be implemented using one dirty ancilla and a circuit with 
    $\Theta(m+ n\log n)$ gates and 
    $\Theta(\log m+\log^2n)$ depth.
     The cost of classical precomputation for computing all atom positions~$\{\bm{\mathrm{x}}_a\}$ is simply $\Theta(N_\text{atm})$.
    \item The equality test~$U_\text{test}$ for each atom is essentially a $3n$-controlled NOT gate. By~\cref{lem:multi_controlled_not}, this can be implemented by one dirty ancilla and a circuit with $\Theta(n)$ gates and $\Theta(\log n)$ depth.
    \item \texttt{CADD} is a 1-controlled addition of the classically known $m$-bit value~$-a$ to the $m$-qubit register \texttt{idx}. By~\cref{lem:classical_quantum_adder}, this can be implemented by one dirty ancilla and a circuit with $\Theta(m\log m)$ gates and $\Theta(\log^2m)$ depth.
    \item The shift operation $U_\text{shift}$ is a 1-controlled increment operator in each direction, so it can be implemented by one dirty ancilla and a circuit with~$\Theta(n)$ gates and $\Theta(\log n)$ depth as per~\cref{lem:multi_controlled_increment}.
    \item The reflection~$R_t$ can be implemented using a dirty ancilla and a circuit with $\Theta(n+m)$ gates and $\Theta(\log(n+1))$ depth as per~\cref{cor:multi_controlled_reflection}.
    The reflection~$R_\text{G}$ is implemented similarly, but with $\Theta(1)$ gates and $\Theta(1)$ depth.
\end{itemize}

Altogether with $m=\ceil{\log_2N_\text{atm}}$, the overall resource scaling for loading the initial state is
\begin{align}
\label{eq:load_gate_cost}
    \gate_\text{load} &= \Theta\left(N_\text{atm}\left(\log N_\text{atm} + n\log n\right)\right),\\
    \depth_\text{load}&=
    \Theta\left(N_\text{atm}\left(\log\log N_\text{atm} + \log^2 n\right)\right),\\
    \cost^\text{classical}_\text{load}
    &=\Theta(N_\text{atm}\log N_\text{atm}),
\end{align}
using $m+5$ clean ancilla qubits and one dirty ancilla. The clean ancillas are used in the $m$-qubit atom-index register, three selector qubits, one flag, and one additional clean ancilla for implementation.

\subsubsection{Loading the initial state with beyond-nearest-neighbor charge distribution}
\label{subsubsec:loading_beyond_NN}

Our approach can be extended to discretizations in which the charge of an atom is distributed over a larger set of neighboring grid points. For instance, in 1D, to distribute the charge of atom~$a$ over three points, we map the Dirac delta onto three grid points $\mathrm{x}_a=\round{x_a/h}$ and $\mathrm{x}_a\pm1$ as
\begin{equation}
    \delta(x-x_a)\mapsto \frac1{4h}
    \left(
    \delta_{j,\mathrm{x}_a-1}+
    2\delta_{j,\mathrm{x}_a}+
    \delta_{j,\mathrm{x}_a+1}
    \right),
\end{equation}
where the weight $w_s=2^{1-\abs{s}}/4$ for $s\in\{0,\pm1\}$ is chosen to preserve the continuum normalization of the Dirac delta. The corresponding normalized initial state in 1D is then
\begin{equation}
\label{eq:initial_state_beyond_NN}
    \ket{\bm{b}} =
    \frac1{\norm{\bm{z}}} \sum_{a=0}^{N_\text{atm}-1}z_a \ket{\psi_a}
\end{equation}
with the weighted atom-position state
\begin{equation}
    \ket{\psi_a}=\frac{1}{\sqrt{6}} 
    \sum_{s\in\{0,\pm1\}}
    2^{1-\abs{s}}
    \ket{\mathrm{x}_a+s}.
\end{equation}
In $d$ dimensions, the weight for $\bm{s}\in\{0,\pm1\}^d$ is
\begin{equation}
\label{eq:dD_weight_unifom_3points}
    w_{\bm{s}}=\frac{2^{d-\norm{\bm{s}}_1}}{4^d} = \frac{1}{2^{d+\norm{\bm{s}}_1}},
\end{equation}
where $\norm{\bm{s}}_1=\sum_\nu\abs{s_\nu}$ is the 1-norm of vector~$\bm{s}$.
The initial state for the $d$-dimensional problem is simply obtained by the tensor product of the 1D atom-position states, in which case we have
\begin{equation}
    \ket{\psi_a}=\frac{1}{\sqrt{6^d}} 
    \sum_{s\in\{0,\pm1\}^d}
    2^{d-\norm{\bm{s}}_1}
    \ket{\bm{\mathrm{x}}_a+\bm{s}}.
\end{equation}
To ensure atoms have disjoint supports needed for mutually orthogonal atom-position states~$\ket{\psi_a}$, we choose grid spacing sufficiently small so that atoms on grid coordinates are separated by at least~$3h$ in at least one spatial direction. A simple and sufficient condition for disjoint supports in~$d$ dimensions is to choose~$h$ so that $\norm{\bm{\mathrm{x}}_a-\bm{\mathrm{x}}_{a'}}\geq 3h\sqrt{d}$.

The modification needed in the previously discussed approach for state loading is that instead of preparing the selector register in the uniform state~$\ket{+}^d$, we prepare it in the selector state defined as
\begin{equation}
    \ket{\text{sel}}=(U\ket{00})^{\otimes d}
    =\frac{1}{2^d}\sum_{\bm{s}}
    \sqrt{2^{d-\norm{\bm{s}}_1}}\ket{\bm{s}}
\end{equation}
where
\begin{align}
    U\ket{00}
    &= \frac12\sum_{s=-1}^{1}\sqrt{2^{1-\abs{s}}}\ket{s}
    \nonumber\\
    &=\frac12(\ket{11}+\sqrt{2}\ket{00}+\ket{01})
    \nonumber\\
    &=C(H)(H\otimes\id)\ket{00}
\end{align}
is the selector state for each direction
with $\ket{11}\leftrightarrow \ket{s=-1}$, $\ket{00}\leftrightarrow \ket{s=0}$ and $\ket{01}\leftrightarrow \ket{s=1}$, and where $C(H)$ is the controlled Hadamard gate.
Note that the selector register now has two qubits per direction.

For the above modification, we replace the right Hadamard in~$W$ given in~\cref{eq:W} with the two-qubit operation~$U$ and the left Hadamard with~$U^\dagger$. Then, as in \cref{eq:W_action}, applying the modified~$W$ on $\ket{0^{2d}}\ket{\text{pos}}$ yields the initial state $\ket{\bm{b}}$ in~\cref{eq:initial_state_beyond_NN} with the success probability~$(3/8)^d$, where the success is conditioned on the $2d$-qubit selector register being in the all-zero state.
This follows as
\begin{align}
    (\bra{0^{2d}}\otimes\id)
    W\ket{0^{2d}}\ket{\text{pos}}
    &= (\bra{0^{2d}}\otimes\id) (U^{\dagger\otimes d}\otimes\id) \frac1{2^d}
    \sum_{\bm{s}}\sqrt{2^{d-\norm{\bm{s}}_1}}
    \sum_a \bar{z}_a \ket{\bm{s}}\ket{\bm{\mathrm{x}}_a+\bm{s}}\\
    &=\frac{1}{4^d}\sum_{\bm{s},\bm{s}'}
    \sqrt{2^{d-\norm{\bm{s}'}_1}}
    \sqrt{2^{d-\norm{\bm{s}}_1}}
    \langle \bm{s}'|\bm{s}\rangle
    \sum_a \bar{z}_a
    \ket{\bm{\mathrm{x}}_a+\bm{s}}\\
    &=\frac1{4^d}\sum_{\bm{s}}
    2^{d-\norm{\bm{s}}_1}
    \sum_a \bar{z}_a
    \ket{\bm{\mathrm{x}}_a+\bm{s}}\\
    &=\frac{\sqrt{6^d}}{4^d}\ket{\bm{b}},
\end{align}
and so the success probability is~$(3/8)^d$.

As the success probability is exactly known, we can amplify it to exactly~$1$ by the technique discussed in~\cref{subsubsec:loading_uniform}. 
For our $3$D problem, the initial success probability is~$p=(3/8)^3>p^\star_3=\sin^2(\pi/14)$, so we choose the rotation angle~$\alpha$ for the ancilla qubit, where
\begin{equation}
    \cos(\alpha/2)=\sqrt{(8/3)^3}\sin(\pi/14),
\end{equation}
to reduce the initial success probability to~$\sin^2(\pi/14)$. Then three rounds of amplitude amplifications generate the desired initial state.
The resource scaling in this case is therefore the same as the resource scaling for the initial state with uniform charge distribution.

\subsubsection{Loading the initial state with position-dependent charge distribution}
\label{eq:subsubsec:loading_nonuniform}
We now extend the charge distribution with equal weights in~\cref{eq:1D_delta_equal_weights} to the general case with position-dependent weights. To this end, we map the 1D Dirac delta as
\begin{equation}
    \delta(x-x_a)\mapsto \frac1{h}
    \left[ (1-\eta_a)\delta_{j,\mathrm{x}_a} + \eta_a \delta_{j,\mathrm{x}_a+1}\right],
\end{equation}
where $\eta_a:= x_a/h-\mathrm{x}_a$ depends on the atom's position and measures the distance between the actual position~$x_a$ of atom~$a$ and its left grid point~$\mathrm{x}_a=\floor{x_a/h}$ as a fraction of~$h$.
Note that the charge~$z_a$ of atom~$a$ is now distributed over the nearest-neighbor points $\mathrm{x}_a+s$ with position-dependent weights $w_{a,s} := (1-\eta_a)^{1-s}\eta^s_a$; the uniform charge distribution is a special case with~$\eta_a=1/2$ that gives $w_{a,s}=1/2$ for each $s\in\{0,1\}$.
In~$d$ dimensions, we have
\begin{equation}
\label{eq:dD_approximate_delta}
    \delta(\bm{x}-\bm{x}_a)\mapsto
    \tilde{\delta}(\bm{x}-\bm{x}_a)=\frac1{h^d}
    \sum_{\bm{s}\in\{0,1\}^d}
    w_{a,\bm{s}}\,
    \delta_{\bm{q},\bm{\mathrm{x}}_a+\bm{s}},
\end{equation}
where the weight
\begin{equation}
\label{eq:dD_weight}
    w_{a,\bm{s}} = \prod_{\nu=0}^{d-1} w_{a,s_\nu},
    \enspace 
    w_{a,s_\nu} := (1-\eta_{a,\nu})^{1-{s_\nu}}\eta^{s_\nu}_{a,\nu},
\end{equation}
is a product of 1D weights $w_{a,s_\nu}$ for each direction~$\nu$ with $\eta_{a,\nu}:= x_{a,\nu}/h-\mathrm{x}_{a,\nu}$.

As weights are now position-dependent, the weighted atom-position state is now
\begin{equation}
    \ket{\psi_a}=
    \frac{1}{\sqrt{w_a}}\sum_{\bm{s}\in\{0,1\}^d} w_{a,\bm{s}}\ket{\bm{\mathrm{x}}_a+\bm{s}},
    \enspace w_a:=\sum_{\bm{s}} w^2_{a,\bm{s}},
\end{equation}
where the normalization~$w_a$ is atom-dependent, unlike the uniform case in~\cref{eq:initial_state_uniform}.
The normalized initial state with position-dependent weights is then
\begin{equation}
    \ket{\bm{b}} =
    \frac1{\norm{\bm{z}'}} \sum_{a=0}^{N_\text{atm}-1} z'_a \ket{\psi_a},
    \enspace
    z'_a:= z_a \sqrt{w_a},
\end{equation}
where~$z'_a$ is the rescaled charge of atom~$a$ and $\norm{\bm{z}'}^2= \sum_a z^2_a w_a$.

To prepare this initial state, we first prepare the state
\begin{equation}
    \ket{\bm{z}'}:=
    \frac1{\norm{\bm{z}'}}\sum_a z'_a\ket{a}
    = \frac1{\norm{\bm{z}'}}\sum_a z_a\sqrt{w_a}\ket{a}
\end{equation}
on the index register.
Then controlled on the state~$\ket{a}$ of this register, we prepare the atom-dependent selector state
\begin{equation}
    \ket{\text{sel}_a}=U_a\ket{0^d}=\frac{1}{\sqrt{w_a}}\sum_{\bm{s}}w_{a,\bm{s}}\ket{\bm{s}}
\end{equation}
on the selector register, where~$U_a$ is a unitary that prepares~$\ket{\text{sel}_a}$. That is to say, we perform the selector operation
\begin{equation}
\label{eq:U_sel}
    U_\text{sel}:=
    \sum_a\ket{a}\!\!\bra{a}\otimes U_a
\end{equation}
on the index and selector registers so that
\begin{equation}
    U_\text{sel} \ket{a}_\texttt{idx}
    \ket{0^d}_\texttt{sel}
    =
    \frac{1}{\sqrt{w_a}}\sum_{\bm{s}}w_{a,\bm{s}}
    \ket{a}_\texttt{idx}\ket{\bm{s}}_\texttt{sel}.
\end{equation}
We now apply \texttt{POS} in~\cref{eq:pos} to load the position of atom~$a$ into \texttt{sys} and then erase \texttt{idx} as before by $U_\text{erase}$ in~\cref{eq:U_erase}.
By the operations so far, we effectively replace $U_\text{pos}$ in~\cref{eq:U_pos} with
\begin{equation}
    U_{\text{pos}'} := 
    U_\text{erase} \texttt{POS}\, U_\text{sel}\, U_{\bm{z}'}
\end{equation}
that prepares the state
\begin{equation}
\label{eq:pos_prime}
    \ket{\text{pos}'}:=
    \frac{1}{\norm{\bm{z}'}}
    \sum_{a} z_a \sum_{\bm{s}} w_{a,\bm{s}}
    \ket{\bm{s}}
    \ket{\bm{\mathrm{x}}_a}
\end{equation}
on the selector and system registers.
That is,
\begin{equation}
\label{eq:U_pos_prime}
    U_{\text{pos}'}:
    \ket{0^d}_\texttt{sel}
    \ket{0^n}^{\otimes d}_\texttt{sys}
    \mapsto
    \ket{\text{pos}'}_{\texttt{sel}\otimes\texttt{sys}},
\end{equation}
where the action on the flag qubit and the index register is omitted as they start and finish in the zero state.

Next we apply the shift operator in~\cref{eq:Ushift} to perform the map $\ket{\bm{\mathrm{x}}_a} \mapsto \ket{\bm{\mathrm{x}}_a+\bm{s}}$.
Notice that we cannot now apply the inverse of~$U_\text{sel}$ on the selector register to conditionally prepare the initial state as in~\cref{eq:W_action}, because~$U_\text{sel}$ depends on the atom index and that the index register has been erased.
However, applying $H^{\otimes d}$ on the selector register does prepare the initial state on \texttt{sys} conditioned on \texttt{sel} being in the all zero state. Specifically, we have  
\begin{equation}
    (H^{\otimes d}\otimes\id)U_\text{shift}
    \ket{\text{pos}'}_\texttt{sys}
    = \frac{1}{\sqrt{2^d}}\ket{0^d}_\texttt{sel}\ket{\bm{b}} + \ket{\perp}_{\texttt{sel}\otimes\texttt{sys}}
\end{equation}
with $(\bra{0^d}\otimes\id)\ket{\perp}=0$.
To show this, observe that
\begin{align}
    (\bra{0^d}\otimes\id)(H^{\otimes d}\otimes\id)U_\text{shift}
    \ket{\text{pos}'}
    &=
    (\bra{0^d}\otimes\id)(H^{\otimes d}\otimes\id)
    \frac{1}{\norm{\bm{z}'}}
    \sum_{a} z_a \sum_{\bm{s}} w_{a,\bm{s}}
    \ket{\bm{s}}
    \ket{\bm{\mathrm{x}}_a+\bm{s}}
    \nonumber\\
    &=\frac{1}{\sqrt{2^d}}
    \frac{1}{\norm{\bm{z}'}}
    \sum_{a} z_a \sum_{\bm{s},\bm{s}'} w_{a,\bm{s}}
    \langle\bm{s}'|\bm{s}\rangle
    \ket{\bm{\mathrm{x}}_a+\bm{s}}
    \nonumber\\
    &=\frac{1}{\sqrt{2^d}}\ket{\bm{b}}.
\end{align}
The success probability is known and is exactly~$1/2^d$, as in the case with uniform charge distribution.
This probabilistic state loading can therefore be made deterministic using the same steps discussed in~\cref{subsubsec:loading_uniform}.
Specifically, for the 3D problem, we prepare the flag qubit in the state~$R(\alpha)\ket{0}$ with the rotation angle~$\alpha$ given in~\cref{eq:alpha}.
Then by replacing
\begin{equation}
    U_\text{prob} \rightarrow U_{\text{prob}'} := R(\alpha) H^{\otimes d} U_\text{shift} U_{\text{pos}'},
\end{equation}
we have the same probabilistic state loading in \cref{eq:U_prob_action} with the initial success probability of~$\sin^2(\pi/10)$.
Then by two rounds of amplitude amplification, with the operator
\begin{equation}
    \mathcal{A}' := U_{\text{prob}'}\, R_t\, U^\dagger_{\text{prob}'}\,
    R_\text{G},
\end{equation}
the initial state is generated with unit success probability.

Finally, we decompose the unitary~$U_a$ for preparing the atom-dependent selector state $\ket{\text{sel}_a}$ as a tensor product of single-qubit rotations.
This follows because the weight~$w_{a,\bm{s}}$ is a product of 1D weights as in \cref{eq:dD_weight}.
To see this, observe that
\begin{align}
    \ket{\text{sel}_a}
    &=\frac{1}{\sqrt{w_a}} \sum_{\bm{s}}
    \prod_{\nu=0}^{d-1} w_{a,s_\nu}\ket{\bm{s}}
    \nonumber\\
    &= \frac{1}{\sqrt{w_a}} \bigotimes_{\nu=0}^{d-1}
    \sum_{s_\nu=0}^1 w_{a,s_\nu}\ket{s_\nu}
    \nonumber\\
    &= \bigotimes_{\nu=0}^{d-1}
    \frac{1}{\sqrt{\mathcal{N}_{a,\nu}}}
    \sum_{s_\nu=0}^1 w_{a,s_\nu}\ket{s_\nu}
    \nonumber\\
    &=\bigotimes_{\nu=0}^{d-1} R\left(\theta_{a,\nu}\right)\ket{0},
\end{align}
where, in the third line, we used
\begin{equation}
    w_a = \sum_{\bm{s}} w^2_{a,\bm{s}} =
    \sum_{s_0,\ldots,s_{d-1}}\prod_{\nu=0}^{d-1} w^2_{a,s_\nu}
    = \prod_{\nu=0}^{d-1} \mathcal{N}_{a,\nu}
\end{equation}
with
\begin{equation}
    \mathcal{N}_{a,\nu} :=
    \sum_{s_\nu=0}^1 w^2_{a,s_\nu}
    = \left(1-\eta_{a,\nu}\right)^2 + \eta^2_{a,\nu},
\end{equation}
and where the rotation angles satisfy
\begin{equation}
\label{eq:theta_a_nu}
    \tan\left(\frac{\theta_{a,\nu}}{2}\right) = \frac{\eta_{a,\nu}}{1-\eta_{a,\nu}}.
\end{equation}
The unitary~$U_a$ is therefore a tensor product of $y$-rotation gates with angles~$\theta_{a,\nu}$.
Consequently, $U_\text{sel}$ in \cref{eq:U_sel} for preparing the selector state is a uniformly controlled-rotation gate, which can be implemented by $\Theta(N_\text{atm})$ single-qubit rotations and $\Theta(N_\text{atm})$ CNOT gates.
The cost of classical precomputation of the rotation angles $\{\theta_{a,\nu}\}$ is also $\Theta(N_\text{atm})$.
The resource scaling in this case is also the same as the resource scaling for the initial state with uniform charge distribution.

\subsection{The quantum solution and tailored implementation for the operator oracle}
\label{subsec:tailoredBE}
In this section, we develop a tailored quantum algorithm for generating a quantum state that encodes the solution of the discretized LPB equation for simulating the biomolecular system.
Our approach builds on the quantum PDE algorithm introduced in~\cite{BNW+26} that uses wavelet preconditioning to solve a broad class of PDEs with a cost independent of the condition number of the discretized differential operator.
The differential operator of the LPB equation in
\cref{eq:differential_operator} has bounded coefficients,
with~$\varepsilon(\bm{x})$ strictly positive and~$k(\bm{x})$ nonnegative, and belongs to the class of elliptic operators considered that can be optimally preconditioned by wavelets~\cite{BNW+26}.
We therefore adopt the distributed quantum solution introduced in~\cite{BNW+26} to construct a quantum solution of the discretized system.

We begin by describing the distributed quantum solution in \cref{subsubsec:distributed_quantum_solution}, tailor it for our application, and show how it can be used to extract properties of the system.
We then discuss our approach for generating the distributed quantum solution in~\cref{subsec:generating_distributed_solution}.
We continue in~\cref{subsubsec:cost_error_analysis} to analyze the error in generating the solution state and its gate cost in terms of the gate cost of a block encoding of the preconditioned matrix.
Finally, we construct a block encoding for the PDE operator in~\cref{subsubsec:operator_oracle_implementation} and a block encoding for the preconditioned matrix in~\cref{subsubsec:block_encoding_Aprec}, and discuss their gate cost, block encoding normalization and the number of ancillas used for block encodings.

\subsubsection{The distributed quantum solution}
\label{subsubsec:distributed_quantum_solution}
Let $A\bm{u}=\bm{b}$ be the system of linear equations obtained from the finite-difference discretization of the LPB equation, where $\bm{b}$ is the normalized vector discussed in~\cref{subsec:initial_state},
and let $\ket{\bm{u}}$ be the unnormalized solution vector.
Following~\cite{BNW+26}, let us consider the preconditioned matrix
\begin{equation}
    A_\precond = PWAW^\dagger P,
\end{equation}
where~$W$ is the wavelet transform matrix and~$P$ is the diagonal wavelet preconditioner; here we consider the~1D problem; extension to the~3D follows similarly~\cite{BNW+26}.
Then the solution can be expressed as $\ket{\bm{u}}=W^\dagger P A^{-1}_\precond PW\ket{\bm{b}}$ with $\ket{\bm{b}}$ the normalized initial state in~\cref{subsec:initial_state}.
The preconditioned matrix $A_\precond$ has a condition number, $\kappa_\precond\in\Theta(1)$, independent of the matrix size or the grid resolution, so $A^{-1}_\precond$ can be implemented efficiently by a quantum linear-system algorithm.
The transformation by~$W$ and its inverse can also be implemented efficiently by the quantum wavelet-transform algorithm~\cite{BA24}. However, directly applying the wavelet preconditioner~$P$ to prepare the conventional solution state $\ket{\bm{u}}/\norm{\bm{u}}$ would make the cost of the algorithm linear in the matrix size. 

To bypass this bottleneck, the algorithm in~\cite{BNW+26} expresses~$P$ as a combination of two unitaries $U^\pm=\exp(\pm\i \arccos P)$ as $P=(U^++U^-)/2$, where each unitary can be efficiently implemented by controlled-rotation gates.
By this decomposition, we obtain $\ket{\bm{u}}=\frac14\sum_{a,b}\ket{\psi_{ab}}$,
where
\begin{equation}
    \ket{\psi_{ab}}:=W^\dagger U^a A^{-1}_\precond U^b W\ket{\bm{b}},
    \enspace a,b\in\{0,1\},
\end{equation}
and $U^{0/1}\equiv U^{+/-}$.
Ref.~\cite{BNW+26} then constructs the distributed quantum solution state in the extended space by two qubits as
\begin{equation}
    \ket{\psi} = \frac{1}{2\xi} \sum_{a,b\in\{0,1\}} \ket{ab}\ket{\psi_{ab}},
    \enspace
    \xi^2:=\frac14\sum_{a,b}\norm{\ket{\psi_{ab}}}^2,
\end{equation}
which provides an alternative to the conventional solution state~$\ket{\bm{u}}/\norm{\bm{u}}$.
The central property of this representation is that~$\ket{\bm{u}}$ can be obtained by projecting the two qubits onto the uniform state~$\ket{++}$ as $\xi\left(\bra{++}\otimes\id\right)
\ket{\psi}=\ket{\bm{u}}$.
This identity allows us to extract the solution property $\langle\bm{u}|M|\bm{u}\rangle$ from the distributed quantum solution. 
Specifically, we have
\begin{equation}
    \langle\bm{u}|M|\bm{u}\rangle
    =\xi^2\langle\psi|\left(\ket{++}\!\!\bra{++}\otimes M\right)|\psi\rangle
    =\xi^2\langle\psi|M'|\psi\rangle
\end{equation}
where
\begin{equation}
    M' := \ket{++}\!\!\bra{++}\otimes M
    = \frac14\sum_{a,b,c,d\in\{0,1\}} \ket{ab}\!\!\bra{cd}\otimes M,
\end{equation}
is the observable in the extended space.
The observable~$M'$ is a $4\times4$ block matrix whose blocks are all equal to~$M/4$.
The block encoding~$U_{M'}$ of $M'$ is then constructed by a single query to the block encoding of~$M$ as
\begin{equation}
    U_{M'}= V\otimes U_M,
\end{equation}
where
\begin{equation}
    V = (H\otimes H\otimes\id_f) X_f C^2_0(X_f)(H\otimes H\otimes\id_f) 
\end{equation}
is a $(1,1,0)$ block encoding of $\ket{++}\!\!\bra{++}$ using an ancilla qubit~$f$, and $C^2_{0}(X_f)$ flips~$f$ when the two control qubits are~$\ket{00}$.
Consequently, the distributed quantum solution enables access to properties of the original solution without directly preparing the original solution state.

For our application, we also discuss how the distributed quantum solution can be used to extract the expectation value of the form $\langle\phi|M|\phi\rangle$ with the unnormalized state $\ket{\phi}$ and the observable~$M$ defined as
\begin{equation}
\label{eq:phi_M}
    \ket{\phi}:=\frac{1}{\sqrt{2}}\left(
    \ket{0}\ket{\bm{u}}+\ket{1}\ket{\text{pos}}
    \right),
    \enspace
    M=:\ket{0}\!\!\bra{1}\otimes T + \ket{1}\!\!\bra{0}\otimes T^\dagger,
\end{equation}
where $T$ is a known operator, and $\ket{\text{pos}}$ is a normalized quantum state with a known preparation unitary~$U_\text{pos}:\ket{0}\mapsto \ket{\text{pos}}$; we discuss these operators later in~\cref{subsec:obs}.
To this end, let us define the normalized state
\begin{equation}
    \ket{\Psi}:=\frac{1}{\sqrt{2}}\left(
    \ket{0}\ket{\psi}+\ket{1}\ket{\chi}
    \right),\enspace
    \ket{\chi}:=\ket{++}\ket{\text{pos}}
\end{equation}
where $\ket{\psi}$ is the distributed quantum solution and $\ket{++}$ is the uniform state on the first two qubits of $\ket{\psi}$.
This state can be prepared by applying a Hadamard on the leftmost qubit and then applying controlled preparation of~$\ket{\psi}$ and~$\ket{\chi}$. 
By the identity $\xi\left(\bra{++}\otimes\id\right)
\ket{\psi}=\ket{\bm{u}}$, we then have
\begin{equation}
\label{eq:phi_M_prime}
    \langle\phi|M|\phi\rangle=
    \xi\langle\Psi|M'|\Psi\rangle,
    \enspace
    M'=\ket{0}\!\!\bra{1}\otimes\id_2\otimes T + \ket{1}\!\!\bra{0}\otimes\id_2\otimes T^\dagger.
\end{equation}
Notice that $M'$ here only has $\id_2$ on the first two qubits of~$\ket{\psi}$.
This construction is particularly useful for computing properties of the biomolecular system as we discuss in~\cref{subsec:obs}.

\subsubsection{Generating the distributed quantum solution}
\label{subsec:generating_distributed_solution}
We now describe a quantum circuit for generating the distributed solution state.
We derive a lower bound on the success amplitude for generating this state and discuss the fixed-point amplitude amplification.
To generate the state, we use Quantum Matrix Inversion, \texttt{QMI}, as the block encoding of $A^{-1}_\precond$. Specifically, \texttt{QMI} is an $(\alpha_\text{inv},a_\text{inv},\varepsilon_\text{inv})$ block encoding of $A^{-1}_\precond$ such that
\begin{equation}
    \norm{A^{-1}_\precond-\alpha_\text{inv}
    \left(\bra{0^{a_\text{inv}}}\otimes\id\right)
    \,\texttt{QMI}\,
    \left(\ket{0^{a_\text{inv}}}\otimes\id\right)
    }\leq \varepsilon_\text{inv},
\end{equation}
where $a_\text{inv}$ is the number of ancilla qubits in the block encoding and the normalization factor satisfies $\alpha_\text{inv}\geq \norm{A^{-1}_\precond}$.
Later we construct \texttt{QMI} with the parameters
\begin{equation}
\label{eq:QMI_params}
    \alpha_\text{inv}=\frac{4\kappa_\precond}{3\norm{A_\precond}},
    \enspace
    a_\text{inv}=a_\precond+1,
    \enspace
    \varepsilon_\text{inv}=\frac{\varepsilon}{4\norm{A_\precond}},
\end{equation}
where $a_\precond$ is the number of ancillas in the block encoding of $A_\precond$ and $\varepsilon$ denotes the target error in generating the state.

We first prepare the $(n+1)$-qubit state
\begin{equation}
    \ket{\phi_\text{in}}:=
    \frac1{\sqrt{2}}
    \sum_{b\in\{0,1\}}\ket{b}U^b W\ket{\bm{b}}.
\end{equation}
This state is obtained by preparing~$\ket{\bm{b}}$, applying the quantum wavelet transform~$W$, and using a selector qubit in the~$\ket{+}$ state to coherently apply~$U^b$ conditioned on its value~$b$.
We then apply \texttt{QMI} on the system register (with no action on the selector qubit). The action of \texttt{QMI} is then
\begin{equation}
\label{eq:QMI_action}
    \texttt{QMI} \ket{0^{a_\text{inv}}}
    \ket{\phi_{\text{in}}}=
    \frac1{\alpha_\text{inv}}\ket{0^{a_\text{inv}}}
    \ket{\wt{\phi}_\text{out}} + \ket{\perp},
\end{equation}
where $\ket{\wt{\phi}_\text{out}}$ is an $\varepsilon_\text{inv}$-approximation of the ideal output state $\ket{\phi_\text{out}}:=(\id_1\otimes A^{-1}_\precond)\ket{\phi_\text{in}}$ with $A^{-1}_\precond$ replaced with its approximation~$\wt{A}^{-1}_\precond$. The state $\ket{\perp}$ is orthogonal to the all-zero state of the block encoding ancillas, $(\bra{0^{a_\text{inv}}}\otimes\id)\ket{\perp}=0$.
The all-zero state of the ancilla qubits then determines the successful application of the approximate inverse.

To obtain a lower bound on the success amplitude, let us write~\cref{eq:QMI_action} as
\begin{equation}
    \ket{\Phi}:=\texttt{QMI} \ket{0^{a_\text{inv}}}\ket{\phi_{\text{in}}} = \sqrt{p_\text{succ}}\ket{0^{a_\text{inv}}}\ket{\wt{\text{G}}}
    +\sqrt{1-p_\text{succ}}\ket{\wt{\text{B}}},
\end{equation}
where $\ket{\wt{\text{G}}}$ is an approximation of the normalized output state $\ket{\text{G}}:=\ket{\phi_{\text{out}}}/\norm{\ket{\phi_\text{out}}}$, and where $\ket{\wt{\text{B}}}$ has no support on the zero state of the block encoding ancillas.
The success probability $p_\text{succ}$ satisfies
\begin{equation}
\label{eq:psucc_lowerbound}
    \sqrt{p_\text{succ}}=
    \frac{\norm{\ket{\wt{\phi}_\text{out}}}}{\alpha_\text{inv}} \geq \frac{9}{16 \kappa_\precond} \in \Theta(1).
\end{equation}
This is because
\begin{equation}
    \frac{1}{\norm{A_\precond}}\leq \norm{\ket{\phi_\text{out}}}\leq \norm{\ket{\wt{\phi}_\text{out}}}+\varepsilon_\text{inv},
\end{equation}
where the right-hand side follows from the triangle inequality, and so we obtain the lower bound on success amplitude as
\begin{equation}
    \sqrt{p_\text{succ}}
    \geq\frac{1}{\alpha_\text{inv}}\left(\frac{1}{\norm{A_\precond}}-\varepsilon_\text{inv}\right)
    \geq\frac{3}{4}\frac{1-\varepsilon/4}{\kappa_\precond} \geq \frac{9}{16\kappa_\precond},
\end{equation}
where we used \cref{eq:QMI_params}.
After amplitude amplification, we use a second selector qubit in the~$\ket{+}$ state to coherently apply~$U^a$ conditioned on its value~$a$, and then apply the inverse quantum wavelet transform~$W^\dagger$ on the system register. The generated state becomes an~$\varepsilon$-approximation of the distributed solution state.

\paragraph{Fixed-point amplitude amplification.}
As the exact value of the success probability~$p_\text{succ}$ is unknown, the ordinary amplitude amplification cannot be applied.
We instead use the fixed-point amplitude amplification~(FPAA)~\cite{YLC14} as it only needs a guaranteed lower bound on~$p_\text{succ}$, as in our case in~\cref{eq:psucc_lowerbound}, to make the final success probability arbitrarily close to one.
To apply the FPAA, let us define the generalized reflection operators about~$\ket{\wt{\text{G}}}$ and $\ket{\Phi}$ respectively as
\begin{equation}
\label{eq:reflections}
    R_{\wt{\text{G}}}(\beta):=\id - (1-\e^{\i\beta})\Pi_{\wt{\text{G}}},
    \enspace
    R_{\Phi}(\alpha):=\id- (1-\e^{-\i\alpha})\Pi_\Phi,
\end{equation}
with the projections $\Pi_{\wt{\text{G}}}:=\ket{{\wt{\text{G}}}}\!\!\bra{{\wt{\text{G}}}}$ and $\Pi_\Phi:=\ket{\Phi}\!\!\bra{\Phi}$.
Here $\alpha$ and $\beta$ are some rotation angles; the ordinary reflection is a particular case with $\alpha=\beta=\pm\pi$.
Let $L=2\ell+1$ be an odd integer, then the FPAA is a sequence of~$\ell$ generalized Grover iterates $G(\alpha,\beta)$ as
\begin{equation}
    \A_L = G(\alpha_\ell,\beta_\ell)\cdots G(\alpha_1,\beta_1),
    \enspace G(\alpha,\beta) := -R_{\Phi}(\alpha)R_{\wt{\text{G}}}(\beta)
\end{equation}
with some angles~$\{(\alpha_j,\beta_j)\}$ to be specified, which transforms the state~$\ket{\Phi}$ as
\begin{equation}
    \A_L \ket{\Phi} = \sqrt{p_L}\,\e^{\i\theta_{\wt{\text{G}}}} \ket{0^{a_\text{inv}}}\ket{{\wt{\text{G}}}}
    +\sqrt{1-p_L}\,\e^{\i\theta_{\wt{\text{B}}}}
    \ket{\wt{\text{B}}},
\end{equation}
where $\theta_{\wt{\text{G}}}$ and $\theta_{\wt{\text{B}}}$ are phases of $\ket{{\wt{\text{G}}}}$ and $\ket{\text{B}}$, respectively, and~$p_L$ is the amplified success probability.
That is to say, only the amplitudes and phases of these states are changed, but the states remain unchanged.
Following the FPAA construction in~\cite{YLC14}, the angles in the sequence~$\A_L$ are chosen as
\begin{equation}
    \alpha_j=-\beta_{\ell-j+1} = 2\cot^{-1}\left[\tan\left(\frac{2\pi j}{L}\right)\sqrt{1-\gamma^2}\right],
\end{equation}
where
\begin{equation}
    \gamma^{-1}
    =\cosh\left[\frac{1}{L}\operatorname{arccosh}
    \left(\frac{1}{\delta}\right)\right]
\end{equation}
for some $0<\delta<1$.
The FPAA guarantees $p_L\geq1-\delta^2$ when $p_\text{succ}\geq1-\gamma^2$.

We now determine a sufficient length~$L$ for the sequence using the lower bound on the initial success amplitude.
Let $\sqrt{p_\text{min}}=9/16\kappa_\precond$ be the lower bound for $\sqrt{p_\text{succ}}$ in~\cref{eq:psucc_lowerbound}.
To have~$p_\text{succ}\geq 1-\gamma^2$, we take~$p_\text{min}\geq 1-\gamma^2$, from which we obtain $L\geq L_\star$ with
\begin{equation}
    L_\star=\frac{\operatorname{arccosh}(1/\delta)}{\operatorname{arctanh}(\sqrt{p_\text{min}})}
    \leq \frac{1}{\sqrt{p_\text{min}}}\ln\left(\frac{2}{\delta}\right),
\end{equation}
where the upper bound is achieved using $\operatorname{arccosh}(1/x)\leq \ln(2/x)$
and $\operatorname{arctanh}(x)\geq x$ for any $x\in(0,1)$. We then take the smallest odd integer~$L$ that satisfies
\begin{equation}
\label{eq:L_FPAA}
    L\geq \frac{1}{\sqrt{p_\text{min}}}\ln\left(\frac{2}{\delta}\right)=
    \frac{16}{9}
    \kappa_\precond\ln\left(\frac{2}{\delta}\right),
\end{equation}
from which we obtain $\ell=(L-1)/2$.
With this~$L$, the FPAA guarantees the approximate good state $\ket{\wt{\text{G}}}$ is prepared with the amplified success probability of $p_L\geq 1-\delta^2$.
Because $\kappa_\precond\in\Theta(1)$, the length of the sequence scales as~$L\in\Theta(\log(1/\delta))$.
Finally, we remark that when the phase of the amplified good state matters, one can instead use the QSVT-based FPAA~\cite[Theorem~27]{GSL+19}, which avoids the unknown phase issue and achieves the same same scaling for~$L$.

\begin{figure}[ht]
    \centering
    \includegraphics[width=.76\linewidth]{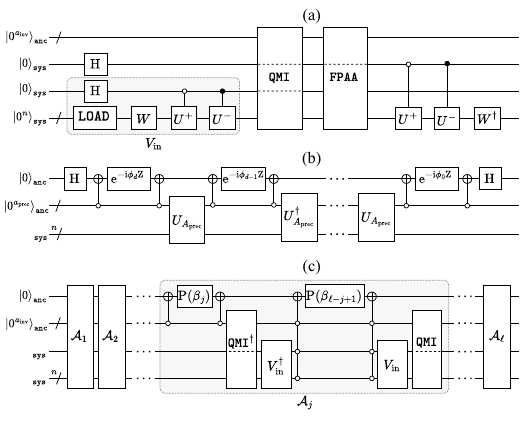}
    \caption{Quantum circuits for generating the distributed quantum solution state. (a)~The main circuit based on high-level operations. (b)~Implementation of \texttt{QMI} using the block encoding of the preconditioned matrix, its inverse, and low-level operations. (c)~Implementation of \texttt{FPAA} by the operations used in the main circuit.}
    \label{fig:qsolution}
\end{figure}

\paragraph{Implementing FPAA.} 
To implement the FPAA, note that~$\ket{\Phi}$ is a $t$-qubit state with $t=a_\text{inv}+n+1$ as $a_\text{inv}$ ancillas are used in \texttt{QMI} and $\ket{\phi_\text{in}}$ is a $(n+1)$-qubit state.
Also, note that the good subspace is determined by the ancilla qubits, so we have $\Pi_{\wt{\text{G}}}=\Pi_{a_\text{inv}}\otimes\id_{n+1}$, where $\Pi_m:=\ket{0^m}\!\!\bra{0^m}$ is the projection onto the $m$-qubit zero state.
Let $V_\text{in}$ be the unitary that prepares $\ket{\phi_\text{in}}$, then the projection onto $\ket{\Phi}$ can be decomposed as $\Pi_\Phi=\ket{\Phi}\!\!\bra{\Phi}=
\texttt{QMI}\, V_\text{in}\,\Pi_t\, V_\text{in}^\dagger\, \texttt{QMI}^\dagger$.
Let $R_m(\theta):=
    \id_m - (1-\e^{\i\theta})\Pi_m$
be the generalized reflection about the $m$-qubit zero state, then we have the decompositions
\begin{equation}
    R_{\wt{\text{G}}}
    (\beta)= R_{a_\text{inv}}(\beta),
    \enspace
    R_\Phi(\alpha)=\texttt{QMI}\, V_\text{in}\,
    R_t(-\alpha)\,
    V_\text{in}^\dagger\, \texttt{QMI}^\dagger
\end{equation}
for the reflection operators in~\cref{eq:reflections}.
With $\alpha_j=-\beta_{\ell-j+1}$ we achieve
\begin{equation}
\label{eq:A_L}
    \mathcal{A}_L= \prod_{j=\ell}^{1}
    \texttt{QMI}\, V_\text{in}\,
    R_t(\beta_{\ell-j+1})\,
    V_\text{in}^\dagger\, \texttt{QMI}^\dagger R_{a_\text{inv}}(\beta_j),
\end{equation}
where $L=2\ell+1$. We use this decomposition to construct the quantum circuit in~\cref{fig:qsolution}~(c) for implementing~\texttt{FPAA}. The generalized reflections here are implemented in the circuit using the phase gate $\text{P}(\theta)=\operatorname{diag}(1,\exp(\i\theta))$ based on the approach discussed in~\cref{subsec:controlled_reflection}. 

\subsubsection{Cost and error analysis for the distributed quantum solution}
\label{subsubsec:cost_error_analysis}
Here we first analyze the query cost for implementing \texttt{QMI} using the QSVT approach.
Then we analyze how the matrix-inversion error and the failure amplitude after FPAA contribute to the error in the generated solution state.
Finally, we use the query cost and the gate cost of other operations to obtain the overall gate cost of generating the quantum solution in terms of the gate cost of a block encoding of the preconditioned matrix.

\paragraph{Query cost.}
To analyze the query cost, first we consider a general linear system $A\bm{x}=\bm{b}$ with an invertible Hermitian matrix~$A$, and then specialize the result for the preconditioned matrix $A_\precond$ in our case.
As a standard assumption in quantum linear-system solvers, suppose that $\norm{A}\leq 1$ and that the spectrum of~$A$ satisfies $\operatorname{spec(A)}\subseteq I_\Delta:=[-1,-\Delta]\cup[\Delta,1]$, where $\Delta$ is a lower bound on the absolute eigenvalues of~$A$.
In particular, for $\norm{A}\leq1$, we can choose $\Delta= \norm{A}/\kappa$ with~$\kappa$ the condition number of~$A$.

Let $p(x)$ be an odd real polynomial of degree $d\in\order{\Delta^{-1}\log(1/\varepsilon')}$ that satisfies $\abs{p(x)}\leq 1$ for all $x\in[-1,1]$ and $\abs{p(x)-3\Delta/(4x)}\leq \varepsilon'$ for all~$x\in I_\Delta$; the factor here ensures~$1/x$ is appropriately rescaled for a bounded polynomial approximation. 
Given a $(1,a_A,0)$-block encoding~$U_A$ of~$A$, i.e., a perfect block encoding with zero error, QSVT constructs a $(1,a_A+1,\varepsilon')$-block encoding $U_{p(A)}$ of $p(A)$ using~$d$ queries to~$U_A$ and~$U^\dagger_A$.
The polynomial approximation guarantees $\norm{p(A)-(3\Delta/4)A^{-1}}\leq \varepsilon'$, and by rescaling we obtain
\begin{equation}
\label{eq:QSVT_error}
    \norm{\frac{4}{3\Delta} p(A)- A^{-1}}\leq \frac{4\varepsilon'}{3\Delta}.
\end{equation}
Consequently, $U_{p(A)}$ is an $(\alpha_\text{inv},a_A+1,\varepsilon_\text{inv})$ block encoding of~$A^{-1}$ with
\begin{equation}
    \alpha_\text{inv}:=\frac{4}{3\Delta},
    \enspace
    \varepsilon_\text{inv}:=
    \frac{4\varepsilon'}{3\Delta}.
\end{equation}
For $\Delta=\norm{A}/\kappa$, these become $\alpha_\text{inv}=4\kappa/3\norm{A}$ and
$\varepsilon_\text{inv}=4\kappa\varepsilon'/3\norm{A}$. Therefore, we can express the polynomial degree in terms of properties of~$A$ and the error $\varepsilon_{\text{inv}}$ in approximating~$A^{-1}$ as
$d\in\order{(\kappa/\norm{A})\log(\kappa/(\norm{A}\varepsilon_\text{inv}))}$.
If~$\varepsilon$ is the error in generating an approximation of the normalized solution state $\ket{x}/\norm{\ket{x}}$ of the linear system with $\ket{x}:=A^{-1}\ket{b}$, then choosing $\varepsilon_\text{inv}$ such that $2\norm{A}\varepsilon_\text{inv}\leq \varepsilon$ guarantees that the approximate state is within~$\varepsilon$ error of the ideal state. The query cost then becomes $\order{(\kappa/\norm{A})\log(\kappa/\varepsilon)}$.
If $\norm{A}=1$, the familiar scaling $\order{\kappa\log(\kappa/\varepsilon)}$ for the query cost is obtained.

We now consider the more general case in which the block encoding of~$A$ is not exact. Suppose $U_A$ is a $(1,a_A,\varepsilon_A)$ block encoding of~$A$ with tunable error $\varepsilon_A$.
By the robustness of QSVT~\cite[Lemma~22]{GSL+19}, the imperfect block encoding introduces an additional error bounded by $4d\sqrt{\varepsilon_A}$ in the polynomial approximation; it makes the total error in approximating $(3\Delta/4)A^{-1}$ to increase from $\varepsilon'$ to $\varepsilon'+4d\sqrt{\varepsilon_A}$. By imposing the condition $4d\sqrt{\varepsilon_A}\leq \varepsilon'$ on~$\varepsilon_A$, the total error is then at most $2\varepsilon'$, and hence $\varepsilon_\text{inv}$ now becomes $\varepsilon_\text{inv}=8\varepsilon'/3\Delta$.

In our case, \texttt{QMI} block encodes $A^{-1}_\precond$ using a block encoding of $A_\precond$ with the norm $\norm{A_\precond}\in\Theta(1)$ and the condition number $\kappa_\precond\in\Theta(1)$.
In~\cref{subsubsec:block_encoding_Aprec}, we construct an $(\alpha_\precond,a_\precond,\varepsilon_\precond)$ block encoding $U_{A_\precond}$ of $A_\precond$ with the normalization $\alpha_\precond\in\Theta(n)$ and the error~$\varepsilon_\precond$ that satisfies the condition $4d\sqrt{\varepsilon_\precond/\alpha_\precond}\leq \varepsilon'$.
Equivalently,~$U_{A_\precond}$ is a $(1,a_\precond,\varepsilon_\precond/\alpha_\precond)$ block encoding of the rescaled matrix $A_\precond/\alpha_\precond$ whose norm is~$\leq1$ and whose eigenvalues are greater than $\Delta=\norm{A_\precond}/(\alpha_\precond\kappa_\precond)$. The \texttt{QMI} built by the QSVT approach, i.e., $\texttt{QMI}=U_{p(A_\precond/\alpha_\precond)}$, then provides an $(\alpha_\text{inv},a_\precond+1,\varepsilon_\text{inv})$ block encoding of $A^{-1}_\precond$, where
\begin{equation}
    \alpha_\text{inv}=\frac{4}{3\Delta\alpha_\precond}=\frac{4\kappa_\precond}{3\norm{A_\precond}},
    \enspace
    \varepsilon_\text{inv}= \frac{8\varepsilon'}{3\Delta\alpha_\precond}=
    \frac{8\kappa_\precond\varepsilon'}{3\norm{A_\precond}}
\end{equation}
by \cref{eq:QSVT_error} with $A\rightarrow A_\precond/\alpha_\precond$ and $\varepsilon'\rightarrow2\varepsilon'$.
By expressions of $\Delta$ and $\varepsilon'$, it follows that \texttt{QMI} makes
\begin{equation}
\label{eq:query_cost}
    \mathcal{Q}_\texttt{QMI}\in
    \order{
    \left(\frac{\alpha_\precond\kappa_\precond}{\norm{A_\precond}}\right)
    \log\left(
    \frac{\kappa_\precond}{\norm{A_\precond}\varepsilon_\text{inv}}
    \right)}
\end{equation}
queries to $U_{A_\precond}$ and $U^\dagger_{A_\precond}$
In the following error analysis, we show that choosing $\norm{A_\precond}\varepsilon_\text{inv}\leq \varepsilon/4$ guarantees that the normalized distributed solution state is obtained with error~$\leq\varepsilon$.
With $\alpha_\precond\in\Theta(n)$ and $\kappa_\precond, \norm{A_\precond}\in\Theta(1)$, we have
\begin{equation}
    \mathcal{Q}_\texttt{QMI}
    \in\order{n\log(1/\varepsilon)}
\end{equation}
for the query cost.
Note that the factor~$n$ in the query cost is due to the normalization factor in the block encoding of~$A_\precond$.

\paragraph{Error analysis.}
We next bound the error in the normalized distributed solution state.
The two sources of error are the approximation of $A^{-1}_\precond$ by \texttt{QMI} and the failure amplitude after FPAA.
The error in the normalized output state of \texttt{QMI} conditioned on $\ket{0^{a_\text{inv}}}$ is bounded as
\begin{equation}
    \norm{\ket{\wt{\text{G}}}-\ket{\text{G}}}
    \leq
    2\frac{\norm{\ket{\wt{\phi}_\text{out}}-\ket{\phi_\text{out}}}}
    {\norm{\ket{\phi_\text{out}}}}
    \leq 2\varepsilon_\text{inv}\norm{A_\precond}
\end{equation}
where we used $\norm{\ket{\phi_\text{out}}}\geq1/\norm{A_\precond}$ and that the numerator is $\leq \varepsilon_\text{inv}$.
The output state of FPAA is 
\begin{equation}
    \ket{\wt{\Phi}_\text{out}}= \sqrt{p_L}\ket{0^{a_\text{inv}}}\ket{\wt{\text{G}}} +\sqrt{1-p_L}\ket{\wt{\text{B}}},
\end{equation}
and the target output of FPAA is $\ket{\Phi_\text{out}}=\ket{0^{a_\text{inv}}}\ket{\wt{\text{G}}}$, so we have
\begin{equation}
    \norm{\ket{\wt{\Phi}_\text{out}}-\ket{\Phi_\text{out}}}^2\leq 2-2\sqrt{p_L} \leq 2(1-p_L)\leq 2\delta^2
\end{equation}
using $1-\delta^2\leq p_L\leq 1$ for $L$ in \cref{eq:L_FPAA}.
Let us now define
\begin{equation}
    \ket{\wt{\Psi}_\text{out}} :=
    \frac{1}{\sqrt{2}}\sum_{a\in\{0,1\}}\ket{a} W^\dagger U^a\ket{\wt{\Phi}_\text{out}},
\end{equation}
which is an approximation of
\begin{equation}
    \ket{\Psi_\text{out}}=\ket{0^{a_\text{inv}}}\ket{\psi},
    \enspace
    \ket{\psi} =
    \frac{1}{\sqrt{2}}\sum_{a\in\{0,1\}}\ket{a} W^\dagger U^a\ket{\text{G}}
\end{equation}
with $\ket{\psi}$ the ideal distributed solution state.
We then have
\begin{align}
    \norm{\ket{\wt{\Psi}_\text{out}}-\ket{\Psi_\text{out}}}
    &\leq
    \norm{\ket{\wt{\Phi}_\text{out}}-\ket{0^{a_\text{inv}}}\ket{\text{G}}}\\
    &\leq \norm{\ket{\wt{\Phi}_\text{out}}-\ket{\Phi_\text{out}}}
    + \norm{\ket{\wt{\text{G}}}-\ket{\text{G}}}\\
    &\leq \sqrt{2}\delta+
    2\varepsilon_\text{inv}\norm{A_\precond},
\end{align}
where the second line follows by adding and subtracting $\ket{\Phi_\text{out}}=\ket{0^{a_\text{inv}}}\ket{\wt{\text{G}}}$ and using the triangle inequality.
By setting $2\varepsilon_\text{inv}\norm{A_\precond}=\varepsilon/2$ and $\delta\sqrt{2}=\varepsilon/2$, we obtain the distributed solution state $\ket{\psi}$ with error~$\leq\varepsilon$.
Notice that the smallest odd integer $L$ used for FPAA is then $L\in\Theta\left(\log(1/\varepsilon)\right)$.

\paragraph{Gate cost.}
Let $\mathcal{G}(U)$ denote the gate cost of implementing a unitary~$U$ or its inverse and let $\gate_\varepsilon(\psi)$ be the gate cost of generating an $\varepsilon$-approximation of the solution state.
The solution state is generated by applying $\A_L\,\texttt{QMI}\,V_\text{in}$ on the all-zero state followed by controlled~$U^{\pm}$ and the inverse quantum wavelet transform.
Then by~\cref{eq:A_L} we have
\begin{equation}
    \gate_\varepsilon(\psi)=
    L\left(\gate(\texttt{QMI})+
    \gate(V_\text{in})\right)+
    \frac12(L-1)\left(\gate(R_t)+\gate(R_{a_\text{inv}})\right) + \gate(U^{\pm}) + \gate(W),
\end{equation}
where $R_m$ denotes the generalized reflection about the $m$-qubit zero state.
We first analyze the cost of \texttt{QMI}.
As discussed before, it makes $\mathcal{Q}_\text{QMI}\in\order{n\log(1/\varepsilon)}$ queries to $U_{A_\precond}$ and $U^\dagger_{A_\precond}$. 
Moreover, it uses $\mathcal{Q}_\text{QMI}$ generalized reflections about the $a_\precond$-qubit zero state, implemented by $\ket{0^{a_\precond}}$-controlled $X$ gate followed by a single-qubit $Z$ rotation and $\ket{0^{a_\precond}}$-controlled $X$ gate. Its gate cost is then
\begin{equation}
    \gate(\texttt{QMI})=
\order{n\log(1/\varepsilon)[
\gate(U_{A_\precond})+\gate(R_{a_\precond})]}.
\end{equation}

We now analyze the gate cost of other operations.
Note that $V_\text{in}=C_1(U^-)C_0(U^+)(H\otimes W)U_\text{load}$, where $U_\text{load}$ prepares the initial state $\ket{\bm{b}}$. The gate cost of~$U_\text{load}$ is~$\Theta(n\log n)$ by~\cref{eq:load_gate_cost}; the gate cost of $W$ is~$\Theta(n^2)$~\cite{BA24}; and the gate cost of controlled~$U^\pm$ is $\order{n}$~\cite{BNW+26}, so we obtain $\mathcal{G}(V_\text{in})\in\order{n^2}$.
As discussed in~\cref{subsec:controlled_reflection}, the generalized reflection about the $m$-qubit zero state can be implemented with $\Theta(m)$ gates, so we have
$\mathcal{G}(R_t)\in\Theta(t)$; $\mathcal{G}(R_{a_\text{inv}})\in\Theta(a_\text{inv})$; and $\mathcal{G}(R_{a_\precond})\in\Theta(a_\precond)$.
With $t=a_\text{inv}+n+1$ and $a_\text{inv}=a_\precond+1\in\Theta(n+\log(n/\varepsilon))$, all three reflection costs scale as $\Theta(n+\log(n/\varepsilon))$.
Combining these bounds with $L\in\Theta\left(\log(1/\varepsilon)\right)$ gives the overall gate cost
\begin{equation}
\label{eq:gate_cost_solution_state}
    \gate_\varepsilon(\psi)
    \in \order{n\log^2(1/\varepsilon) \gate(U_{A_\precond})}+ 
    \order{n^2\log(1/\varepsilon)}+
    \order{n\log^2(1/\varepsilon)[n+\log(n/\varepsilon)]}
\end{equation}
for generating an $\varepsilon$-approximation of the solution state.

\subsubsection{Implementing the operator oracle by inequality tests}
\label{subsubsec:operator_oracle_implementation}
The differential operator~$\L$ of the linearized Poisson-Boltzmann equation in~\cref{eq:differential_operator} is a variable-coefficient operator of the form in \cref{eq:variable_coeff_PDE}.
The discretized version of such an operator in $d$ dimensions has the form (see \cref{eq:A3D} for the 3D case)
\begin{equation}
\label{eq:mtxA}
    A = \sum_{\nu=0}^{d-1} D^T_\nu E_\nu D_\nu + K,
\end{equation}
where only the diagonal matrices $E_\nu$ and $K$ are problem dependent. By the oracle implementation in~\cref{subsec:block_encoding_variable_operators}, we only need to construct a tailored implementation for the block encodings~$U_K$ and $U_E=\sum_\nu \ket{\nu}\!\!\bra{\nu}\otimes U_{E_\nu}$ to achieve a tailored implementation for the block encoding~$U_A$ of the discretized operator~$A$.
As the Poisson–Boltzmann equation is naturally a 3D problem, we focus on implementation for~$d=3$, in which case $\nu\in\{x,y,z\}$. Generalization of our implementation here to higher dimensions is straightforward. 

In the grid discretization of the Poisson–Boltzmann equation, the diagonal elements of~$E_\nu$ encode the dielectric coefficient~$\varepsilon(\bm{x})$ on grid's face points in the~$\nu$ direction and the diagonal elements of~$K$ encode the screening function~$k(\bm{x})$ on grid's node points; see~\cref{eq:diagonal_encodings}.
These functions are piecewise constant functions as in~\cref{eq:epsKappa}, and their values depend on whether the associated grid point is inside or outside the molecular cavity.
Specifically, let $\bm{q}:=(i,j,k)$ denote the 3D index of grid nodes and let $\ket{\bm{q}}:=\ket{k}\ket{j}\ket{i}$ denote the computational basis.
Also, let $\bme_\nu$ denote the 3D unit vector in the~$\nu$ direction. Then $(\bm{q}+\bme_\nu/2)$ is the index of the face point adjacent to~$\bm{q}$ in the $\nu$ direction, and
\begin{equation}
    \bm{x}_{\bm{q}} := h\bm{q} \enspace\text{and}\enspace
    \bm{x}_{\bm{q},\nu}:=h\left(\bm{q}+\frac12\bme_\nu\right)
\end{equation}
are the physical positions of the node point~$\bm{q}$ and its adjacent face point in the $\nu$ direction, respectively.
Moreover, we can write
\begin{equation}
    E_\nu = \varepsilon_\text{out} \id + (\varepsilon_\text{cav}-\varepsilon_\text{out}) P^\nu_\text{cav},
\end{equation}
where
\begin{equation}
    P^\nu_\text{cav} =\sum_{\bm{q}:\bm{x}_{\bm{q},\nu}\in \Omega_\text{cav}}\ket{\bm{q}}\!\!\bra{\bm{q}}
\end{equation}
is the projector onto cavity face points in the~$\nu$ direction.
Similarly, we have
\begin{equation}
    K = \varepsilon_\text{out} \kappa^2_\text{out}
    \left(\id - P_\text{cav}\right),
\end{equation}
where
\begin{equation}
    P_\text{cav} =\sum_{\bm{q}:\bm{x}_{\bm{q}}\in \Omega_\text{cav}}\ket{\bm{q}}\!\!\bra{\bm{q}}
\end{equation}
is the projector onto cavity nodes.
With the reflection operator defined as $R:=\id-2P$ for each projector~$P$, we write
\begin{equation}
    E_\nu = \alpha^+ \id + \alpha^- R^\nu_\text{cav},
    \quad
    \alpha^\pm := \frac12\left(\varepsilon_\text{out}\pm\varepsilon_\text{cav}\right),
\end{equation}
and
\begin{equation}
    K = \frac12
    \varepsilon_\text{out}\kappa^2_\text{out}
    \left(\id + R_\text{cav}\right).
\end{equation}
The reflection operators act on basis states as
\begin{align}
\label{eq:node_reflection_operator}
    R_\text{cav} \ket{\bm{q}} &=(-1)^{f(\bm{q})}\ket{\bm{q}},\\
    \label{eq:face_reflection_operator}
    R^\nu_\text{cav} \ket{\bm{q}} &=(-1)^{f_\nu(\bm{q})}\ket{\bm{q}},
\end{align}
where $f_\nu(\bm{q})=1$ if the face point $(\bm{q}+\bme_\nu/2)h$ is inside the cavity and $f_\nu(\bm{q})=0$ otherwise.
Similarly, $f(\bm{q})=1$ if the node point $\bm{q}h$ is inside the cavity and $f(\bm{q})=0$ otherwise.
Constructing a block encoding for $E_\nu$ and $K$ then reduces to implementing these reflection operators.

Let us first focus on implementing the reflection~$R_\text{cav}$.
To this end, we can initialize an ancilla qubit in $\ket{0}$ and apply the unitary
\begin{equation}
    U_f:\ket{\bm{q}}\ket{0} \mapsto \ket{\bm{q}}\ket{f(\bm{q})},
\end{equation}
which flips the ancilla to $\ket{1}$ only if $\bm{q}$ is a cavity node.
A $Z$ gate on the ancilla produces the desired phase, and another application of $U_f$ resets the ancilla back to $\ket{0}$.
Similarly, the reflection~$R^\nu_\text{cav}$ can be implemented using the unitary
\begin{equation}
    U_{f_\nu}:\ket{\nu}\ket{\bm{q}}\ket{0}
    \mapsto
    \ket{\nu}\ket{\bm{q}}\ket{f_\nu(\bm{q})},
\end{equation}
which flips the ancilla only if $\bm{q}+\bme_\nu/2$ is a face point inside the cavity.
A naive approach to implement these unitaries is to apply a sequence of multi-controlled bit-flip operations (one for each cavity pint), each conditioned on that~$\ket{\bm{q}}$ encodes the index of the cavity point.
However, this approach would be inefficient because its cost scales linearly with the number of cavity nodes, and so does with the total number of nodes, as cavity nodes is a fraction of the total nodes.

\textbf{Inequality test for cavity nodes.}
To construct an implementation with a logarithmic cost in the number of nodes, we use the fact that the molecular cavity is determined by the set $\{\bm{x}_a,r_a\}$ of positions and radii of the atoms.
Therefore, to determine whether a point~$\bm{q}h$ is inside the cavity, we only need to test whether~$\bm{q}h$ is inside at least one of the~$N_\text{atm}$ atom balls in \cref{eq:cav}.
That is to say, we use the fact that
\begin{equation}
    f(\bm{q}) = f_0(\bm{q})\lor f_1(\bm{q})\lor\cdots f_{N_\text{atm}-1}(\bm{q})
\end{equation}
where~$\lor$ is the logical~\textsc{OR} and $f_a(\bm{q})$ with $a\in\{0,\ldots,N_\text{atm}-1\}$ indicates whether the node point~$\bm{q}h$ is inside the ball of atom~$a$. 
Similarly, we have
\begin{equation}
    f_\nu(\bm{q}) = f_{\nu,0}(\bm{q})\lor f_{\nu,1}(\bm{q})\lor\cdots f_{\nu,N_\text{atm}-1}(\bm{q}),
\end{equation}
where $f_{\nu,a}(\bm{q})$ indicates whether the face point~$(\bm{q}+\bme_\nu/2)h$ is inside the ball of atom~$a$. 
Specifically,
\begin{equation}
\label{eq:original_tests}
    f_a(\bm{q}) =
    \begin{cases}
        1& \norm{\bm{x}_a-\bm{q}h}_2\leq r_a\\
        0&\text{otherwise}
    \end{cases}
    \quad
    f_{\nu,a}(\bm{q}) =
    \begin{cases}
        1& \norm{\bm{x}_a-(\bm{q}+\bme_\nu/2)h}_2\leq r_a\\
        0&\text{otherwise}
    \end{cases}
\end{equation}
where $\bm{x}_a=(x_a,y_a,z_a)$ is position of atom~$a$.
We can therefore implement the unitaries $U_f$ and $U_{f_\nu}$ with~$N_\text{atm}$ inequality test (one per atom).

Instead of using one ancilla for each inequality test~($N_\text{atm}$ ancillas for all tests), we use an ancilla register with
\begin{equation}
\label{eq:n_cnt}
    n_\texttt{cnt}:=\ceil{\log_2(N_\text{atm}+1)}
\end{equation}
qubits as a counter register, which we denote by~\texttt{cnt}. This register starts in the all-zero state, and we increment the counter by~$1$ when each inequality test yields~$1$.
A non-zero counter then indicates the point is inside the cavity.
This approach exponentially reduces the number of ancilla qubits with respect to the number of atoms.
For efficient implementation, we also modify the inequality tests so that we only use integer numbers for comparison and avoid computing square roots on the quantum computer needed for computing the norm.
To this end, first we multiply both sides of the inequalities in~\cref{eq:original_tests} by $2^b/h$, for some positive integer~$b$ to be determined.
We then use the rounding $2^b\bm{x}/h\mapsto \bm{\mathrm{x}}_a:=(\mathrm{x}_a, \mathrm{y}_a, \mathrm{z}_a)$ and $2^br_a/h\mapsto \mathrm{r}_a$ with integers
\begin{equation}
    \mathrm{x}_a:=  \round{2^b\frac{x_a}{h}},\,
    \mathrm{y}_a:= \round{2^b\frac{y_a}{h}},\,
    \mathrm{z}_a:= \round{2^b\frac{z_a}{h}},\,
    \mathrm{r}_a := \round{2^b\frac{r_a}{h}},
\end{equation}
and take the square of both sides. The inequalities are then mapped as
\begin{align}
    \norm{\bm{x}_a-\bm{q}h}_2 \leq r_a
    &\mapsto \norm{\bm{\mathrm{x}}_a-2^b\bm{q}}_2^2 \leq \mathrm{r}_a^2,\\
    \norm{\bm{x}_a-(\bm{q}+\bme_\nu/2)h}_2\leq r_a
    &\mapsto\norm{\bm{\mathrm{x}}_a-2^b(\bm{q}+\bme_\nu/2)}^2_2\leq \mathrm{r}^2_a.
\end{align}
By expanding the norm, we separate the atom-dependent and atom-independent parts as
\begin{align}
\label{eq:modified_node_inequality}
    &4^b\norm{\bm{q}}^2 - 2^{b+1} \bm{q}\cdot\bm{\mathrm{x}}_a
    \leq
    \underbrace{\mathrm{r}_a^2-\norm{\bm{\mathrm{x}}_a}^2-4^{b-1}}_{:=C_a}\\
    \label{eq:modified_face_inequality}
    &4^b(\norm{\bm{q}}^2+\bm{q}\cdot\bme_\nu) - 2^{b+1} \bm{q}\cdot\bm{\mathrm{x}}_a - 2^b \bm{\mathrm{x}}_a\cdot\bme_\nu
    \leq C_a,
\end{align}
where the integer~$C_a$ depends only on the atomic data and can be classically precomputed.
The left-hand sides here are integers to be computed on the quantum computer.

We note that~$b$ is a small constant determined by the grid spacing~$h$ and the precision of the input atomic data.
Atomic coordinates in the standard Protein Data Bank format~\cite{PDB19} are specified to three decimal places in \aa ngestrom, so requiring the rounding error to be smaller than the input precision gives~$b=4\text{--}9$ for $h\in(0.01,0.5)$\AA, where $h=0.5$\AA\ is commonly used in the classical PBE solvers~\cite{HBF13}.

We now implement the inequality tests in~\cref{eq:modified_face_inequality} for the face points by the following steps.
The implementation of inequality tests in~\cref{eq:modified_node_inequality} for the node points follow similarly.
\begin{enumerate}
    \item Compute the atom-independent term
    \begin{equation}
    \label{eq:atom_dependent_value}
        4^b(\norm{\bm{q}}^2+\bm{q}\cdot\bme_\nu) = 4^b(i^2+j^2+k^2) + 4^b q_\nu
    \end{equation}
    into a temporary register, where $q_\nu=i,j,k$ for $\nu=x,y,z$, respectively.
    We denote this register by~\texttt{tmp} and use it to store the value of the left-hand side in~\cref{eq:modified_face_inequality}.
    The magnitude of this value is no greater than~$4^{n+b+1}$, so a temporary register with
    \begin{equation}
    \label{eq:n_tmp}
        n_\texttt{tmp}:=2(n+b)+3 = 2n+\Theta(1)
    \end{equation}
    qubits suffices to represent the stored value in two's complement. The constant here is between 11--21 for $b=4\text{--}9$.
    To compute this term, we first compute the direction-dependent part~$4^bq_\nu$ into \texttt{tmp} controlled on the direction register~$\ket{\nu}$. As \texttt{tmp} is initially in the all-zero state and multiplying by~$4^b$ is equivalent to shifting the $n$-bit number~$q_\nu$ to the left by~$2b$ bits, we simply copy~$q_\nu$ into qubits~$2b,\ldots,2b+n-1$ of \texttt{tmp} conditioned on the value of~$\nu$.
    We then compute the square terms, such as~$4^bi^2$, by single- and double-controlled constant additions as follows. Let us use the binary expansion $i=\sum_{r=0}^{n-1}i_r2^r$, where $i_r$ are bits of~$i$, then
    \begin{equation}
        4^bi^2=\sum_{r}i_r4^{b+r} + 2\sum_{r<s}i_ri_s2^{2b+r+s}
    \end{equation}
    and so $4^bi^2$ can be added by~$n$ single-controlled additions of~$4^{b+r}$ controlled on qubit~$r$ of~$\ket{i}$, and $n(n-1)/2$ double-controlled additions of~$2^{2b+r+s+1}$ controlled on qubits~$r$ and~$s$ of~$\ket{i}$.
    These terms are $\nu$-independent, so no control on~$\ket{\nu}$ is needed.
    \item For each atom~$a=0,\ldots,N_\text{atm}-1$:
    \begin{enumerate}
        \item Add the atom-dependent terms
        \begin{equation}
            -2^{b+1}\bm{q}\cdot \bm{\mathrm{x}}_a - 2^b \bm{\mathrm{x}}_a\cdot\bme_v =
            -2^{b+1}(i\mathrm{x}_a+j\mathrm{y}_a
            +k\mathrm{z}_a) - 2^b \mathrm{x}_{a,\nu}
        \end{equation}
        to the temporary register. Here $\mathrm{x}_{a,\nu}=\mathrm{x}_a, \mathrm{y}_a \mathrm{z}_a$ depending on~$\nu$, so we add the constant $-2^b\mathrm{x}_{a,\nu}$ to \texttt{tmp} controlled on~$\ket{\nu}$.
        Other terms are $\nu$-independent and we add them by single-controlled constant additions.
        For instance, we add~$-2^{b+1}i\mathrm{x}_a$ by $n$ single-controlled additions of~$-2^{b+1+r}\mathrm{x}_a$, controlled on qubit~$r$ of~$\ket{i}$.
        \item Compare the value of temporary register with the classical value~$C_a$ and write the result into a flag qubit. This is achieved by a classical-quantum comparator operation that implements that map
        \begin{equation}
        \label{eq:classical_quantum_comparator}
        \ket{T}_\texttt{tmp}\ket{0}_f \mapsto
            \begin{cases}
                \ket{T}_\texttt{tmp}\ket{1}_f & \text{if } T\leq C_a\\
                \ket{T}_\texttt{tmp}\ket{0}_f & \text{if } T> C_a
            \end{cases}
        \end{equation}
        where $T$ is the value stored in the temporary register.
        \item Controlled on $\ket{1}$ of the flag qubit, increment the counter register by~1.
        \item Apply the inverse of step~(b) to reset the flag qubit.
        \item Apply the inverse of step~(a) to restore the temporary register to the atom-independent value in~\cref{eq:atom_dependent_value}.
    \end{enumerate}
    \item Test whether the counter register \texttt{cnt} is nonzero and write the result into the flag qubit; note that this qubit is reset to~$\ket{0}$ at the end of step~2 and can be reused. To this end, first we apply~$X$ on the flag qubit and then flip it conditioned on \texttt{cnt} being in the all-zero state. To implement the latter, we apply a multi-controlled~$X$ gate with~$\ket{0}$ controls on all qubits of \texttt{cnt}. The flag qubit is then in~$\ket{1}$ only if the counter is nonzero.
    \item Apply $Z$ on the flag qubit to produce the desired phase for the cavity points. Then reset the flag by applying the multi-controlled~$X$ gate followed by~$X$ gate. Step~4 is only applied to implement the reflection operator in~\cref{eq:face_reflection_operator}.
    \item  Apply the inverse of step~2 to reset the counter register to the all-zero state.
    \item Apply the inverse of step~1 to reset the temporary register to the all-zero state.
\end{enumerate}

\textbf{Cost analysis.}
Let $\cost_s$ denote the cost of step~$s$ above. Then~$\cost = 2\cost_1 + 2N_\text{atm} \cost_2 +\cost_3 + \cost_4$ is the total cost of implementing all steps, and we have~$\cost_2=2\cost_{2a}+2\cost_{2b}+\cost_{2c}$.
Below we analyze the cost of the unique steps.
We use the following lemma for the cost of the classical-quantum comparator in~\cref{eq:classical_quantum_comparator}.
\begin{lemma}[{\cite[Thm.~3]{vandaele26}}]
\label{lem:classical_quantum_comparator}
    The $n$-qubit classical-quantum comparator can be implemented with one dirty ancilla qubit by a circuit with~$\Theta(n)$ gates from $\{X,\mathrm{CNOT}, \mathrm{Toffoli}\}$ and $\Theta(\log n)$ depth.
\end{lemma}

\begin{itemize}
    \item Step~1: Copies the $\nu$-dependent part into \texttt{tmp} and adds the $\nu$-independent part.
    For each direction~$\nu$, the copy needs $n$ 3-controlled~$X$ gates, two controls on the direction register and one control on one of the qubits of the $n$-qubit basis state~$\ket{q_\nu}$, where~$\ket{q_\nu}=\ket{i},\ket{j}, \ket{k}$ depending on~$\nu$.
    Adding each term ($i,j \text{ or } k$) of the $\nu$-independent part needs $n$ single-controlled and $n(n-1)/2$ double-controlled additions of classical constants into \texttt{tmp}.

    By \cref{lem:classical_quantum_adder}, the single- and double-controlled classical-quantum addition on~$n_\texttt{tmp}$ qubits can be implemented using one dirty ancilla qubit and a circuit with $\Theta(n_\texttt{tmp}\log n_\texttt{tmp})$ gates and~$\Theta(\log^2n_\texttt{tmp})$ depth.
    Therefore, Step~1 can be implemented using one dirty ancilla and a circuit with $\Theta(n^2 n_\texttt{tmp}\log n_\texttt{tmp})$
    gates and $\Theta(n^2\log^2 n_\texttt{tmp})$ depth.
    \item Step~2a: Adds the $\nu$-dependent part~$-2^b\mathrm{x}_{a,\nu}$ to \texttt{tmp} by~$3$ double-controlled constant additions (one for each direction~$\nu$) and adds the $\nu$-independent part by $3n$ single-controlled constant additions.
    So, by~\cref{lem:classical_quantum_adder}, this step can be implemented using one dirty ancilla and a circuit with~$\Theta(nn_\texttt{tmp}\log n_\texttt{tmp})$ gates and $\Theta(n\log^2 n_\texttt{tmp})$ depth.
    \item Step~2b: Compares the value stored in the $n_\texttt{tmp}$-qubit register with the classical value~$C_a$.
    This $n_\texttt{tmp}$-qubit classical-quantum comparator can be implemented using one dirty ancilla and a circuit with $\Theta(n_\texttt{tmp})$ gates and $\Theta(\log n_\texttt{tmp})$ depth as per~\cref{lem:classical_quantum_comparator}.
    \item Step~2c: Increments the $n_\texttt{cnt}$-qubit counter register by~+1. By~\cref{lem:multi_controlled_increment}, this can be implemented using one dirty ancilla and a circuit with $\Theta(n_\texttt{cnt})$ gates and $\Theta(\log n_\texttt{cnt})$ depth.
    \item Step~3: Applies one~$X$ gate and one $n_\texttt{cnt}$-controlled~$X$ gate controlled on the all-zero state of the counter register.
    By~\cref{lem:multi_controlled_not},
    this multi-controlled~$X$ gate can be implemented using one dirty ancilla and a circuit with $\Theta(n_\texttt{cnt})$ gates and $\Theta(\log n_\texttt{cnt})$ depth.
    \item Step~4: Applies one~$Z$, one~$X$, and the $n_\texttt{cnt}$-controlled~$X$ in Step~3, so its cost is similar to that of Step~3.
\end{itemize}

Putting all together, all inequality tests, and hence the reflection operators, can be implemented by a circuit with $\mathcal{G}$ gates and depth~$\mathcal{D}$ scaling as
\begin{align}
\label{eq:gate_cost_diagonal_encodings}
    \mathcal{G} &= \Theta(n^2\,n_\texttt{tmp}\log n_\texttt{tmp})+
    N_\text{atm}\left(\Theta(n\,n_\texttt{tmp}\log n_\texttt{tmp}) + 
    \Theta(n_\texttt{cnt})\right) + \Theta(n_\texttt{cnt})\nonumber\\
    &=\Theta(n^3\log n) +
    N_\text{atm}\left(\Theta(n^2\log n) +
    \Theta(\log N_\text{atm})\right)\\
    \mathcal{D} &= \Theta(n^2\log^2 n_\texttt{tmp}) +
    N_\text{atm}\left(\Theta(n\log^2 n_\texttt{tmp})
    +\Theta(\log n_\texttt{cnt}) 
    \right) +
    \Theta(\log n_\texttt{cnt})\nonumber\\
    &=\Theta(n^2\log^2 n) +
    N_\text{atm}\left(\Theta(n\log^2 n) +
    \Theta(\log\log N_\text{atm})\right),
\end{align}
where we used~$n_\texttt{tmp}\in\Theta(n)$ by~\cref{eq:n_tmp} and~$n_\texttt{cnt}\in\Theta(\log N_\text{atm})$ by~\cref{eq:n_cnt}.

\textbf{Ancilla count.}
Two ancillas are used for the direction register. With the flag qubit and the qubits used for the temporary and counter register, the total number of clean ancilla qubits used is
\begin{align}
\label{eq:ancilla_count_diagonals}
    n_\texttt{anc} &= n_\texttt{tmp}+n_\texttt{cnt} + 3\nonumber\\
    &= 2(n+b)+\ceil{\log_2(N_\text{atm}+1)}+6\nonumber\\
    &=2n+\ceil{\log_2(N_\text{atm}+1)}+\Theta(1),
\end{align}
where we used~\cref{eq:n_cnt} and~\cref{eq:n_tmp}.
The constant~$\Theta(1)$ here is in the range~$14\text{--}24$ depending on the value of~$b=4\text{--}9$.
We note that only a single dirty ancilla is used, as it can be shared between all steps.

\subsubsection{Block encoding of the preconditioned matrix with logarithmic normalization}
\label{subsubsec:block_encoding_Aprec}

A block encoding of the preconditioned matrix
$A_\precond=PWAW^\dagger P$ can be constructed by the LCU technique using the block encoding $U_A$ of the original matrix~$A$ and the decomposition $P=(U^++U^-)/2$ for the preconditioner as linear combination of two unitaries $U^\pm$. Specifically, if we are given a $(\alpha_A,a_A,0)$ block encoding of $A$, then an $(\alpha_\precond,a_\precond,0)$ block encoding of $A_\precond$ can be constructed with $\alpha_\precond=\alpha_A$ and $a_\precond=a_A$ by a single call to $U_A$~\cite[Appendix~E.1]{BNW+26}.
This approach is useful only if the normalization $\alpha_A$ is at most polylogarithmic in the dimension $N_g=N^3=2^{3n}$ of~$A$, as the query cost $Q_\texttt{QMI}$ in \cref{eq:query_cost} depends linearly on~$\alpha_\precond$.
As discussed in~\cref{subsubsec:operator_oracle_implementation}, $\alpha_A\in\Theta(N)$ in our application, so this approach is not appropriate to obtain an algorithm with a polylogarithmic cost.
The block-encoding normalization in an important factor in quantum preconditioning, as a large normalization factor can offset the benefits of reducing the condition number by preconditioning~\cite{LS25,JLM+25}.
We note that simply rescaling the original linear system by $1/\alpha_A$ does not remove the linear scaling of $Q_\texttt{QMI}$ in~$\alpha_A$ by this approach as $Q_\texttt{QMI}$ depends inversely on $\norm{A_\precond}$ and $\norm{A_\precond}\mapsto \norm{A_\precond}/\alpha_A$ by rescaling.

Here we present a direct block encoding of~$A_\precond$ with the normalization $\alpha_\precond\in\Theta(n)$, i.e., logarithmic in the matrix dimension. We present our construction for block encoding of the preconditioned matrix in 1D, as extension to the 3D case is straightforward.
Using $A=D^TED+K$ in 1D, we express the 1D preconditioned matrix as
\begin{equation}
\label{eq:mtxA_prec}
    A_\precond =
    Q^T E Q + K_\precond,
    \enspace
    Q := DW^\dagger P,
\end{equation}
where $K_\precond=PWKW^TP$.
A block encoding of $A_\precond$ is then obtained by block encodings of $Q$, $E$ and $K_\precond$ using the standard product and summation rules for block encodings.
The inequality-test approach in~\cref{subsubsec:operator_oracle_implementation} provides block encodings $U_E$ and $U_K$ for $E$ and $K$ with normalizations $\alpha_E,\alpha_K\in\Theta(1)$, respectively. With $\alpha_K\in\Theta(1)$, a block encoding of $K_\precond$ with $\Theta(1)$ normalization is obtained by the LCU technique using $U_K$ and $P=(U^++U^-)/2$.
In the following, we construct a block encoding for $Q$ with normalization $\alpha_Q\in\Theta(\sqrt{n})$, from which we obtain a block encoding for $A_\precond$ with normalization $\alpha_\precond\in\Theta(n)$ by the standard product and summation rules for block encodings.

To construct a block encoding for $Q$, we note that $W$ is an $n$-level wavelet transform matrix and is specified by a sequence of numbers $(h_0,\ldots,h_{M-1})$ known as the wavelet filter coefficients; the sequence depends on the wavelet used and its length~$M$ is the wavelet order. See Refs.~\cite{BA24,BNW+26,mbm,BSB+22,GSM+22,GMB+26} for details of wavelets and their applications in quantum physics and computing.
For $n$-level wavelet transform, $W$ has the row block matrix form 
\begin{equation}
    W =
    \left[
    \begin{array}{c}
        \bm{1}\\
        \hline
        W_0\\
        \hline
        \vdots\\
        \hline
        W_{n-1}
     \end{array}
     \right]
\end{equation}
where the top block is the constant vector $\bm{1}:=(1,1,\ldots,1)/\sqrt{2^n}$ of size $2^n$
and the block $W_s$ with $0\leq s<n$ is a $2^s\times2^n$ matrix obtained as $W_s = G_s H_{s+1}\cdots H_{n-1}$. Here $H_r$ is known as the refinement matrix of size $2^{r}\times 2^{r+1}$ with entries $[H_r]_{ij}=h_{(j-2i)\bmod 2^{r+1}}$, and $G_r$ has the same size with entries $[G_r]_{ij}=g_{(j-2i)\bmod 2^{r+1}}$, where $g_\ell:=(-1)^\ell h_{M-1-\ell}$.
Note that each row of these matrices is obtained from the previous row by a cyclic shift of two positions to the right.
It then follows that each row of $W_s$ is obtained from the previous row by a cyclic shift of $2^{n-s}$ positions to the right. 

As $W$ is a row block matrix, $W^\dagger$ is a column block matrix as
$W^\dagger = [\bm{1}^\dagger|W^\dagger_0|\cdots |W^\dagger_{n-1}]$.
By the structure of $W^\dagger$ and~$P$, it then follows that
$Q = \left[\bm{0} | C_0 |\cdots |C_{n-1}\right]$ is a column block matrix with blocks $C_s = 2^{-s} D\mathrm{W}_s^{\dagger}
    =\frac{1}{L} A_s$, where the $2^n\times 2^s$ matrix
\begin{align}
\label{eq:As_expansion}
    A_s:= 2^{n-s} (S_n-\id_n) W^\dagger_s = 2^{n-s} (S_n-\id_n) H^\dagger_{n-1}\cdots H^\dagger_{s+1} G^\dagger_s
\end{align}
is obtained by expanding $D$ and $W^\dagger_s$. Note that the leftmost column block of $Q$ is the zero column vector.
Now let $Q_s$ be the matrix obtained from $Q$ with all blocks replaced with zero except the column block $C_s$. Then $Q=\sum_{s=0}^{n-1} Q_s$, and so a block encoding of $Q$ can be constructed by a block encoding of~$Q_s$ with a fixed scale~$s$. 
We first construct a block encoding of~$Q_s$ with a constant normalization by a direct implementation.
Then we extend the fixed-scale construction to obtain a block encoding for the full matrix~$Q$ with normalization $\alpha_Q\in\Theta(\sqrt{n})$.

To construct a block encoding of~$Q_s$, let us write the first column of $A_s$ as
\begin{equation}
    A_s\ket{0^s} = \sum_{m=0}^{2^n-1} a_{s,m}\ket{m},
\end{equation}
where $m$ is the row index of $A_s$ and the coefficient $a_{s,m}$ is the entry at row $m$ of the first column.
We use the following lemmas in our implementation; see~\cref{apx:lemmas_prex_mtx} for proofs for these lemmas.

\begin{lemma}
\label{lem:nonzero_entries}
     Let $n_s:=n-s-1$ for $0\leq s<n $. The coefficient $a_{s,m}$ is nonzero only for $m=2^{n_s}c+b$, where $0\leq b<2^{n_s}$ and $0\leq c< c_\text{max}$ with $c_\text{max}$ a constant that depends only on the wavelet used.
\end{lemma}
\begin{lemma}
\label{lem:bounded_coefficient}
    The magnitude of the coefficient $a_{s,m}$ is bounded as $\sqrt{2^{n-s}}\abs{a_{s,m}}\leq c_w$, where $c_w$ is a constant that depends only on the wavelet used.
\end{lemma}

We also note that each column $0\leq k<2^s$ of $A_s$ is obtained from the previous column by a cyclic shift of $2^{n-s}$ positions downward.
This is because $A_s$ is obtained from $W^\dagger_s$.
The action of $A_s$ on the $s$-qubit state $\ket{k}$ is then
\begin{equation}
    A_s\ket{k} = \sum_{m=0}^{2^n-1} a_{s,m}\ket{(m+2^{n-s}k) \bmod 2^n},
\end{equation}
where the output basis state is an $n$-qubit state. By this action, we obtain the action of $Q_s$ on the $n$-qubit basis state $\ket{j}$ as
\begin{equation}
    Q_s \ket{j}=
    \begin{cases}
        \frac{1}{L} A_s\ket{k} & \text{if $j=2^s+k$}\\
        0 & \text{else.}
    \end{cases}
\end{equation}
To implement this action, we note that $a_{s,m}$ is nonzero only for $m=2^{n_s}c+b$ by \cref{lem:nonzero_entries}, and express the action of $A_s$ as
\begin{equation}
\label{eq:action_As_new}
    A_s\ket{k} = \sum_{c=0}^{c_\text{max}-1}\sum_{b=0}^{2^{n_s}-1} a_{s,m}\ket{(2k+c) \bmod 2^{s+1}}\ket{b}.
\end{equation}
This equation follows because with $m=2^{n_s}c+b$ the output basis state can be decomposed as $\ket{(2^{n-s}k+m) \bmod 2^n}=\ket{(2k+c)\bmod 2^{s+1}}\ket{b}$ by observing that
\begin{align}
    (2^{n-s}k+m) \bmod 2^n
     &= 2^{n-s-1}[(2k+c)\bmod 2^{s+1}]+b\nonumber\\
     &= ([(2k+c)\bmod 2^{s+1}] b)_2,
\end{align}
where the second line is in binary; $b$ is the lower $n_s=n-s-1$ bits and $2k+c$ is the upper $s+1$ bits. 
We now utilize the action of $A_s$ in \cref{eq:action_As_new} to build a direct block encoding of $A_s$, from which we obtain a block encoding of $Q_s$.
We summarize our approach in the following steps and analyze the cost of each step.

\begin{enumerate}
    \item \textit{(Test the scale of $j$.)} First we test if the input basis state $\ket{j}$ is at scale $s$; i.e., we test if $j=2^s+k$ for some $k$. If $j$ is at scale~$s$, then $\ket{j}=\ket{0^{n_s}}\ket{1}\ket{k}$, where $\ket{k}$ has $s$ qubits and $n_s=n-s-1$. To test, we initialize a flag qubit~$t$ at~$\ket{0}_t$ and flip it if the leftmost $n_s+1$ qubits of $\ket{j}$ are in the state $\ket{0^{n_s}1}$. We then flip the flag so that its state is~$\ket{0}_t$ when~$\ket{j}$ is at scale~$s$.
    Therefore, for any basis state~$\ket{j}$, the transformation in this step is
    $$
    \ket{0}_t\ket{j} \mapsto
    \begin{cases}
        \ket{0}_t\ket{0^{n_s}}\ket{1k} & j=2^s+k,\\
        \ket{1}_t\ket{j} & \text{otherwise}.
    \end{cases}
    $$
    The following steps are conditioned on the flag being in the zero state.
    The gate cost of this step is equivalent to the cost of $(n_s+1)$-controlled-NOT gate, which is $\Theta(n_s)$ as per~\cref{lem:multi_controlled_not}.
    With $n_s<n$, the cost of this step is $\order{n}$.
    \item (\textit{Swap blocks of system register}.) We now perform the map $\ket{0^{n_s}}\ket{1k}\mapsto\ket{1k}\ket{0^{n_s}}$. That is, we swap the right block of $s+1$ qubits with the left block of $n_s$ qubits. To implement this, we reverse the order of qubits in each block and then reverse the order of all qubits. Reversing a block $\ket{q_1q_2,\ldots q_k}$ with $k$ qubits to $\ket{q_k\ldots q_2q_1}$ can be implemented by $\floor{k/2}$ SWAP gates. Reversing the block $\ket{1k}$ needs $\floor{(s+1)/2}$ SWAP gates, reversing the block $\ket{0^{n_s}}$ needs $\floor{n_s/2}$ SWAP gates, and reversing all qubits needs~$\floor{n/2}$ SWAP gates. With $s<n$, the overall gate cost is then $\order{n}$. 

    \item (\textit{Prepare uniform states over $c$ and $b$}). We now use an ancilla register \texttt{creg} with $n_c=\log_2c_\text{max}$ qubits initialized at the all-zero state and map it to uniform state over~$c$ by $n_c$ Hadamards; for convenience here we assume $c_\text{max}$ is a power of two.
    We also map $\ket{0^{n_s}}$ to uniform state over $b$. The transformation in this step is then
    $$\ket{0^{n_c}}_\texttt{creg}\ket{0^{n_s}}\mapsto
    \frac{1}{\sqrt{2^{n_s} c_\text{max}}} \sum_{c,b}\ket{c}_\texttt{creg}
    \ket{b}.
    $$
    Note that $2^{n_s}$ can be exponentially large in $n$ depending on~$s$.
    This step can be implemented with~$\order{n}$ Hadamards.
    \item (\textit{Compute the scaled coefficients}.)
    We now compute the scaled coefficient $a'_{s,m}=\sqrt{2^{n-s}}a_{s,m}$ into a temporary register by implementing the operation 
    \begin{equation}
    \texttt{COEF}: \ket{c}\ket{b}
    \ket{0^{n_\texttt{tmp}}}
    \mapsto \ket{c}\ket{b}
    \ket{a'_{s,m}},
    \end{equation}
    where the temporary register \texttt{tmp} has $n_\texttt{tmp}=p+\Theta(1)$ qubits. This register stores the scaled coefficient as a fixed-point binary number in two's complement representation, where~$p$ is the number of bits for the fractional part, and $\Theta(1)$ is a constant number of bits for the integer part.
    Note that by~\cref{lem:bounded_coefficient}, the magnitude of the scaled coefficient is bounded by a constant, so only a constant number of bits is sufficient for the integer part. In~\cref{apx:error_for_Aprec}, we show $p\in\Theta(\log(n/\varepsilon))$ is sufficient.
    The coherent transformation so far if the flag qubit $t$ is at zero state is then
    $$
    \frac{1}{\sqrt{2^{n_s}c_\text{max}}} \sum_{c,b}\ket{c}\ket{1k}\ket{b}
    \ket{0^{n_\texttt{tmp}}}
    \mapsto
    \frac{1}{\sqrt{2^{n_s}c_\text{max}}} \sum_{c,b}\ket{c}\ket{1k}\ket{b}
    \ket{a'_{s,m}}.
    $$
    As we show later the gate cost of implementing \texttt{COEF} is $\order{n\log^2(n/\varepsilon)}$.
    \item (\textit{Perform amplitude transduction}.)
    In this step, we perform the map $\ket{a'_{s,m}}\mapsto \alpha_{s,m}\ket{a'_{s,m}}$, where $\alpha_{s,m}=a'_{s,m}/c_w$ and $\abs{\alpha_{s,m}}\leq 1$ by \cref{lem:bounded_coefficient}. To implement this map, we use inequality testing.
    To this end, we compute the $p$-bit unsigned integer $\alpha^{(p)}_{s,m}=\round{2^p \abs{\alpha_{s,m}}}$ into a $p$-qubit data register \texttt{data}, prepare a $p$-qubit reference register \texttt{ref} in the uniform state, and perform the transformation
    \begin{equation}
        \frac{1}{\sqrt{2^p}}
        \sum_{r=0}^{2^p-1}
        \ket{0}_f\ket{r}_\texttt{ref}
        \ket{\alpha^{(p)}_{s,m}}_\texttt{data}
        \xrightarrow{\texttt{COM}}
        \frac{1}{\sqrt{2^p}}
        \left(
        \sum_{r<\alpha^{(p)}_{s,m}}
        \ket{0}_f \ket{r}_\texttt{ref}+
        \sum_{r\geq \alpha^{(p)}_{s,m}}
        \ket{1}_f \ket{r}_\texttt{ref}
        \right)
        \ket{\alpha^{(p)}_{s,m}}_\texttt{data}
    \end{equation}
    where \texttt{COM} is the quantum-quantum comparison operation and flips the flag if $r\geq \alpha^{(p)}_{s,m}$. We now unprepare \texttt{ref} register by applying~$H^{\otimes p}$.
The amplitude of the state where the flag~$f$ and \texttt{ref} is zero becomes $\alpha^{(p)}_{s,m}/2^p\approx \abs{\alpha_{s,m}}$. To observe this, note that $\langle 0^p |H^{\otimes p}| r\rangle=1/\sqrt{2^p}$ for each $r$, so with $\sum_{r<\alpha^{(p)}_{s,m}}=\alpha^{(p)}_{s,m}$ we obtain
\begin{align}
(\prescript{}{f}{\bra{0}}
\prescript{}{\texttt{ref}}{\bra{0^p}}\otimes\id_p)
    H^{\otimes p}
    \,\texttt{COM}\,
    H^{\otimes p}
    \ket{0}_f\ket{0^p}_\texttt{ref}
    \ket{\alpha^{(p)}_{s,m}}_\texttt{data}
    &= \frac{\alpha^{(p)}_{s,m}}{2^p}
    \ket{\alpha^{(p)}_{s,m}}_\texttt{data}
    \nonumber\\
    &\approx \abs{\alpha_{s,m}}
    \ket{\alpha^{(p)}_{s,m}}_\texttt{data},
\end{align}
where $H^{\otimes p}$ acts on the middle state.
Note that $\alpha_{s,m}$ and $a'_{s,m}$ have same signs, and because $\ket{a'_{s,m}}$ encodes $a'_{s,m}$ in two's complement, its leftmost qubit is the sign bit of $\alpha_{s,m}$. To produce the sign of $\alpha_{s,m}$ in the amplitude, we flip the flag~$f$, apply a controlled~$Z$ with control on the leftmost qubit of $\ket{a'_{s,m}}$ and the flag qubit as the target, and then flip the flag again. With these operations $\abs{\alpha_{s,m}}\mapsto\alpha_{s,m}$.
Finally, we erase $\alpha^{(p)}_{s,m}$ in the data register. 
The output state at this step is
$$\frac{1}{\sqrt{2^{n_s}c_\text{max}}} \sum_{c,b}\alpha_{s,m}\ket{c}\ket{1k}\ket{b} \ket{a'_{s,m}} =
\frac{\sqrt{2}}{c_w\sqrt{c_\text{max}}} \sum_{c,b}a_{s,m}\ket{c}\ket{1k}\ket{b}
\ket{a'_{s,m}}$$
if the flags $t$ and $f$, and the \texttt{ref} register are at the zero state.

We now discuss the gate cost of this step. For convenience, we consider $c_w$ to be a power of two as $c_w=2^q$ with $q<p$.
To compute the unsigned integer $\alpha^{(p)}_{s,m}$ into the data register, temporarily map $\ket{a'_{s,m}} \mapsto\ket{|a'_{s,m}|}$ as discussed in Step~2 in~\cref{subsec:sparse}. Then copy the integer bits of $|a'_{s,m}|$ and its first $p-q$ fractional bits to data register, and increment the value of data register by~$+1$ if the $(p-q+1)$th fractional bit is~$1$. Finally, perform $\ket{|a'_{s,m}|}\mapsto \ket{a'_{s,m}}$.
The gate cost of these operations is $\order{p}$ as the integer part only has $\Theta(1)$ bits.
Preparing and unpreparing \texttt{ref} register needs $2p$ Hadamards and \texttt{COM} can be implemented by $\Theta(p)$ gates.
With $p\in\Theta(\log(n/\varepsilon))$, the overall cost of this step is $\order{\log(n/\varepsilon)}$.

\item (\textit{Uncompute the scaled coefficient}). We now erase the \texttt{tmp} register by the inverse of the \texttt{COEF} operation. 
\item (\textit{Prepare the output basis state and unprepare the \texttt{creg} register}).
To prepare $\ket{(2k+c)\bmod 2^{s+1}}\ket{b}$, we perform the map
\begin{equation}
    \ket{c}_\texttt{creg}\ket{1k}\ket{b}
    \mapsto\ket{c}_\texttt{creg}\ket{(2k+c)\bmod 2^{s+1}}\ket{b}
\end{equation}
by first implementing $\ket{1k}\mapsto\ket{k0}=\ket{2k}$ and then adding the values of the first and second registers. 
Here $\ket{1k}$ is a $(s+1)$-qubit state and $\ket{1k}\mapsto\ket{k0}$ follows by flipping the leftmost qubit and a cyclic shift of all qubits to the left.
To unprepare $\ket{c}$, we apply $H^{\otimes n_c}$ on \texttt{creg}. The overall output state conditioned on the flags $t$ and $f$ and the \texttt{ref} and \texttt{creg} registers being in the zero state is
\begin{equation}
    \frac{\sqrt{2}}{c_wc_\text{max}}\sum_{c,b}  a_{s,m}\ket{(2k+c)\bmod 2^{s+1}}\ket{b} =
    \frac{\sqrt{2}}{c_\text{max}c_w}A_s\ket{k} =
    \frac{L\sqrt{2}}{c_\text{max}c_w} Q_s\ket{j},
\end{equation}
so we achieve a block encoding for $Q_s$ with the normalization $\alpha_{Q_s}=c_\text{max}c_w/L\sqrt{2}\in\Theta(1)$ independent of~$s$.

As $n_c$ is a constant, the cost of unpreparing $\ket{c}$ is $\Theta(1)$.
The cost of the first part is $\order{s}$, so by $s<n$, the overall cost of this step is $\order{n}$.
\end{enumerate}

Considering the gate cost of each step, the overall gate cost for block encoding of $Q_s$ is $\order{n\log^2(n/\varepsilon)}$.
The flag qubits $t$ and $f$, the $n_c$-qubit \texttt{creg} register, the $p$-qubit \texttt{ref} register are used for block encoding of $Q_s$; the rest of ancilla qubits start and return to the zero state. With $p\in\Theta(\log(n/\varepsilon))$ and $n_c\in\Theta(1)$, the number of block encoding ancillas~$a$ for~$Q_s$ is $a=p+n_c+2\in\Theta(\log(n/\varepsilon))$.

\paragraph{Block encoding of the full matrix.}
A simple approach to construct a block encoding of the full matrix~$Q=\sum_s Q_s$ using the block encoding~$U_{Q_s}$ of~$Q_s$ constructed above is by the LCU technique. However, the block encoding constructed by this approach would have a $\Theta(n)$ normalization.
We present an extension of the fixed-scale construction above that achieves a block encoding of~$Q$ with $\Theta(\sqrt{n})$ normalization.

For a fixed~$s$, the block encoding $U_{Q_s}$ we constructed first tests (in Step~1 above) whether the input basis state $\ket{j}$ belongs to scale~$s$.
The result of this test is stored in the flag qubit~$t$, and the remaining steps~(Steps~$\geq2$ above) implement the action of~$Q_s$ conditioned on~$t=0$.
For the full matrix~$Q$, instead of performing this scale test inside $U_{Q_s}$ for each $s=0,\ldots,n-1$, we first determine the scale of~$\ket{j}$ and store it on an $m$-qubit register \texttt{scl} with $m=\ceil{\log_2(n+1)}$.
To this end, we implement the map
 \begin{equation}
\label{eq:write_scale}
    \ket{0^m}_\texttt{scl}\ket{j}\mapsto \ket{s(j)}_\texttt{scl}\ket{j},
\end{equation}
where $s(j)$ is the scale of~$j$, which satisfies $j=2^{s(j)}+k$ with $0\leq k<2^{s(j)}$; for $j=0$, we use $s(0)=n$ as unused scale.
As the scale of~$\ket{j}$ is now stored in \texttt{scl}, we no longer need to perform Step~1 in $U_{Q_s}$ to test the scale of $\ket{j}$.
We therefore remove the flag~$t$, the initial scale test, and the controls on $t=0$ from the remaining steps.
Let $V_s$ denote the resulting circuit, consisting of Steps~$\geq2$ in the construction of~$U_{Q_s}$ above.
When $\ket{j}$ is at scale~$s$, we have
\begin{equation}
    V_s\ket{0^{a}}\ket{j} = \ket{0^{a}}\frac{Q_s}{\alpha}\ket{j}+\ket{\perp}
\end{equation}
with $(\langle0^{a}|\otimes\id) \ket{\perp}=0$, where we used $a=p+n_s+1$ ancillas in block encoding of $U_{Q_s}$ after removing the scale-test flag~$t$.
We now use the scale register to select the appropriate~$V_s$.
That is, for $s=0,\ldots,n-1$, we apply $V_s$ controlled on $s(j)=s$.
Note that this performs $\texttt{SELECT}(V)=\sum_s \ket{s}\!\!\bra{s}\otimes V_s$ and maps the basis state as
\begin{equation}
    \ket{s(j)}\ket{j} \mapsto
    \frac{1}{\alpha} \ket{s(j)} Q_{s(j)}\ket{j}
\end{equation}
when the~$a$ ancillas are in the all-zero state.
Now let \texttt{UNIF} prepare the uniform state $\texttt{UNIF}\ket{0^m}=(1/\sqrt{n}) \sum_{s=0}^{n-1} \ket{s}$, then applying $\texttt{UNIF}^\dagger$ on the scale register gives a block encoding of~$Q$ with normalization $\alpha_Q=\alpha\sqrt{n}\in\Theta(\sqrt{n})$. 
That is, $Q/\alpha_Q$ is applied on $\ket{j}$ conditioned on the scale register being in the all-zero state. To see this, observe that
\begin{align}
    (\bra{0^m}\otimes\id)(\texttt{UNIF}^\dagger\otimes\id)
    \frac{1}{\alpha} \ket{s(j)} Q_{s(j)}\ket{j}
    &= \frac{1}{\alpha} \bra{0^m}\texttt{UNIF}^\dagger\ket{s(j)}Q_{s(j)}\ket{j}\nonumber\\
    &= \frac{1}{\alpha\sqrt{n}}
    Q_{s(j)}\ket{j}\nonumber\\
    &= \frac{1}{\alpha\sqrt{n}} \sum_s Q_s \ket{j} =\frac{1}{\alpha\sqrt{n}} Q \ket{j},
\end{align}
where we used $Q_s\ket{j}=0$ for $s\neq s(j)$.
The resulting circuit is then a block encoding~$U_Q$ of~$Q$ with normalization~$\alpha\sqrt{n}$.
The ancilla count for block encoding of~$U_Q$ is $a_Q= m+a\in\Theta(\log(n/\varepsilon))$ using $m=\ceil{\log (n+1)}$ and $a\in\Theta(\log(n/\varepsilon))$.

\paragraph{Cost of the full block encoding.}
We use the scale test to implement the map in~\cref{eq:write_scale} that writes the scale of~$j$ into the scale register. Specifically, for each $s=0,\ldots,n-1$, we test whether the leftmost $n_s+1$ qubits of $\ket{j}$, with $n_s=n-s-1$, are in the state $\ket{0^{n_s}1}$ and flip a temporary flag qubit. Then, controlled on $\ket{1}$ of the flag, we write the classically known $m$-bit binary representation of~$s$ into \texttt{scl}, and finally uncompute the flag.
The scale test for each $s$ has gate cost $\order{n_s}$, and writing $s$ into \texttt{scl} needs at most~$m=\ceil{\log n}$ CNOTs. For all~$s$, the gate cost is then~$\order{n^2}$.

To implement $\texttt{SELECT}(V)$, we compute a temporary flag for each $s=0,\ldots,n-1$ to indicate whether the scale register is in the state~$\ket{s}$, i.e., to test if $s(j)=s$. We then apply~$V_s$ controlled on this flag, and finally uncompute the flag.
This implementation replaces the direct $m$-controlled application of~$V_s$ by a single-controlled~$V_s$.
Computing and uncomputing the flag has gate cost~$\order{m}$ for each~$s$.
The gate cost of~$V_s$ is simply the gate cost $\order{n\log^2(n/\varepsilon)}$ of~$U_{Q_s}$ discussed before. So, applying controlled~$V_s$ for all~$n$ possible values of~$s$ has gate cost~$\order{n^2\log^2(n/\varepsilon)}$.
Altogether, we obtain $\mathcal{G}(Q)\in\order{n^2\log^2(n/\varepsilon)}$ for the overall gate cost of implementing the block encoding~$U_Q$ of~$Q$.

Having the block encoding~$U_Q$, we construct the block encoding~$U_{A_\precond}$ of $A_\precond$ by the block encodings $U_Q$, $U_E$ and $U_{K_\precond}$ using the standard product and summation rules for block encodings.
The block encoding of $K_\precond$ is obtained as $U_{K_\precond}=U_P(\id\otimes W)U_K(\id\otimes W^\dagger) U_P$, where the block encoding~$U_P=(H\otimes\id_n)C_0(U^+)C_1(U^-)(H\otimes\id_n)$ for~$P$ in terms of controlled application of~$U^\pm$ is obtained by the LCU technique and the decomposition $P=(U^++U^-)/2$~\cite{BNW+26}.
By the inequality-test approach in~\cref{subsubsec:operator_oracle_implementation}, the gate cost of $U_E$ and $U_K$ is $\Theta(n^3\log n)$ as per~\cref{eq:gate_cost_diagonal_encodings}.
With the gate cost $\order{n}$ for controlled $U^\pm$~\cite{BNW+26} and the gate cost $\mathcal{G}(Q)$ for~$Q$, we have
\begin{equation}
\label{eq:gate_cost_Aprec}
    \mathcal{G}(U_{A_\precond}) \in\order{n^3\log n} + \order{n^2\log^2(n/\varepsilon)}
    \subseteq \order{n^3\log^2(n/\varepsilon)}
\end{equation}
for the gate cost of implementing  the block encoding of the preconditioned matrix.
The number of ancillas used for block encoding of the diagonals~$E$ and~$K$ is~$\Theta(n)$ as per~\cref{eq:ancilla_count_diagonals} and $a_Q\in\Theta(\log(n/\varepsilon))$ ancillas are used for block encoding of~$Q$.
The total number of ancilla qubits~$a_\precond$ for block encoding of $A_\precond$ then scales as $a_\precond\in\Theta(n+\log(n/\varepsilon))$. 

\paragraph{Computing the scaled coefficients.}
We now present a procedure for computing the scaled coefficients with a logarithmic time and space.
We use the following lemma for our implementation.

\begin{lemma}
\label{lem:As_prime_expression}
    Let $A'_s= \sqrt{2^{n-s}} A_s$ with $A_s$ given in \cref{eq:As_expansion}. 
    Then we can express $A'_s$ as
    \begin{equation}
        A'_s= \sqrt{8} H'^{\dagger}_{n-1}
    \cdots H'^{\dagger}_{s+1} (S_{s+1}-\id_{s+1}) G^\dagger_s,
    \end{equation}
    where $H'_r$ is the modified version of the refinement matrix $H_r$ with entries $[H'_r]_{ij}=h'_{(j-2i)\bmod 2^{r+1}}$ obtained from the modified filter coefficients $h'_\ell:= \sqrt{8}\sum_{k=0}^{\ell} (-1)^{\ell-k}h_k$.
\end{lemma}

The scaled coefficient $a'_{s,m}=\langle m|A'_s|0^s\rangle$ associated with $c$ and $b$ is at the index $m=2^{n_s}c+b$, where $n_s=n-s-1$ and $0\leq b<2^{n_s}$. Here $n_s$ is the number of refinements, i.e., the number of applications of $H'^\dagger$.
Each refinement maps an input vector $\bm{x}$ to the output vector $\bm{y}=H'^\dagger\bm{x}$ with elements $y_m=\sum_j h'_{m-2j}x_j$; we omit the modulo operation for notational simplicity. Let $D_s:=(S_{s+1}-\id_{s+1})G^\dagger_s$, the initial vector is the nonzero elements of $\sqrt{8}D_s\ket{0^s}$.
As shown in~\cref{apx:lemmas_prex_mtx}, these elements are $d_r=\langle r|D_s|0^s\rangle$ for $r=0,1,\ldots,M$, where $M$ is the length of the original wavelet filter, so the initial vector is the fixed-size vector $x=\sqrt{8}(d_0,\ldots,d_M)$.

To describe our approach for computing the desired coefficient, let us first consider the case with one refinement ($n_s=1$). In this case, 
$m=2c+b_0$ with $b_0\in\{0,1\}$ and the desired coefficient $y_{2c+b_0} = \sum_{\ell} h'_{2\ell+b_0} x_{c-\ell}$ can be computed conditioned on the bit~$b_0$.
Let us define the vector $\bm{x}^1=(x^1_0,x^1_1,\ldots,x^1_{M'-1})$ with $x^1_\ell:=x_{c-\ell}$, where $M'$ is the length of the vector of modified filters $\bm{h}'=(h'_0,\ldots,h'_{M'-1})$.
If $b_0=0$, the desired coefficient $y_{2c}=\sum_\ell h'_{2\ell}x^1_\ell$ is the inner product of the even-indexed filter coefficients $(h'_0,h'_2,\ldots)$ and $\bm{x}^1$.
If $b_0=1$, the desired coefficient $y_{2c+1}$ is obtained by the inner product of the odd-indexed filter coefficients $(h'_1,h'_3,\ldots)$ and $\bm{x}^1$.

Let us now consider two refinements ($n_s=2$). In this case, $m=2^2c+b$ with $0\leq b<2^2$.
In binary $b=(b_1b_0)_2=2b_1+b_0$, so we can write $m=2(2c+b_1)+b_0=2c_1+b_0$ with $c_1=2c+b_1$.
Let $\bm{x}^2=(x^2_0,x^2_1,\ldots,x^2_{M'-1})$ be the vector obtained from the output vector $\bm{y}$ of the first refinement as $x^2_\ell:=y_{c_1-\ell}=y_{2c+b_1-\ell}$; the bit $b_1$ determines how this vector is obtained from $\bm{x}^1$.
The second refinement then gives the desired coefficient at $m=2c_1+b_0$. Let $\bm{z}$ be the output vector of the second refinement, the desired coefficient is then $z_{2c_1+b_0} = \sum_{\ell} h'_{2\ell+b_0} y_{c_1-\ell} = \sum_{\ell} h'_{2\ell+b_0} x^2_\ell$. Notice that just as the one-refinement case, the last bit $b_0$ determines the even-indexed filter coefficients when $b_0=0$ and the odd-indexed filter coefficients when $b_0=1$.

The same procedure can be used to extend the approach to $n_s$ refinements. That is, for refinement $r=1,\ldots,n_s$, the bit $b_{n_s-r}$ of $b=(b_{n_s-1}\ldots b_1b_0)_2$, from left to right, controls whether to use the even- or odd-indexed filter coefficients.
For all intermediate refinements, we compute a fixed-size vector of $M+1$ coefficients needed for the next refinement. At the final refinement, only the single coefficient $a'_{s,m}$ is computed.
This approach gives the following steps for computing the desired coefficient on a quantum register for $n_s\geq1$.
For $n_s0$, no refinement is needed, and the desired coefficient is directly obtained from the initial vector. 

\begin{enumerate}
    \item (\textit{Load the initial vector}).
    Initialize $M+1$ registers, each with $n_\texttt{tmp}$ qubits, in the all-zero state and
    perform the map $\ket{c}\ket{0^{n_\texttt{tmp}}}\cdots \ket{0^{n_\texttt{tmp}}}
        \mapsto \ket{c} \ket{x^1_0} \ket{x^1_1}\cdots\ket{x^1_{M}}$,
    where $x^1_\ell=x_{c-\ell}$ are known classical values from the initial vector $\bm{x}=\sqrt{8}(d_0,\ldots,d_M)$.
    In compact notation, we write this map as $\ket{c}\ket{\bm{0}}\mapsto\ket{c}\ket{\bm{x}^1}$.
    \item (\textit{Perform the intermediate refinements}).
    For each refinement $r=1,\ldots,n_s-1$, compute the refined vector into a fresh set of $M+1$ $n_\texttt{tmp}$-qubit registers controlled on qubit $\ket{b_{n_s-r}}$. That is, perform the map $\ket{b_{n_s-r}}\ket{\bm{x}^r}\ket{\bm{0}}
        \mapsto \ket{b_{n_s-r}} \ket{\bm{x}^r}\ket{\bm{x}^{r+1}}$.
    The control qubit determines whether the even- or odd-indexed filter coefficients are used for computing the refined vector $\bm{x}^{r+1}$.
    \item (\textit{Compute the coefficient by the final refinement}). Now controlled on qubit $\ket{b_0}$, perform the final refinement to obtain the desired coefficient. Specifically, we perform the map $\ket{b_0}\ket{\bm{x}^{n_s}}
        \ket{0^{n_\texttt{tmp}}}
        \mapsto
        \ket{b_0}\ket{\bm{x}^{n_s}}
        \ket{a'_{s,m}}$.
\end{enumerate}
Finally, we uncompute the intermediate vectors by reversing Steps~$1$ and~$2$, so that only the desired coefficient is in the output register.

The quantum circuit by the above steps acts on $\ket{c}\ket{b}$ and computes $a'_{s,m}$ on the fly for $m=2^{n_s}c+b$ on an $n_\texttt{tmp}$-qubit register.
The number of ancilla registers used for storing intermediate vectors and the final coefficient is $n_s(M+1)+1$ and each register contains $n_\texttt{tmp}$ qubits. The overall number of ancilla qubits is then $n_\texttt{anc}=(n_s(M+1)+1)n_\texttt{tmp}$.
Using $n_s< n$, $n_\texttt{tmp}\in\Theta(\log(n/\varepsilon))$, and $M\in\Theta(1)$ for a fixed wavelet, we achieve $n_\texttt{anc}\in\order{n\log(n/\varepsilon)}$.

The overall cost of implementing \texttt{COEF} is as follows.
 The first step writes $M+1$ fixed-point numbers on $n_\texttt{tmp}$-qubit registers controlled on a constant number of qubits. As $M\in\Theta(1)$, the cost of loading the initial vector is then $\order{n_\texttt{tmp}}$. The total number of refinements in Step~2 and Step~3 above is $n_s$ and we show in the following that the gate cost of each refinement is $\Theta(n^2_\texttt{tmp})$.
 With $n_s<n$ and $n_\texttt{tmp}\in\Theta(\log(n/\varepsilon))$, the overall gate cost of \texttt{COEF} is then $\order{n\log^2(n/\varepsilon)}$.

\paragraph{Implementing one refinement.}
Each refinement is a map of the form $\ket{\beta}\ket{\bm{x}}\ket{\bm{0}}\mapsto
\ket{\beta}\ket{\bm{x}}\ket{\bm{y}}$, where $\beta\in\{0,1\}$, and $\bm{x}$ and $\bm{y}$ are fixed-length vectors that satisfy $y_k=\sum_{\ell}h'_{2\ell+\beta-k}x_\ell$.
Let us consider the even $2k$ and odd $2k+1$ entries of~$\bm{y}$ based on~$\beta$.
For $\beta=0$, the even entries are $y_{2k}=\sum_j h'_{2j} x_{k+j}$, where we set $\ell-k=j$, and similarly the odd entries are $y_{2k+1}=\sum_j h'_{2j+1} x_{k+j+1}$.
Therefore, when $\beta=0$, the even entries of $\bm{y}$ are computed using even-indexed filter coefficients and the odd entries are obtained using the odd-indexed filter coefficients.
It then follows that when $\beta=1$, the even entries are computed using the odd-indexed filter coefficients as $y_{2k}=\sum_jh'_{2j+1}x_{k+j}$ and the odd entries are computed using even-indexed filter coefficients as $y_{2k+1}=\sum_jh'_{2j}x_{k+j}$.  

We compute each output entry using a sequence of multiply-and-accumulate operation defined as 
\begin{equation}
    \texttt{MAC}(h'_j)\ket{x}\ket{y} = \ket{x}\ket{y+h'_j x},
\end{equation}
which multiplies the first register by the classically known fixed-point constant $h'_j$ and accumulates it to the second register.
For example, for $\beta=0$, $y_{2k}=h'_0x_k+h'_2x_{k+1}+\cdots$, so to compute $y_{2k}$ we initialize the corresponding output register as $\ket{y_{2k}}=\ket{0}$ and apply $\texttt{MAC}(h'_0)$ on $\ket{x_k}$ and $\ket{y_{2k}}$ to obtain $\ket{y_{2k}}=\ket{h'_0x_k}$. We then apply $\texttt{MAC}(h'_2)$ on $\ket{x_{k+1}}$ and the same accumulator to obtain $\ket{y_{2k}}=\ket{h'_0x_k+h'_2x_{k+1}}$. This process is continued over the nonzero even-indexed filter coefficients until $y_{2k}$ is computed.
The other entries of $\bm{y}$ are computed similarly.

\paragraph{Cost of one refinement.}
The modified filter vector $\bm{h'}$ has length $M'=M-1$, so each output entry can be computed by at most $\ceil{M'/2}$ \texttt{MAC} operations. The output vector $\bm{y}$ has $M+1$ entries, so the overall number of \texttt{MAC} operations used in one refinement is $\order{M^2}$.
For a fixed wavelet, $M$ is constant, so the number of \texttt{MAC} operations used in one refinement is~$\order{1}$.
As each $h'_j$ is a classically known fixed-point constant, its multiplication does not need a general quantum multiplier. Using the binary representation of $h'_j$, multiplication by $h'_j$ can be implemented using at most $n_\texttt{tmp}$ predetermined shifted additions~\cite[Appendix~A]{HRS18}.
Each addition of two $n_\texttt{tmp}$-qubit quantum registers can be implemented using $\Theta(n_\texttt{tmp})$ gates, so the overall cost of implementing \texttt{MAC} is $\order{n^2_\texttt{tmp}}=\order{\log^2(n/\varepsilon)}$.

\subsection{Properties of interest and tailored implementation for the observable oracle}
\label{subsec:obs}
Having developed a quantum algorithm for generating a quantum solution state that encodes the electrostatic potential of the biomolecular system, we now discuss extracting key properties of the solution. In particular, we consider two properties of interest: electrostatic free energy and solvation free energy. We show how these quantities can be expressed in terms of expectation values of a suitable observable with respect to the solution state.
We then construct a block encoding of the observable, present a quantum circuit implementation, and analyze its gate cost and ancilla count.

\subsubsection{Electrostatic free energy and its observable construction}
A key quantity of interest for the biomolecular system is the electrostatic free energy, which represents the work needed to assemble the atomic charge distribution of the biomolecule. This energy is given by~\cite{BSH+01,KFS+20}
\begin{equation}
\label{eq:electostatic_free_energy}
    E = \frac12 \sum_{a=0}^{N_\text{atm}-1} z_a \bar{u}_a,
\end{equation}
where $\bar{u}_a$ is the average electrostatic potential acting on atom~$a$ with charge~$z_a$.
The average $\bar{u}_a$ is over nearby grid points and is obtained from the discretization of the Dirac delta.
Specifically, by the continuum form of electrostatic free energy $E=\frac12\int_\Omega \rho(\bm{x})u(\bm{x})\,\mathrm{d}\bm{x}$ and the atomic charge density $\rho(\bm{x})=\sum_az_a\delta(\bm{x}-\bm{x}_a)$, the average potential is
\begin{equation}
\label{eq:u_average}
    \bar{u}_a=\int_\Omega u(\bm{x}) \tilde{\delta}(\bm{x}-\bm{x}_a)\,\mathrm{d}\bm{x}
    =\sum_{\bm{s}\in\{0,1\}^d} w_{a,\bm{s}} u_{\bm{\mathrm{x}}_a+\bm{s}},
\end{equation}
where $\tilde{\delta}(\bm{x}-\bm{x}_a)$ is the discretized Dirac delta in~\cref{eq:dD_approximate_delta} in the position-dependent charge distribution with weight~$w_{a,\bm{s}}$ in \cref{eq:dD_weight}.
For the uniform charge distribution over the nearest-neighbor points discussed in~\cref{subsubsec:loading_uniform}, the weight is~$1/2^d$, independent of~$\bm{s}\in\{0,1\}^d$ and the position of atom~$a$. Similarly, the weight $w_{\bm{s}}=1/2^{d+\norm{\bm{s}}_1}$ with $\bm{s}\in\{0,\pm1\}^d$ in \cref{eq:dD_weight_unifom_3points} for the charge distribution over three nearby points is independent of the position of atom~$a$.
We first construct an observable for these position-independent cases and then discuss the general position-dependent average potential. 

For the position-independent cases, let us define the translation operator
\begin{equation}
\label{eq:T}
    T = \sum_{\bm{s}} w_{\bm{s}} T_{\bm{s}},
    \enspace
    T_{\bm{s}}:= \sum_{\bm{q}}
    \ket{\bm{q}+\bm{s}}\!\!\bra{\bm{q}},
\end{equation}
where $\bm{q}$ is the index of grid points and $T_{\bm{s}}$ shifts the grid states as $\ket{\bm{q}}\mapsto\ket{\bm{q}+\bm{s}}$. Also, let $\ket{\bm{u}}=\sum_{\bm{q}}u_{\bm{q}}\ket{\bm{q}}$ be the vector (an unnormalized state) of approximate potential on grid points~$\bm{q}$.
Then notice that $T^\dagger$ maps~$\ket{\bm{u}}$ to the vector of average potentials on grid points as
\begin{equation}
    T^\dagger: \ket{\bm{u}} \mapsto \ket{\bar{\bm{u}}}:=\sum_{\bm{q}} \bar{u}_{\bm{q}}\ket{\bm{q}},
\end{equation}
where $\bar{u}_{\bm{q}}=\sum_{\bm{s}}w_{\bm{s}}u_{\bm{q}+\bm{s}}$ is average potential on grid point~$\bm{q}$.
The average potential over atom~$a$ is then obtained as
\begin{equation}
    \bra{\bm{\mathrm{x}}_a} T^\dagger\ket{\bm{u}} =
    \sum_{\bm{s}}w_{\bm{s}} u_{\bm{\mathrm{x}}_a+\bm{s}} = \bar{u}_a,
\end{equation}
and by the atom-position state $\ket{\text{pos}}$ in~\cref{eq:ket_pos}
\begin{equation}
    \bra{\text{pos}}T^\dagger\ket{\bm{u}}=
    \frac{1}{\norm{\bm{z}}}
    \sum_{a} z_a \bar{u}_a=\frac{2E}{\norm{\bm{z}}}.
\end{equation}
To turn this overlap into an expectation value, we define the unnormalized state $\ket{\phi}$ and the measurement operator~$M$ as
\begin{equation}
\label{eq:observable_M}
    \ket{\phi} := \frac{1}{\sqrt{2}}
    \left(\ket{0}\ket{\bm{u}}+\ket{1}\ket{\text{pos}}\right),
    \enspace
    M:=\ket{0}\!\!\bra{1}\otimes T + \ket{1}\!\!\bra{0}\otimes T^\dagger,
\end{equation}
then $E=\tfrac12\norm{\bm{z}}\langle\phi |M|\phi\rangle$.
Note that $T$ is symmetric for the uniform charge distribution over three nearby points, and in this case $M=X\otimes T$.
The state $\ket{\phi}$ and the operator~$M$ here are the same as those in~\cref{eq:phi_M}, so by~\cref{eq:phi_M_prime} we have
\begin{equation}
    E=\frac12\xi\norm{\bm{z}}
    \langle\Psi |M'|\Psi\rangle,
\end{equation}
where
\begin{align}
    \ket{\Psi}&=\frac{1}{\sqrt{2}}\left(
    \ket{0}\ket{\psi}+\ket{1}\ket{\chi}
    \right),\enspace
    \ket{\chi}:=\ket{++}\ket{\text{pos}}\\
    \label{eq:M_prime_obs}
    M'&=\ket{0}\!\!\bra{1}\otimes\id_2\otimes T + \ket{1}\!\!\bra{0}\otimes\id_2\otimes T^\dagger.
\end{align}

\begin{figure}[ht]
    \centering
    \includegraphics[width=.56\linewidth]{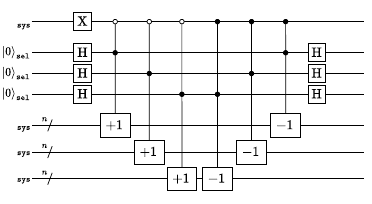}
    \caption{Quantum circuit for implementing the block encoding of the observable~$M$ in~\cref{eq:observable_M} used for estimating the electrostatic free energy.}
    \label{fig:UM}
\end{figure}

We now discuss block encoding of~$T$ for the case of uniform charge distribution over the nearest-neighbor points with weights $w_{\bm{s}}=1/2^d$; the block encoding for the symmetric position-independent charge distribution over three points follows similarly.
Let
\begin{equation}
    \texttt{PREP}\ket{0^d}=
    \sum_{\bm{s}\in\{0,1\}^d}\sqrt{w_{\bm{s}}}\ket{\bm{s}}
    =\ket{+}^d,
\end{equation}
where $\texttt{PREP}=H^{\otimes d}$, and let
\begin{equation}
\label{eq:select}
    \texttt{SEL}=
    \sum_{\bm{s}}
    \ket{\bm{s}}\!\!\bra{\bm{s}} \otimes T_{\bm{s}},
\end{equation}
then $U_T = \left(\texttt{PREP}^\dagger\otimes\id\right) \texttt{SEL} \left(\texttt{PREP}\otimes\id\right)$ is an~$(1,d,0)$-block encoding of~$T$, and
\begin{equation}
    U_M=\ket{0}\!\!\bra{1}\otimes U_T + \ket{1}\!\!\bra{0}\otimes U_T^\dagger
    = \left(\ket{0}\!\!\bra{0}\otimes U_T + \ket{1}\!\!\bra{1}\otimes U_T^\dagger\right)\left(X\otimes\id\right)
\end{equation}
is a~$(1,d,0)$-block encoding of $M$.
This block encoding can be simplified as
\begin{equation}
    U_M
    = 
    \left(
    \id_1\otimes\texttt{PREP}^\dagger\otimes\id
    \right)
    \left(
    \ket{0}\!\!\bra{0}\otimes \texttt{SEL} + \ket{1}\!\!\bra{1}\otimes \texttt{SEL}^\dagger
    \right)
    \left(X\otimes\texttt{PREP}\otimes\id
    \right),
\end{equation}
which we use in our implementation in~\cref{fig:UM}.
The block encoding of $M'$ is similar but with an extra~$\id_2$ as in~\cref{eq:M_prime_obs}.
In our application $d=3$, and so the number of ancillas in~$U_M$ is constant.

To implement \texttt{SEL}, we note that $T_{\bm{s}}$ with $\bm{s}=s_1s_2\cdots s_d$ in binary can be decomposed as product of 1D shifts as $T_{\bm{s}}= S_1^{s_1}\cdots S_d^{s_d}$, where $S_j$ is the shift operator in direction~$j$, so we obtain $\texttt{SEL}=C_{s_1}(S_1) \cdots C_{s_d}(S_d)$ as product of controlled shift (increment) operations. Then $U_M$ is implemented by~$d$ double-controlled increment operations and their inverses, and \texttt{PREP} and its inverse. The gate cost of implementing $U_M$ is therefore $\mathcal{G}(U_M)\in\Theta(dn)$
by \cref{lem:multi_controlled_increment}.
We have the same gate cost for $U_{M'}$.

To extend our construction to the general position-dependent case, we use the trick in~\cref{eq:subsubsec:loading_nonuniform} that the position-dependent weight~$w_{a,\bm{s}}$ is encoded in the state $\ket{\text{pos}'}$ in \cref{eq:pos_prime}.
Notice that
\begin{equation}
    \texttt{SEL}\ket{\text{pos}'} = \frac{1}{\norm{\bm{z}'}}
    \sum_{a,\bm{s}} z_a w_{a,\bm{s}}\ket{\bm{s}}\ket{\bm{\mathrm{x}}_a+\bm{s}},
\end{equation}
so by~\cref{eq:u_average} we have
\begin{equation}
    \bra{+}^{\otimes d} \bra{\bm{u}}
    \texttt{SEL}\ket{\text{pos}'} =
    \frac{1}{\norm{\bm{z}'}\sqrt{2^d}}\sum_{a,\bm{s}} z_a w_{a,\bm{s}} u_{\bm{\mathrm{x}}_a+\bm{s}}
    = \frac{1}{\norm{\bm{z}'}\sqrt{2^d}}\sum_{a} z_a \bar{u}_a
    =\frac{2E}{\norm{\bm{z}'}\sqrt{2^d}}.
\end{equation}
To turn this overlap into an expectation value as the uniform case, we modify the unnormalized state $\ket{\phi}$ and the measurement operator~$M$ as
\begin{align}
    \ket{\phi}
    &\rightarrow \ket{\phi}=
    \frac{1}{\sqrt{2}}
    \left(\ket{0}\ket{+}^d\ket{\bm{u}}
    +\ket{1}\ket{\text{pos}'}\right)\\
    M &\rightarrow M = \ket{0}\!\!\bra{1}\otimes \texttt{SEL} + \ket{1}\!\!\bra{0}\otimes \texttt{SEL}^\dagger.
\end{align}
With these changes, we have
\begin{equation}
    E=\frac{\sqrt{2^d}}2\norm{\bm{z}'}\langle\phi |M|\phi\rangle
    =\frac{\sqrt{2^d}}2\xi\norm{\bm{z}'}\langle\Psi |M'|\Psi\rangle.
\end{equation}
The second identity here follows as before, but with the modified state
\begin{equation}
    \ket{\Psi} \rightarrow
    \ket{\Psi} =
    \frac{1}{\sqrt{2}}\left(
    \ket{0}\ket{+}^d\ket{\psi}
    +\ket{1}\ket{\chi'}\right),
    \enspace
    \ket{\chi'}:=\ket{++}\ket{\text{pos}'}
\end{equation}
and the modified measurement operator in the extended space
\begin{align}
    &M'\rightarrow M'=
    \ket{0}\!\!\bra{1}\otimes \texttt{SEL}' + \ket{1}\!\!\bra{0}\otimes \texttt{SEL}^{'\dagger},\\
    &\texttt{SEL}\rightarrow
    \texttt{SEL}':=
    \sum_{\bm{s}}
    \ket{\bm{s}}\!\!\bra{\bm{s}}\otimes\id_2 \otimes T_{\bm{s}}.
\end{align}
We note that, as per~\cref{subsubsec:loading_beyond_NN}, encoding the position-dependent weight~$w_{a,\bm{s}}$ in the state $\ket{\text{pos}'}$ does not change the gate cost of preparing the initial state for a fixed~$d$.
Moreover, note that $M$ and $M'$ here are Hermitian and unitary as their select operations are unitary. Their block encodings are therefore trivial.
That is, $M$ and $M'$ are $(1,0,0)$-block encoding of themselves. The cost of these block encodings is~$\Theta(dn)$, which is the cost of implementing the select operation.

\subsubsection{Solvation free energy}
Our approach for computing the electrostatic free energy from the quantum solution of the Poisson-Boltzmann equation can be used to compute other quantities of interest, such as the solvation free energy~\cite{JNM+23,KFS+20}. This quantity is the difference between the electrostatic free energies of the biomolecule in two environments: a homogeneous reference environment, where the dielectric coefficient is constant in the entire domain and equal to that of the biomolecule, and an inhomogeneous solvated environment, where the biomolecule and the surrounding solvent have different dielectric coefficients. 
The solvation free energy can then be obtained by applying our algorithm to estimate the electrostatic free energy in each environment and subtracting the reference energy from the solvated energy.

\subsection{The complete algorithm and its overall cost}
The complete end-to-end quantum algorithm is obtained by composing the algorithms for preparing the initial state, generating the distributed quantum solution state, and extracting a property of the solution state.
As summarized in~\cref{tab:cost}, the overall gate cost of the end-to-end quantum algorithm is determined by the gate cost~$\mathcal{G}_\varepsilon(\psi)$ of generating the quantum solution state $\ket{\psi}$, which includes the cost of preparing the initial state, and the gate cost~$\mathcal{G}(U_M)$ of a block encoding of the observable~$M$ associated with a property of the solution state.

In~\cref{eq:gate_cost_solution_state}, we obtained the gate cost~$\mathcal{G}_\varepsilon(\psi)$ in terms of the gate cost $\mathcal{G}(U_{A_\precond})$ of a block encoding of the preconditioned matrix.
For our construction of the block encoding, we established in~\cref{eq:gate_cost_Aprec} that $\mathcal{G}(U_{A_\precond})\in\order{n^3\log^2(n/\varepsilon)}$. By these equations, we obtain the overall gate cost 
\begin{equation}
\label{eq:cost_distributed_qsolution}
    \mathcal{G}_\text{dist}\in \order{n^4\log^4(n/\varepsilon)}
\end{equation}
for generating the distributed quantum solution, where $n=\log_2N$ with~$N$ the number of grid points per dimension.
Moreover, as shown in~\cref{subsec:obs}, our block encoding of the observable~$M$ has the gate cost~$\mathcal{G}(U_M)\in\Theta(n)$.
Considering these costs and the number of uses of~$U_\psi$ (the unitary for preparing the solution state) and the observable block encoding~$U_M$, as specified for each property-estimation method in~\cref{tab:cost}, we obtain the overall gate cost reported in~\cref{tab:cost_tailored_algorithm} for our tailored end-to-end algorithm for simulating the biomolecular system.

We now consider the number of ancilla qubits used in the algorithm.
As we established in~\cref{subsubsec:block_encoding_Aprec}, the number of ancilla qubits for block encoding of the preconditioned matrix is $a_\precond\in\Theta(n+\log(n/\varepsilon))$.
To block encode the inverse of this matrix, the QSVT construction uses one additional ancilla qubit; i.e., $a_\text{inv}=a_\precond+1$.
As discussed in~\cref{subsec:obs}, our block encoding of the observable~$M$ uses~$\Theta(1)$ ancilla qubits.
Consequently, the overall ancilla count for block encodings, accounting for both the PDE operator and the observable, scales as~$\Theta(n+\log(n/\varepsilon))$.
In addition, our construction for the block encoding of the preconditioned matrix uses $\Theta(n\log(n/\varepsilon))$ scratchpad qubits that start and return to zero state and are used for intermediate computation.
These ancilla counts are summarized in~\cref{tab:cost_tailored_algorithm} along with the ancilla qubits used by each property-estimation method specified in~\cref{tab:cost}. 

As seen in~\cref{tab:cost_tailored_algorithm}, the three approaches considered for property estimation exhibit different tradeoffs between gate cost, ancilla count, and number of circuit runs, but all approaches have circuit size and overall gate cost polylogarithmic in~$N$. The Hadamard test only needs one ancilla qubit for property estimation and has a gate cost polylogarithmic in~$1/\varepsilon$ per run. Achieving accuracy~$\varepsilon$ needs a quadratic number of runs in~$1/\varepsilon$.
Phase estimation needs $\Theta(\log(1/\varepsilon))$ ancillas for property estimation and has an additional gate cost of $\Theta(n/\varepsilon)$ per run, without reducing the scaling of the number of runs.
In contrast, the amplitude-estimation approach improves the overall gate cost to quasilinear in~$1/\varepsilon$. Specifically, repeating the amplitude-estimation approach $\order{\log(1/\delta)}$ times yields an estimate with success probability~$1-\delta$, with an overall gate cost  of~$\order{(1/\varepsilon)\log^4(1/\varepsilon)}$ per run. Alternatively, the desired estimate can be achieved in a single run, at an overall gate cost that scales linearly in~$1/\delta$. 

\begin{figure}[htb]
    \centering
    \includegraphics[width=\linewidth]{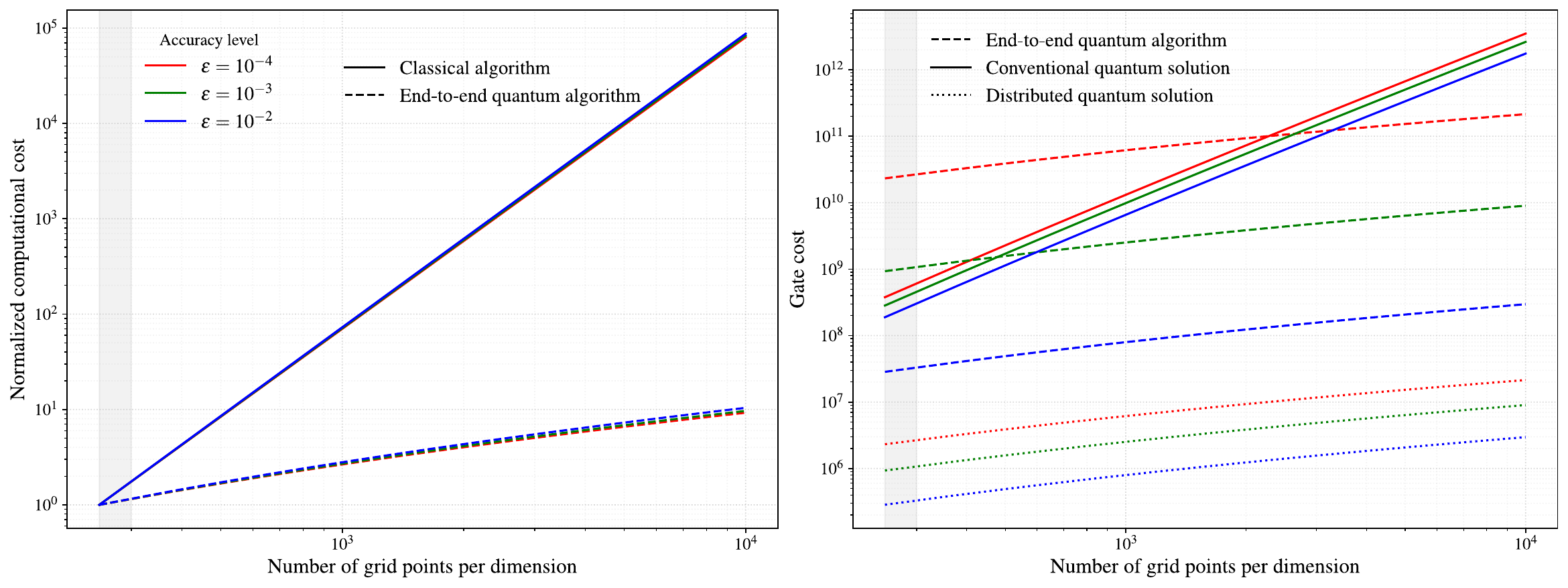}
    \caption{
    Illustrative comparison of computational cost scaling for solving the LPB equation.
    Left: Normalized cost of our end-to-end quantum algorithm, quantified as the number of gates used, compared with that of the classical algorithm based on the geometric multigrid method, quantified as the number of arithmetic operations.
    Each cost is normalized by the cost at the reference value~$N_0=250$ to compare its relative growth with~$N$.
    Three accuracy levels are considered: $\varepsilon=10^{-4}$~(red), $10^{-3}$~(green) and~$10^{-2}$~(blue).
    The highlighted region indicates representative grid sizes of $N=250-300$ points per dimension used in classical computations for the LPB equation~\cite{QL18}.
    Right: The gate cost of generating the distributed quantum solution~[\cref{eq:cost_distributed_qsolution}] compared with that of generating the conventional quantum solution~[\cref{eq:cost_conventional_qsolution}].
    The gate cost of the end-to-end quantum algorithm [\cref{eq:cost_E2E_qalgorithm}] is also shown.
    The plots illustrate only the cost scaling;
    the number of atoms~$N_\texttt{atm}$ and constant factors are not~considered.
    }
    \label{fig:solution_cost}
\end{figure}

In~\cref{fig:solution_cost}, we provide an illustrative comparison of the computational cost scaling of our end-to-end quantum algorithm with that of a classical algorithm that explicitly computes the solution vector of the LPB equation.
For our quantum algorithm, we consider the amplitude-estimation approach, which achieves the best overall scaling 
\begin{equation}
\label{eq:cost_E2E_qalgorithm}
    \gate_{\text{E2E}}\in \order{(n^4/\varepsilon)\log^4(n/\varepsilon)}
\end{equation}
for the gate cost of the end-to-end quantum algorithm.
For the classical approach, we consider the geometric multigrid method, which has been established as an optimal classical solver for the LPB equation in biomolecular applications~\cite{QL18,HS93,WL10}.
For the 3D problem discretized over~$N^3$ grid points ($N$ grid points per dimension), the computational cost of this method is bounded by $\order{N^3\log(N/\varepsilon)}$ arithmetic operations~\cite{QL18,BHM00}, where~$\varepsilon$ is the relative error between the computed solution vector and the exact solution of the discretized system, measured in the Euclidean norm.
This scaling is consistent with the numerical results in~\cite{WL10}, which demonstrate nearly linear dependence in the total number of grid points.

As the classical arithmetic operations and quantum gates are different computational resources, we compare the relative growth of their computational cost instead of their absolute scalings. Specifically,
we compare the normalized computational cost defined as $\cost_\text{norm}(N,\varepsilon)=\cost(N,\varepsilon)/\cost(N_0,\varepsilon)$, which normalizes the cost scaling $\cost(N,\varepsilon)$ of each algorithm by the cost~$\cost(N_0,\varepsilon)$ at a reference size of~$N_0$.
This normalization allows us to compare how the computational costs increase with grid size without treating classical arithmetic operations and quantum gates as equivalent computational resources.

We also provide an illustrative comparison of the gate cost scaling for generating our distributed quantum solution with that of generating the conventional quantum solution that directly encodes the solution vector in the amplitudes of a quantum state. 
Using the optimal quantum linear-system solver, generating the conventional quantum solution requires $\order{\kappa\log(1/\varepsilon)}$ queries to a block encoding of the discretized LPB operator.
For the 3D problem, the condition number of the discretized operator scales as $\kappa\in\Theta(N^2)$. Combining this with $\Theta(n^3\log n)$ gate cost we established in~\cref{subsubsec:operator_oracle_implementation} for implementing the block encoding of the discretized LPB operator, and the $\Theta(n\log n)$ gate cost for loading the initial state in~\cref{subsec:initial_state}, we obtain a total gate cost
\begin{equation}
\label{eq:cost_conventional_qsolution}
    \gate_\text{conv}\in
    \order{N^2n^3\log n\log(1/\varepsilon)}
\end{equation}
for generating the conventional quantum solution.

The comparison in~\cref{fig:solution_cost} illustrates the improved scaling of our approach with the grid size for three values of~$\varepsilon$.
Although the plots provide only illustrative estimates based on scaling (i.e., constant factors are not considered), they demonstrate a substantial reduction in the scaling of computational cost with grid size for solving the LPB equation using our quantum algorithm. The comparison also highlights the cost of extracting properties of the quantum solution as a key factor in the overall cost of the end-to-end quantum algorithm.

\begin{table}[htb]
\centering
\renewcommand{\arraystretch}{1.26}
\setlength{\tabcolsep}{2.8pt}
  \begin{tabular}{cccccc}
    \toprule
    &\multicolumn{3}{c}{Ancilla count}\\
    \cmidrule{2-4}
    Approach & Scratchpad & Block encodings &Property estimation &Gate cost per run & Number of runs\\
    \midrule
    Hadamard test
    &$\Theta(n\log(\frac n\varepsilon))$
    &$\Theta(n+\log(\frac n\varepsilon))$
    &1
    &$\order{n^4\log^4\left(\frac{n}{\varepsilon}\right)}$
    &$\ceil*{\frac2{\varepsilon^2}\ln\left(\frac2\delta\right)}$\\
    Phase estimation
    &$\Theta(n\log(\frac n\varepsilon))$
    &$\Theta(n+\log(\frac n\varepsilon))$
    &$\Theta(\log(\frac{1}{\varepsilon}))$
    & $\order{n^4\log^4\left(\frac{n}{\varepsilon}\right)} + \Theta\left(\frac{n}{\varepsilon}\right)$
    &$\ceil*{\frac8{\varepsilon^2}\ln\left(\frac2\delta\right)}$\\
    \multirow{2}{9em}{Amplitude estimation}
    &$\Theta(n\log(\frac n\varepsilon))$
    &$\Theta(n+\log(\frac n\varepsilon))$
    &$\Theta(\log(\frac{1}{\varepsilon}))$
    &$\order{\frac{n^4}{\varepsilon}\log^4\left(\frac{n}{\varepsilon}\right)}$
    &$\ceil*{\frac{\ln(1/\delta)}{2\left(\frac8{\pi^2}-\frac12\right)^2}}$\\
    &$\Theta(n\log(\frac n\varepsilon))$
    &$\Theta(n+\log(\frac n\varepsilon))$
    &$\Theta(\log(\frac{1}{\varepsilon\delta}))$
    &$\order{\frac{n^4}{\varepsilon \delta}\log^4\left(\frac{n}{\varepsilon}\right)}$
    &1\\
    \bottomrule
  \end{tabular}
  \caption{\label{tab:cost_tailored_algorithm}
  The gate cost and ancilla count of our end-to-end quantum algorithm for simulating the biomolecular system based on three different approaches for estimating a property of the quantum solution with accuracy~$\varepsilon$ and success probability at least~$1-\delta$.
  Scratchpad qubits are ancillary qubits that start from and return to zero. The ancilla count for block encodings is the sum of ancilla qubits used in block encodings of the PDE operator and observable.
  }
\end{table}

\section{Discussion}
\label{sec:dicuttion}
We have constructed an explicit end-to-end quantum algorithm for the linearized Poisson-Boltzmann equation used in simulating biomolecular systems. Our construction provides tailored and efficient implementations for all steps of an end-to-end quantum algorithm, from preparing an appropriate initial state to generating a quantum encoding of the solution and estimating relevant properties of the system.
Our work demonstrates how to overcome challenges of translating abstract quantum algorithms for differential equations into practical applications using a classical specification of the problem rather than abstract oracle inputs.
Moreover, in constructing the algorithm, we have developed and implemented several algorithmic primitives for an end-to-end quantum algorithm that are applicable beyond the specific application considered in this work.

We developed a new discrete representation for differential operators with variable coefficients, which enables efficient construction of the operator oracles, particularly in more than one spatial dimension.
We also utilized the multiscale structure of the wavelet-transformation matrix to directly implement a block encoding of the preconditioned discretized operator with a logarithmic normalization.
This allows us to construct a concrete realization of the algorithm in~\cite{BNW+26} based on wavelet preconditioning for generating a distributed quantum solution of the LPB equation with a cost independent of the condition number of the discretized PDE operator.
Indeed, with our construction of the block encoding of the preconditioned discretized operator, the algorithm can be realized for PDEs whose operators have the form of the Sturm-Liouville operator with variable and bounded coefficients.

We have also rigorously analyzed the gate cost of the complete end-to-end algorithm based on three approaches for property estimation. 
Our cost analysis for these approaches reveals an important tradeoff between the size of quantum circuits for implementing the algorithm and the overall computational cost.
The approach based on Hadamard test offers simpler circuits with a polylogarithmic size in~$1/\varepsilon$ and fewer ancilla qubits, whereas the approach based on amplitude estimation achieves a quadratic improvement in the dependence of the overall gate cost on~$1/\varepsilon$, at the expense of larger circuits. Phase estimation offers no improvement in the overall gate cost.
These tradeoffs are particularly relevant for practical quantum computing, where circuit size, available qubits and the number of circuit runs impose different resource constraints.
The preferred estimation method can therefore depend on the capabilities of the quantum hardware, even though all three approaches have the favorable polylogarithmic dependence on the number of grid points~$N$.

Importantly, we have also accounted for the cost of classical preprocessing needed for constructing the quantum circuits for implementing the algorithm. For a fixed number of atoms, we showed that the cost of classical preprocessing is at most polylogarithmic in both~$N$ and~$1/\varepsilon$, and preserves the favorable scaling of the gate cost of the quantum algorithm.
In our cost analysis, we treated~$N$ and~$\varepsilon$ as independent parameters, motivated by the fact that the spatial discretization of the biomolecular system is constrained by the molecular geometry, particularly the atom positions and van~der~Waals radii. The gate cost we report is therefore the cost of the quantum algorithm for solving the discretized problem.

The polylogarithmic dependence of the overall cost on~$N$ offers an exponential improvement in scaling with the grid size compared with classical algorithms that explicitly solve the discretized LPB equation, whose computational costs scale at least linearly with the total number of grid points~$N_g=N^3$.
However, the extent of practical quantum advantage also depends on the desired accuracy, the molecular parameters, and other constants involved in the circuit constructions.
In particular, although the algorithm based on amplitude estimation has the best overall scaling in~$\varepsilon$ among the other approaches considered, it has a larger circuit size that may constrain practical implementation. Our analysis therefore highlights the importance of distinguishing the size of quantum circuits for implementation from the overall cost of the algorithm when assessing the advantages of quantum algorithms.

While our work provides explicit circuits and rigorous resource bounds, further steps are needed for hardware implementation. In particular, obtaining concrete resource estimates for representative molecular systems will help determine the problem sizes and accuracy regimes for which the algorithm can become practically competitive.
Overall, our results provide a basis for evaluating the potential for practical quantum advantage and lay the groundwork for future implementations on quantum hardware.
Alongside recent developments in end-to-end quantum algorithms for other PDEs and physical applications~\cite{BJL+26,CJK+26}, our work contributes to bridging the gap between theoretical quantum algorithms for differential equations and their practical applications.
Extending the techniques developed here to other applications and more general variable-coefficient PDEs is another important direction for future research.

\section{Acknowledgments}
M.B. thanks Kouhei Nakaji for insightful discussions.
M.B. and A.A.G. acknowledge the generous support and funding of this project by the Defense Advanced Research Projects Agency (DARPA) under Contract No. HR00112330021. Any opinions, findings and conclusions or recommendations expressed in this material are those of the author(s) and do not necessarily reflect the views of the Defense Advanced Research Projects Agency.
A.A.G. also acknowledges support from the Canada 150 Research Chairs program and NSERC-IRC. AAG also acknowledges the generous support of Anders G. Frøseth.
G.K.B. received support
from the Australian Research Council Discovery Project
DP260104888.
\paragraph{AI usage disclosure.}
The idea and research direction of this project were conceived and developed by the authors.
The original writing of the manuscript, as well as the design of the figures and circuits, was carried out by the authors, mainly by the first author.
AI tools were used to polish the writing, find counterexamples to ideas of the authors, particularly regarding the structure of matrices in wavelet bases utilized for quantum implementations, and assist with the literature search for details of the biomolecular system and the ranges of values for its relevant parameters. The authors take full responsibility for the manuscript.

\appendix
\section{Proofs}
\subsection{Bound for systematic error in estimating a property of the quantum solution by QPE}
\label{subsec:systematic_error}
Here we prove the error bound in~\cref{eq:systematic_error} for the systematic error in estimating
\begin{equation}
    \mu= \bra{\phi} U_M\ket{\phi}
    = \sum_j \abs{\alpha_j}^2\e^{2\pi\i\theta_j}
    = \sum_j \abs{\alpha_j}^2 \cos(2\pi\theta_j)
\end{equation}
by quantum phase estimation, where ~$\ket{\phi}=\sum_j\alpha_j\ket{v_j}$ is the solution state.
Recall that the outcome of measuring the $t$-qubit phase register is a random $t$-bit number~$B$, from which we have the random variable $X:=\cos(2\pi B/2^t)$ for estimating~$\mu$.
Let $\mathbb{E}(X|j)$ be the average value obtained from all possible outcomes of the phase register conditioned on the target register being in the eigenstate~$\ket{v_j}$.
We show that $\mathbb{E}(X|j)$ exponentially concentrates around $\cos(2\pi \theta_j)$ as
\begin{equation}
\label{eq:concentration}
    \mathbb{E}(X|j) = \left(1-2^{-t}\right) \cos(2\pi \theta_j) + 2^{-t}\cos\left(2\pi\left(2^t-1\right)\theta_j\right).
\end{equation}
By this equation, the systematic error due to the $t$-bit approximation for each phase is bounded as
\begin{equation}
    \abs{\mathbb{E}(X|j)-\cos(2\pi\theta_j)} \leq 2^{1-t}.
\end{equation}
As this bound is independent of~$j$, it gives $\abs{\mathbb{E}(X)-\mu}\leq 2^{1-t}$ as
\begin{equation}
    \abs{\mathbb{E}(X)-\mu} =
    \abs{\sum_j \abs{\alpha_j}^2 \left(\mathbb{E}(X|j)-\cos(2\pi\theta_j)\right)}
    \leq \sum_j \abs{\alpha_j}^2 2^{1-t} = 2^{1-t}.
\end{equation}

To show the concentration in \cref{eq:concentration}, we first consider the average $\mathbb{E}(\exp(2\pi \i B/2^t)|j)$ and then take its real part. Note that the joint state of the phase and target registers before the inverse QFT is $\sum_j\alpha_j\ket{\phi_j}\ket{v_j}$, where
\begin{equation}
    \ket{\phi_j} := \sum_{k=0}^{2^t-1} a_k \ket{k} \text{ with } a_k:=\frac1{\sqrt{2^t}}\e^{2\pi \i k \theta_j}
\end{equation}
is the state of the phase register if the target register is in the eigenstate~$\ket{v_j}$.
After the inverse QFT
\begin{equation}
    \ket{\phi_j}\mapsto\ket{\psi_j}:=F^\dagger_{2^t}\ket{\phi_j},
\end{equation}
where
\begin{equation}
    F_{2^t}\ket{k} =  \frac1{\sqrt{2^t}}\sum_{b=0}^{2^t-1} \e^{2\pi\i kb/2^t} \ket{b}
\end{equation}
is the $2^t$-dimensional quantum Fourier transform.
Measuring the phase qubits gives the random variable~$B$, and the probability of outcome $B=b$ conditioned on target register being in the eigenstate~$\ket{v_j}$ is
\begin{equation}
    \Pr\left(B=b|j\right) = \abs{\braket{b|\psi_j}}^2 = 
    \braket{\psi_j|b}\braket{b|\psi_j}.
\end{equation}
We therefore have
\begin{align}
    \mathbb{E}\left(\e^{2\pi\i B/2^t}|j\right)
    &= \sum_{b=0}^{2^t-1} \e^{2\pi \i b/2^t} \braket{\psi_j|b}\braket{b|\psi_j}\\
    &= \braket{\psi_j|D|\psi_j} = \braket{\phi_j|F_{2^t}DF^\dagger_{2^t}|\phi_j}
\end{align}
where $D:=\sum_b \exp(2\pi \i b/2^t)\ket{b}\!\!\bra{b}$ is a diagonal unitary.
Then it simply follows that
\begin{equation}
    F_{2^t}DF^\dagger_{2^t} \ket{k}=\ket{k-1\bmod2^t},
\end{equation}
i.e., $F_{2^t}DF^\dagger_{2^t}$ acts as a cyclic shift operator, and so we have
\begin{align}
    \braket{\Phi_j|F_{2^t}DF^\dagger_{2^t}|\Phi_j}
    &= \sum_{k=0}^{2^t-2}a^*_ka_{k+1} + a^*_{2^t-1}a_0\\
    &= \frac1{2^t}\sum_{k=0}^{2^t-2} \e^{2\pi\i\theta_j}
    + \frac1{2^t} \e^{-2\pi\i(2^t-1)\theta_j}\\
    &=(1-2^{-t})\,\e^{2\pi\i\theta_j} + 2^{-t} \e^{-2\pi\i(2^t-1)\theta_j}.
\end{align}
Taking the real part gives~\cref{eq:concentration}.

\subsection{Combined success probability of the lower and upper phase estimates}
\label{subsec:combined_success}
Here we show the success probability in one run of the QPE in the amplitude-estimation approach~(\cref{subsec:AEA}) for computing solution properties can be increased from~$4/\pi^2$ to~$8/\pi^2$ if we increase the maximum error in the phase estimate from~$1/2^{t+1}$ to~$1/2^t$.
To this end, let $\phi\in\{\theta,1-\theta\}\in[0,1]$ denote either of the two eigenphases of the rotation operator~$S$, with the corresponding eigenvalue~$\exp(2\pi\i\phi)$.
Let us also define
\begin{equation}
    \phi_-:=\frac{\floor{2^t\phi}}{2^t},\enspace
    \phi_+:=\frac{\ceil{2^t\phi}}{2^t}
\end{equation}
as the lower and upper $t$-bit estimates of~$\phi$.
When $2^t\phi$ is not an integer, we have $\phi\in(\phi_-,\phi_+)$ and $\phi_+-\phi_-=1/2^t$.

The outcome of measuring the $t$-qubit phase register is a random variable~$B\in\{0,\ldots,2^{t-1}\}$.
If the measurement outcomes is $B=\floor{2^t\phi}$ or $B=\ceil{2^t\phi}$, the measured phase estimate~$\tilde{\phi}=B/2^t$ equals to~$\phi_-$ or~$\phi_+$, respectively.
That is to day, $\phi_\pm$ are two particular values that the phase estimate~$\tilde{\phi}$ can take. 
The probability of the outcome~$B=b$ is
\begin{equation}
    \Pr(B=b) =
    \abs{\frac1{2^{t}}
    \sum_{k=0}^{2^t-1} \e^{2\pi\i k(\phi-b/2^t)}
    }^2
    = \frac1{2^{2t}} \abs{
    \frac{1-\e^{2\pi\i(2^t\phi-b)}}{1-\e^{2\pi\i(\phi-b/2^t)}}
    }^2 =
    \frac{\sin^2(\pi(2^t\phi-b))}{2^{2t}\sin^2(\pi(2^t\phi-b)/2^t)},
\end{equation}
where we used the geometric series sum in the second identity and $|1-\exp(2\i x)|=2|\sin(x)|$ in the last identity.

If we consider the nearest integer~$b=\lfloor2^t\phi\rceil$ to~$2^t\phi$, as conventional in QPE, we obtain an estimate of the phase with error at most~$1/2^{t+1}$ and the success probability at least~$4/\pi^2$.
To see this, let $\Delta:=2^t\phi-\lfloor2^t\phi\rceil$ and note that $\abs{\Delta}\leq1/2$ as $\lfloor2^t\phi\rceil$ is the nearest integer.
Then the error in the phase estimate $\tilde{\phi}=\lfloor2^t\phi\rceil/2^t$ is
\begin{equation}
    \abs{\tilde{\phi}-\phi} = \frac1{2^t}\abs{\Delta}\leq \frac1{2^{t+1}},
\end{equation}
and the success probability, i.e., the probability that $|\tilde{\phi}-\phi|\leq 1/2^{t+1}$, is
\begin{equation}
    \Pr\left(\abs{\tilde{\phi}-\phi}\leq\frac1{2^{t+1}}\right)=
    \Pr(B=\lfloor2^t\phi\rceil) =
    \frac{\sin^2(\pi\Delta)}{2^{2t}\sin^2(\pi\Delta/2^t)}
    \geq \frac{\sin^2(\pi\Delta)}{\pi^2\Delta^2} \geq \frac4{\pi^2},
\end{equation}
where we used $|\sin x| \leq |x|$ in the first inequality and $|\sin(\pi\Delta)|/|\Delta|\geq 2$ for $\abs{\Delta}\leq1/2$ in the second inequality.

We can increase the success probability if we allow the phase error to be at most~$2^{-t}$.
In this case, both the lower and upper phase estimates~$\phi_\pm$ become successful phase estimates as $\abs{\phi_\pm-\phi}\leq2^{-t}$, so the success probability is the probability that 
the measured phase estimate~$\tilde{\phi}$ is either~$\phi_-$ or~$\phi_+$, which correspond to measurement outcomes $B=\floor{2^t\phi}$ or $B=\ceil{2^t\phi}$, respectively.
Specifically, we have
\begin{equation}
    \Pr\left(\abs{\tilde{\phi}-\phi}\leq\frac1{2^t}\right)=
    \Pr(B=\floor{2^t\phi}) + \Pr(B=\ceil{2^t\phi}).
\end{equation}
Now let $x:=2^t\phi-\floor{2^t\phi}\in[0,1]$, then $2^t\phi-\ceil{2^t\phi}=x-1$, and the combined success probability of obtaining the lower and upper phase estimates is
\begin{equation}
    \Pr(B=\floor{2^t\phi}) + \Pr(B=\ceil{2^t\phi})
    \geq  \frac{\sin^2(\pi x)}{\pi^2 x^2} +
    \frac{\sin^2(\pi (x-1))}{\pi^2 (x-1)^2}
    =\frac{\sin^2(\pi x)}{\pi^2} \left(\frac1{x^2}+\frac1{(x-1)^2}\right).
\end{equation}
The minimum of the right-hand side is at~$x=1/2$, so the success probability becomes at least~$8/\pi^2$.

\subsection{Implementing the multi-controlled reflection operator with one dirty ancilla}
\label{subsec:controlled_reflection}
Let~$R_{n}:=\id_n-2\ket{0^n}\!\!\bra{0^n}$ be the reflection operator about the $n$-qubit zero state~$\ket{0^n}$ up to the global phase~$-1$, and let
\begin{equation}
    C^k(R_{n}):=
    \ket{1^k}\!\!\bra{1^k}\otimes R_{n} +
    (\id_k-\ket{1^k}\!\!\bra{1^k})\otimes \id_n
\end{equation}
be the $k$-controlled reflection about~$\ket{0^n}$ of the the~$n$-qubit target register.
Note that~$R_{0^n}$ only applies the phase~$-1$ to~$\ket{0^n}$ and leaves all other basis states unchanged.
A simple implementation for this multi-controlled reflection is then as follows.

We first map $\ket{0^n}\to\ket{1^n}$ by applying $X$ gate on each qubit of the target register.
Then we apply the $m$-controlled~$Z$ gate denoted by~$C^m(Z)$, with~$m=k+n-1$, where the target is the last qubit of the target register and the~$m$ controls are qubits of the~$k$ qubits of the control register and the first $n-1$ qubits of the target register.
Finally, we reapply the~$X$ gates.
Using $Z=HXH$, we replace~$C^m(Z)$ with~$C^m(X)$ by applying the Hadamard gate~$H$ before and after~$C^m(X)$ on the last qubit of the target register.
Implementing the $k$-controlled reflection is now reduced to implementing the $m$-controlled~NOT gate $C^m(X)$. By~\cref{lem:multi_controlled_not}, $C^m(X)$ can be implemented with $\Theta(k+n)$ gates from $\{X,\mathrm{CNOT}, \mathrm{Toffoli}\}$ and $\Theta(\log (k+n))$ circuit depth, using one dirty ancilla qubit.

For the generalized reflection $R_{n}(\theta):=\id_n-(1-\e^{\i\theta})\ket{0^n}\!\!\bra{0^n}$, we replace $C^k(R_{n})$ with $C^k(R_{n}(\theta))$ in the above equation. In this case, we need to apply the phase gate $P(\theta)=\operatorname{diag(1,\e^{\i\theta})}$. If $\theta=\pi$, we have $P(\theta)=Z$, and we can replace $Z$ with $HXH$, but this trick does not work for $P(\theta)$ with geneal phase $\theta$. In this case, we simply use a clean ancilla qubit $f$ to implement the $k$-controlled generalized reflection.
To this end, after applying $X$ gate on each qubit of the target register, we simply $C^{k+n}(X)$ with the clean ancilla~$f$ as target. We then apply the phase gate $P(\theta)$ on the ancilla and then erase the ancilla by $C^{k+n}(X)$. Finally, we reapply the $X$ gates. Notice that for $\theta=\pi$, this approach implements the ordinary reflection, but using a clean ancilla.
The generalized version then can be implemented with $\Theta(k+n)$ gates from $\{X,\mathrm{CNOT}, \mathrm{Toffoli}\}$ and $\Theta(\log (k+n))$ circuit depth, using one dirty ancilla and one clean ancilla.

\subsection{Uniform bound on the condition number of the preconditioned matrix}
Here we analytically prove a uniform upper bound on the condition number of~$A_P$. 
To this end, recall that
\begin{equation}
    A = \sum_{\nu=0}^{d-1} D^T_\nu E_\nu D_\nu + K,
\end{equation}
where the coefficient matrices~$E_\nu$ and $K$ are bounded as
\begin{equation}
   \varepsilon_\text{min}\id \preceq E_\nu \preceq \varepsilon_\text{max}\id,
   \enspace
   0\preceq K\prec k_\text{max}\id,
\end{equation}
as per~\cref{eq:epsKappa} with 
$(\varepsilon_\text{min},\varepsilon_\text{max})=(\varepsilon_\text{cav},\varepsilon_\text{out})$ and $k_\text{max}=\varepsilon_\text{out}\kappa^2_\text{out}$; here~$\preceq$ denotes the Loewner order for Hermitian matrices.
We then have
    \begin{equation}
    \varepsilon_\text{min}\Laplacian
    \preceq A\preceq
    \varepsilon_\text{max}\Laplacian
    +k_\text{max}\id,
\end{equation}
where $\Laplacian =\sum_\nu D^T_\nu D_\nu$ is the Laplacian matrix that represents the discretized Laplacian operator on the gird with periodic boundaries.
By the lower bound $A\succeq \varepsilon_\text{min} \Laplacian$, we obtain
\begin{equation}
    A_P=PWAW^\dagger P
    \succeq \varepsilon_\text{min} \Laplacian_P
    \rightarrow \lambda_\text{min}(A_P)
    \geq \varepsilon_\text{min} \lambda_\text{min}(\Laplacian_P),
\end{equation}
where $\Laplacian_P$ is the preconditioned Laplacian matrix.
Moreover, by the upper bound on~$A$ and by $P^2\preceq \id$, we obtain
\begin{equation}
    A_P \preceq \varepsilon_{\max}\Laplacian_P
    + k_{\max}\id
    \rightarrow
    \lambda_\text{max}(A_P)
    \leq \varepsilon_\text{max} \lambda_\text{max}(\Laplacian_P) +k_\text{max}.
\end{equation}
We therefore have
\begin{equation}
    \kappa(A_P)=\frac{\lambda_\text{max}(A_P)}{\lambda_\text{min}(A_P)}
    \leq \frac{\varepsilon_\text{max}}{\varepsilon_\text{min}}\kappa(\Laplacian_P) + 
    \frac{k_\text{max}}{8\varepsilon_\text{min}}L^2
\end{equation}
where $L$ is the side length of the computational domain~$\Omega=(0,L)^3$ for the Poisson-Boltzmann equation. Here we used the fact that $\lambda_\text{min}(\Laplacian_P)\geq 8/L^2$.
This provides a uniform bound for $A_P$, independent of the biomolecule used.

\subsection{Proof of lemmas for block encoding of the preconditioned matrix}
\label{apx:lemmas_prex_mtx}

\paragraph{Proof of \cref{lem:nonzero_entries}.}
We first show that the first column of $D_s=(S_{s+1}-\id_{s+1}) G^\dagger_s$ has at most $M+1$ nonzero elements, where $M$ is the order of the wavelet used.
Specifically, let us write $D_s\ket{0^s}=\sum_{r=0}^{2^{s+1}-1} d_r \ket{r}$, then we show the coefficient $d_r\neq0$ only for $0\leq r\leq M$.
To this end, note that
\begin{equation}
    G^\dagger_s\ket{0^{s}} =\sum_{\ell=0}^{M-1} g_\ell \ket{\ell \bmod 2^{s+1}},
\end{equation}
where $g_\ell=(-1)^\ell h_{M-1-\ell}$ and $h_\ell$ are the wavelet filter coefficients.
By this equation, we obtain
\begin{equation}
    D_s\ket{0^{s}}=
    (S_{s+1}-\id_{s+1}) G^\dagger_s
    \ket{0^{s}}
    = \sum_{\ell=0}^{M-1} g_\ell (\ket{\ell+1 \bmod 2^{s+1}}-
    \ket{\ell \bmod 2^{s+1}}).
\end{equation}
If $2^{s+1}>M$, then $D_s\ket{0^{s}}=\sum_\ell g_\ell(\ket{\ell+1}-\ket{\ell})=\sum_r d_r\ket{r}$, where $d_r:=g_{r-1}-g_r$ with $g_\ell\neq0$ only for $0\leq\ell<M$.
The only nonzero $d_r$ is then when $0\leq r\leq M$.
If $2^{s+1}\leq M$, we obtain that $d_r:=\bar{g}_{(r-1)\bmod 2^{s+1}}-\bar{g}_r$, where $\bar{g}_r:=\sum_{k\geq0} g_{r+k2^{s+1}}$.
In this case, $d_r$ is nonzero for $0\leq r< 2^{s+1}\leq M$.
Therefore, $d_r\neq0$ only for $r$ that is at most~$M$.

Now suppose the vector of modified filters is $\bm{h}'=(h'_0,\ldots,h'_{M'-1})$, where $M'=M-1$. Then after one refinement (one application of $H'^\dagger$), an initial nonzero coefficient $d_r$ at index $r$ can affect coefficients with index of the form $m=2r+\ell_1$ with $0\leq\ell_1<M'$. With two refinements, any nonzero coefficient has an index of the form $m=2(2r+\ell_1)+\ell_2$.
In general, after $n_s$ refinements, any nonzero coefficient has an index of the form $m=2^{n_s}r+S$, where $S:=\sum_{j=1}^{n_s}2^{n_s-j}\ell_j$ with $0\leq \ell_j<M'$.

Now let us express $m$ as $m=2^{n_s}c+b$ with $0\leq b<2^{n_s}$. We then have
\begin{equation}
    c = \floor*{\frac{m}{2^{n_s}}}
    = r + \floor*{\frac{S}{2^{n_s}}}
    \leq r+ M'-2,
\end{equation}
where we used $S\leq (M'-1)(2^{n_s}-1)$, and so $\floor{S/2^{n_s}}\leq M'-1$. With $r\leq M$ and $M'=M-1$, we obtain $0\leq c\leq c_\text{max}$ with $c_\text{max}:=2M-3$ a constant that depends only on the wavelet used.

\paragraph{Proof of \cref{lem:bounded_coefficient}.}
The coefficient $a_{s,m}$ is the $m$th entry in the first column of $A_s= 2^{n-s} (S_n-\id_n) W^\dagger_s$. Specifically, we have $a_{s,m}=\langle m |A_s|0\rangle$. Because $A_s$ is a real matrix, $a_{s,m}$ is identical to the first $m$th entry in the first row of $A^\dagger_s$, i.e.,
\begin{equation}
    a_{s,m}=\langle 0 |A^\dagger_s|m\rangle
    = 2^{n-s}
    \left(
    \langle 0 |W_s|m-1\rangle -
    \langle 0 |W_s|m\rangle
    \right),
\end{equation}
where we used $S^\dagger_n\ket{m}=\ket{m-1}$.
The matrix $W_s=G_s H_{s+1}\cdots H_{n-1}$ is the row block~$s$ of the $n$-level wavelet transform matrix~$W$ and its entries are the overlap between the scale-$s$ wavelet functions and the scale-$n$ scaling functions.
More specifically, Let $\phi(x)$ and $\psi(x)$ be the father scaling function and the mother wavelet function, respectively.
Then the $(k,m)$ entry of~$W_s$ is 
\begin{equation}
    [W_s]_{k,m} =
    \langle\psi_{s,k}(x) |\phi_{n,m}(x)\rangle,
\end{equation}
where $f_{s,k}(x)=\sqrt{2^s}f(2^sx-k)$ with $f\in\{\phi,\psi\}$.
Let $w_{s,m}:=[W_s]_{0,m}=
\langle 0 |W_s|m\rangle$ be the $m$th entry in the first row.
We then have
\begin{align}
    w_{s,m}
    &=\langle\psi_{s,0}(x) |\psi_{n,m}(x)\rangle\nonumber\\
    &= \sqrt{2^{s+n}}
    \int\psi(2^sx) \phi(2^nx-m) \,\mathrm{d} x\nonumber\\
    &=\frac{1}{\sqrt{2^{n-s}}}
    \int\psi\left(\frac{u+m}{2^{n-s}}\right) \phi(u) \,\mathrm{d}u
\end{align}
where we used $u=2^nx-m$.
If the wavelet function $\psi(x)$ is Lipchitz continuous with the Lipchitz constant $L_\psi$, i.e., if $\abs{\psi(x)-\psi(y)}\leq L_\psi\abs{x-y}$, then we have
\begin{equation}
    \abs{w_{s,m-1} - w_{s,m}} \leq \frac{1}{\sqrt{2^{n-s}}} \frac{L_\psi}{2^{n-s}}
    \int \abs{\phi(u)} \,\mathrm{d}u.
\end{equation}
The 1-norm $\norm{\phi}_1=\int\abs{\phi(u)}\,\mathrm{d}u$ of the scaling function is a constant.
Let us define the constant $c_w:=\norm{\phi}_1L_\psi$, then
\begin{equation}
    \abs{a_{s,m}} = 2^{n-s} \abs{w_{s,m-1}-w_{s,m}}\leq
    \frac{c_w}{\sqrt{2^{n-s}}},
\end{equation}
and the lemma follows.

\paragraph{Proof of \cref{lem:As_prime_expression}.}
We now prove
\begin{align}
    A'_s:=& 2^{3(n-s)/2} (S_n-\id_n) H^\dagger_{n-1}\cdots H^\dagger_{s+1} G^\dagger_s\nonumber\\
    =& \sqrt{8} H'^{\dagger}_{n-1}
    \cdots H'^{\dagger}_{s+1} (S_{s+1}-\id_{s+1}) G^\dagger_s,
\end{align}
where
$H'_r$ is the modified refinement matrix of size $2^r\times 2^{r+1}$ with entries $[H'_r]_{ij}=h'_{(j-2i)\bmod 2^{r+1}}$, and where
\begin{equation}
\label{eq:modified_filter_coeffs}
    h'_\ell= \sqrt{8}\sum_{k=0}^{\ell} (-1)^{\ell-k}h_k.
\end{equation} 
are the modified filter coefficients obtained from the original filter coefficients $\bm{h}=(h_0,\ldots,h_{M-1})$.
To this end, we show
\begin{equation}
    \sqrt{8} (S_{r+1}-\id_{r+1})H^\dagger_r
    = H'^{\dagger}_r
    (S_r-\id_r),
\end{equation}
which immediately gives the equivalent expression for $A'_s$.
We prove this equation by the action of both side on basis vectors.
Note that the action of $H^\dagger_r$ on basis vector $\ket{i}$ is
\begin{equation}
    H^\dagger_r\ket{i} = \sum_jh_{(j-2i)\bmod 2^{r+1}}\ket{j} = 
    \sum_{\ell} h_\ell \ket{2i+\ell \bmod 2^{r+1}},
\end{equation}
and similar formula holds for $H^{'\dagger}_r$, where $h_\ell$ is replaced with $h'_\ell$.
We then have
\begin{align}
    \sqrt{8} (S_{r+1}-\id_{r+1})
    H^\dagger_r\ket{i}
    &= \sum_\ell \sqrt{8}(h_{\ell-1}-h_\ell)\ket{2i+\ell \bmod 2^{r+1}},\\
    H^{'\dagger}(S_r-\id_r)\ket{i}&=
    \sum_\ell (h'_{\ell-2}-h'_\ell)\ket{2i+\ell\bmod2^{r+1}},
\end{align}
so the two sides are equal if $h'_{\ell-2}-h'_\ell = \sqrt{8}(h_{\ell-1}-h_\ell)$.
Because $h_\ell=0$ for $\ell \notin\{0,\ldots,M-1\}$, we obtain \cref{eq:modified_filter_coeffs} by recursive analysis.
We note that the original coefficients satisfy $\sum_{k=0}^{M-1}(-1)^k h_k=0$, so $h'_{M-1}=0$ and therefore $h'_\ell=0$ for all $\ell>M$. The vector $\bm{h}'=(h'_0,\ldots,h'_{M-2})$ of the modified coefficients then has $M-1$ nonzero elements.

\subsection{Error in block encoding of the preconditioned matrix}
\label{apx:error_for_Aprec}
Here we first show $\varepsilon_\precond\in\order{n^22^{-p}}$ by a direct error analysis and then obtain a sufficient size for the number of fractional bits $p$ in terms of $n$ and $\varepsilon$.
For simplicity, we consider $A_\precond =
    Q^T E Q + K_\precond$ in 1D as the result holds for 3D case.
The error $\varepsilon_\precond$ in block encoding of $A_\precond$ is caused by the imperfect block encoding~$U_Q$ of~$Q$ with error $\varepsilon_Q$.
That is, $U_Q$ block encodes an approximation~$\wt{Q}$ of~$Q$ with error~$\varepsilon_Q$. Let us write $\wt{Q}=Q+\Delta $, then $\norm{\Delta}\leq \varepsilon_Q$ and
\begin{equation}
    \wt{A}_\precond - A_\precond = 
    \wt{Q}^T E \wt{Q} - Q^T E Q = \Delta^TEQ +Q^TE\Delta + \Delta^T E \Delta.
\end{equation}
Because $\norm{A_\precond}\in\Theta(1)$ and $\norm{E}\in\Theta(1)$, we have $\norm{Q}\in\Theta(1)$, and so 
\begin{equation}
    \varepsilon_\precond \leq 2\norm{Q}\norm{E}\varepsilon_Q +\norm{E}\varepsilon^2_Q \in\order{\varepsilon_Q}.
\end{equation}

We obtain the block encoding of $Q=\sum_{s=0}^{n-1}Q_s$ using a direct block encoding of each column block $Q_s$ by its action on basis states.
To this end, we first computed the coefficients $a'_{s,m}$ and then performed amplitude transduction using inequality test.
The coefficient $a'_{s,m}$ is computed by loading an initial vector and performing $n_s=n-s-1$ refinements with each refinement using $\Theta(1)$ fixed-point operations.
As each entries of the initial vector has $p$ fractional bits, the error in loading the initial vector is also at most $\order{2^{-p}}$. The error for each refinement is at most $\order{2^{-p}}$ as it uses a constant number of fixed-point operations. The amplitude transduction gives the amplitude $\alpha_{s,m}=a'_{s,m}/c_w$ with error $\order{2^{-p}}$.
The total error in block encoding of $Q_s$ is then $\varepsilon_{Q_s}\in\order{(n_s+2)2^{-p}}$. We then obtain $\varepsilon_Q\leq \sum_s \varepsilon_{Q_s}\in\order{n^22^{-p}}$, and so $\varepsilon_\precond\in\order{n^22^{-p}}$.

To obtain a sufficient size for the number of fractional bits $p$, note that in \cref{subsec:generating_distributed_solution} we imposed the error $\varepsilon_\precond$ to satisfy $4d\sqrt{\varepsilon_\precond/\alpha_\precond}\leq \varepsilon'$.
 First we use the relations 
 \begin{equation}
     \norm{A_\precond}\varepsilon_\text{inv}\leq \frac{\varepsilon}{4},
     \enspace
     \varepsilon_\text{inv}=\frac{8\kappa_\precond\varepsilon'}{3\norm{A_\precond}},
     \enspace
     d\in\order{n\log(1/\varepsilon)}
 \end{equation}
established in \cref{subsec:generating_distributed_solution} to establish a bound for $\varepsilon_\precond$ in terms of $n$ and $\varepsilon$.
By this equation and $\kappa_\precond\in\Theta(1)$, we have
\begin{equation}
    \varepsilon'= \frac{3\norm{A_\precond}\varepsilon_\text{inv}}{8\kappa_\precond}\leq
    \frac{3\varepsilon}{32\kappa_\precond}\in\Theta(\varepsilon),
\end{equation}
so by $4d\sqrt{\varepsilon_\precond/\alpha_\precond}\leq \varepsilon'$, we obtain
\begin{equation}
    \varepsilon_\precond \leq \frac{\alpha_\precond}{16d^2}\varepsilon'^2
    \in \Theta(n\varepsilon^2)/d^2,
\end{equation}
where we used $\alpha_\precond\in\Theta(n)$.
As $d\in\order{n\log(1/\varepsilon)}$, it is then sufficient to choose 
\begin{equation}
    \varepsilon_\precond\in\order{\frac{\varepsilon^2}{n\log^2(1/\varepsilon)}}.
\end{equation}
We had $\varepsilon_\precond\in\order{n^22^{-p}}$ by the direct error analysis.
It is therefore sufficient to choose
\begin{equation}
    \frac{n^2}{2^p} \lesssim \frac{\varepsilon^2}{n\log^2(1/\varepsilon)},
\end{equation}
where $\lesssim$ suppresses constant factors, from which we obtain
\begin{equation}
    p\gtrsim \log \frac{n^3\log^2(1/\varepsilon)}{\varepsilon^2}
    = 3\log n + 2\log(1/\varepsilon)+ 2\log\log(1/\varepsilon)
    \in\Omega(\log(n/\varepsilon)).
\end{equation}
Note that we are free to choose any $p$ that satisfies
$p\in\Omega(\log(n/\varepsilon))$. It is then  sufficient to choose $p\in\Theta(\log(n/\varepsilon))$.

\newpage
\addcontentsline{toc}{section}{References}
\bibliographystyle{apsrev4-2}
\bibliography{ref.bib}
\end{document}